%% file: paper.tex
\documentclass[acmsmall]{acmart}

\usepackage{graphicx}
\usepackage{multirow}
\usepackage{tikz}
\usepackage[ruled,vlined,linesnumbered,noend]{algorithm2e}
\usepackage[labelfont=bf,textfont=bf]{caption}
\usepackage[labelfont=bf,textfont=bf,singlelinecheck=off,justification=raggedright]{subcaption}
\usepackage{amsmath,amsthm,bm}
\usepackage{color}
\definecolor{myred}{RGB}{189, 52, 67}
\definecolor{mygreen}{RGB}{19, 136, 8}
\definecolor{myblue}{RGB}{16, 52, 166}

\newcommand{\karima}[1]{{\color{black} #1}\normalcolor}
\newcommand{\ilias}[1]{{\color{mygreen} #1}\normalcolor}

\AtBeginDocument{%
  }

\newtheorem{defn}{Definition}

\graphicspath{{./img-png/}}

\setcopyright{authorsopen}
\acmYear{2025} \acmVolume{3} \acmNumber{1 (SIGMOD)} \acmArticle{43} \acmMonth{2}\acmDOI{10.1145/3709693}

\begin{document}
	\title{Graph-Based Vector Search: An Experimental Evaluation of the State-of-the-Art}
	
	\author{Ilias Azizi}
		\affiliation{
		\institution{UM6P, Universit{\'e} Paris Cit{\'e}}
		\country{Morocco - France}
	}
	\email{ilias.azizi@um6p.ma}
	\author{Karima Echihabi}
		\affiliation{%
		\institution{UM6P}
		\country{Morocco}
	}
	\email{karima.echihabi@um6p.ma}
	\author{Themis Palpanas}
		\affiliation{%
		\institution{Universit{\'e} Paris Cit{\'e}}
		\country{France}
	}
	\email{themis@mi.parisdescartes.fr}
\renewcommand{\shortauthors}{Ilias Azizi, Karima Echihabi \& Themis Palpanas.}

\pagenumbering{arabic}

\begin{abstract}
Vector data is prevalent across business and scientific applications, and its popularity is growing with the proliferation of learned embeddings. Vector data collections often reach billions of vectors with thousands of dimensions, thus, increasing the complexity of their analysis. Vector search is the backbone of many critical analytical tasks, and graph-based methods have become the best choice for analytical tasks that do not require guarantees on the quality of the answers. We briefly survey in-memory graph-based vector search, outline the chronology of the different methods and classify them according to five main design paradigms: seed selection, incremental insertion, neighborhood propagation, neighborhood diversification, and divide-and-conquer. We conduct an exhaustive experimental evaluation of twelve state-of-the-art methods on seven real data collections, with sizes up to 1 billion vectors. We share key insights about the strengths and limitations of these methods; e.g., the best approaches are typically based on incremental insertion and neighborhood diversification, and the choice of the base graph can hurt scalability. Finally, we discuss open research directions, such as the importance of devising more sophisticated data-adaptive seed selection and diversification strategies.

\end{abstract}

\ccsdesc[500]{Theory of computation~Graph algorithms analysis}
\ccsdesc[500]{Theory of computation~Nearest neighbor algorithms}
\ccsdesc[500]{Theory of computation~Data structures design and analysis}
\ccsdesc[500]{Information systems~Information retrieval}
\ccsdesc[300]{General and reference~Evaluation}
\ccsdesc[300]{General and reference~Experimentation}

\keywords{Vector similarity search, Approximate nearest neighbor, KNN graph analysis, Seed selection, Neighborhood diversification, Graph algorithms}

\received{July 2024}
\received[revised]{September 2024}
\received[accepted]{November 2024}

\maketitle
\input{src/introduction}
\input{src/preliminaries}

\input{src/survey}
\input{src/experiments}

\input{src/discussion}

\input{src/conclusions}

\begin{acks}
This work used the UM6P African Supercomputing Center (ASCC).
Work partially supported by EU Horizon projects AI4Europe (101070000), TwinODIS (101160009), ARMADA (101168951), DataGEMS (101188416) and RECITALS (101168490).
\end{acks}
	
\bibliographystyle{ACM-Reference-Format}
\bibliography{src/ref,src/pargis,src/icde-tutorial,src/parisinmemory}

\end{document}

%% file: src/introduction.tex
\section{Introduction}
\label{sec:introduction}
 Vector data is common in various scientific and business domains, and its prevalence is expected to grow in the future with the proliferation of learned embeddings~\cite{conf/icde/echihabi2021,palpanas2015data}. The volume and dimensionality of this data, which can exceed multiple terabytes and thousands of dimensions, make its analysis very challenging~\cite{palpanas2015data}. A critical component of these data analysis tasks is vector search ~\cite{conf/icde/echihabi2021,palpanas2015data,conf/sigmod/echihabi2020,DBLP:conf/wims/EchihabiZP20}. It supports recommendation~\cite{conf/kdd/wang2018,amazon}, information retrieval~\cite{conf/williams2014}, clustering~\cite{journal/JMLR/bubeck2009,journal/pattrecog/Warren2005},
classification~\cite{DBLP:conf/icdm/PetitjeanFWNCK14} and outlier detection~\cite{discord,norma,series2graph,landmines,nba} in many fields including bioinformatics, computer vision, security, finance and medicine. 
In data integration, vector search is used for entity resolution~\cite{journal/pvldb/ebraheem2018}, missing values imputation~\cite{retro}, 
and data discovery~\cite{journal/pvldb/zhu2016}. 
Besides, it is exploited in software engineering~\cite{journal/pacml/uri2019,conf/icsec/nguyen2016} 
to automate API mappings and in cybersecurity to profile network usage and detect malware~\cite{cybersecurity}. More recently, vector search has been playing a crucial role in improving the performance and interpretability of Large Language Models and reducing their hallucinations~\cite{retrieval-diffusion-models,dense-passage-retrieval,seq2seq,rag-nlp}. 

A vector search algorithm over a dataset ${S}$ of $n$ $d$-dimensional vectors returns answers in ${S}$ that are similar to a given input vector $V_Q$. 
The brute-force approach (a.k.a. a sequential or serial scan) compares $V_Q$ to every single element in ${S}$. The time complexity of this approach is $O\left(nd\right)$, which becomes impractical when dealing with large collections of data with high dimensionality. 
State-of-the-art approaches typically improve this time complexity by reducing the dimensionality $d$ using concise and precise summarization techniques and/or decreasing the number of processed vectors $n$, by developing efficient indexing data structures and search algorithms that prune unnecessary comparisons to $V_Q$. 
Some approaches compute exact answers, while others $\epsilon$- and $\delta$-$\epsilon$-approximate answers (with deterministic and probabilistic guarantees, respectively, on the accuracy of the answers), or ng-approximate answers (without any theoretical guarantees, but high accuracy in practice)~\cite{hydra1,lernaeanhydra2}. 
One popular vector search algorithm is k-nearest neighbor (k-NN) search, which returns the $k$ vectors in ${S}$ closest to $V_Q$ according to a similarity measure, such as the Euclidean distance. 

The efficiency of exact vector search has significantly improved over the last decade~\cite{hydra1, isaxfamily, dumpy, sing, oddysey}. 
However, exact techniques still do not address the query latency constraints of many analytical tasks. Therefore, 
a large body of work has been dedicated to approximate vector search, which trades off accuracy for efficiency~\cite{lernaeanhydra2,dpg,DBLP:journals/debu/00070P023}. These approaches are based on scans~\cite{vafile,va+file}, trees~\cite{conf/vldb/Wang2013,isax2+,parisplus,coconut,localpairbundle,dpisax,ulisse,messi,flann,hdindex,seanet,wang2024dumpyos,leafi}, graphs~\cite{hnsw,nsg,vamana,hcnng, efanna,SPTAG2}, inverted indexes~\cite{gist,conf/icassp/jegou2011,journal/iccv/xia2013,journal/pami/babenko15}, hashing~\cite{conf/stoc/indyk1998,conf/vldb/sun14,srs}, or a combination of these data structures~\cite{ieh,efanna,SPTAG4,hcnng,elpis,lshapg,DBLP:journals/pvldb/WeiPLP24}. Over the last decade, graph-based techniques have emerged as the method of choice for vector search in many real applications that can relax theoretical guarantees to achieve a query latency of a few milliseconds on terabyte-scale collections~\cite{graphrec2,graphrec1,alibabaknngml,faiss}. 

\begin{figure}[htb] 
		\includegraphics[width=0.8\columnwidth]{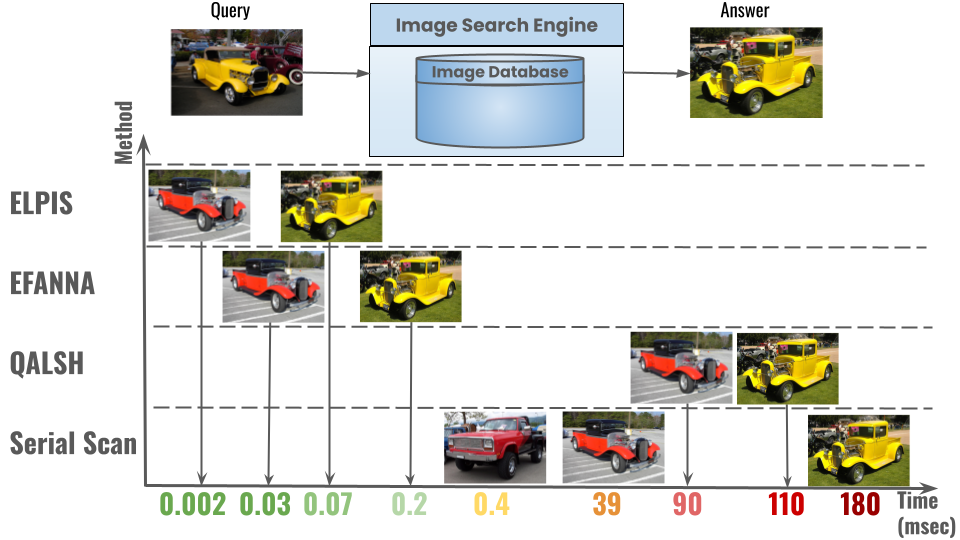}
		\caption{
  Image retrieval using different vector search methods. Each row shows the bsf answer returned by each method (y-axis) at a given timestamp (x-axis). The graph-based approach ELPIS: (i) returns the same answer as the serial scan and QALSH over three orders of magnitude faster; and
  (ii) is 3x faster than the graph-based method EFANNA. 
  }        
		\label{fig:use_case}
 \end{figure}

Figure~\ref{fig:use_case} illustrates vector search in an image retrieval use case. We produce embeddings for ImageNet~\cite{imagenet} using a ResNet50 model~\cite{resnet}. 
We report the time at which a best-so-far (bsf) answer is found (i.e., the image in the database most similar to the query) by vector search techniques from different families: ELPIS~\cite{elpis} and EFANNA~\cite{efanna} for ng-approximate search, QALSH~\cite{qalsh} for $\delta$-$\epsilon$-approximate search, and a serial scan for exact search. We observe that: (i) ELPIS returns the same answer as the serial scan and QALSH over three orders of magnitude faster, which explains the popularity of graph-based vector search in many real applications; and (ii) not all graph-based methods have the same performance, e.g. ELPIS is 3x faster than EFANNA, which motivates this experimental study to help the community better understand the strengths and limitations of the existing graph-based vector search techniques.

Graph-based methods typically structure a dataset ${S}$ as a proximity graph $G\left({V},{E}\right)$ where ${V}$ is the set of $n$ vertices representing the $n$ data points in ${S}$, and ${E}$ is the set of edges that connect similar vertices.
Answering a query $V_Q$ usually consists of running an ng-approximate beam search algorithm that warms up the results queue with an initial list of candidate nodes in $G$ known as seed nodes, selects one of them as an entry node, i.e., the node that will be traversed first, and performs a greedy graph traversal following the edges of the traversed nodes to return all nodes that are similar to $V_Q$. State-of-the-Art (SotA) graph-based vector search methods, such as HNSW~\cite{hnsw}, NSG~\cite{nsg}, ELPIS~\cite{elpis}, Vamana~\cite{vamana}, and NSSG~\cite{nssg} adopt the same beam search algorithm to retrieve ng-approximate answers from the graph. 
However, they differ 
in how they construct the graph and initiate the beam search.

A number of studies have evaluated vector search methods. Some focus on exact search~\cite{hydra1} while others cover approximate search with and without guarantees~\cite{journal/tkde/li19, conf/sisap/martin17, journal/pvld/naidan2015}. 
Given the increased interest in graph-based vector search, a recent experimental evaluation~\cite{graph-survey-vldb} has been dedicated entirely to this family of techniques. However, the results of this study are inconclusive, because the experimental evaluation was conducted on very small datasets, not exceeding 1M vectors (we will show in Section~\ref{sec:experiments} that trends change as dataset size increases, e.g. some methods that are efficient on 1M datasets do not perform well on large collections and vice-versa).  
Besides, our study proposes a new taxonomy that classifies SotA graph-based vector search methods according to five key design paradigms. It also highlights the impact of these design choices on the indexing and search performance, and how this impact changes with dataset size. 
In parallel, some benchmarks~\cite{aumuller2017ann,neurips-2021-ann-competition} also evaluate vector search methods, but they do not focus on graph-based approaches, and thus, do not shed light on why the best methods have superior query performance. In contrast, our study covers real datasets reaching one billion vectors from different domains such as deep network embeddings, computer vision, neuroscience, and seismology, includes more recent approaches, and uncovers interesting insights that have not been reported before, such as the impact of seed selection and neighborhood diversification on indexing and search performance. 

In this paper, we make the following contributions.

\noindent$\bullet$ We identify five main paradigms exploited by state-of-the-art vector search methods: 
(1) Seed Selection (SS) determines the node(s) in the graph where the search initiates. 
(2) Neighborhood Propagation (NP) approximates the k-NN graph, where each node is connected to its exact k nearest neighbors, by propagating neighborhood lists between connected nodes.
(3) Incremental Insertion (II) builds a proximity graph incrementally connecting vertices with short links to their closest neighbors and long links that model small-world phenomena. 
(4) Neighborhood Diversification (ND) sparsifies the graph by pruning edges that lead to redundant directions at each node.
(5) Divide-and-Conquer (DC) splits a large dataset into partitions and builds individual graphs separately. 
We provide a brief overview of these five paradigms, and study in depth the two key paradigms that have a strong impact on query performance (as shown in Section~\ref{sec:experiments}), namely SS and ND. 

\noindent$\bullet$ We propose a new taxonomy that categorizes graph-based vector search methods according to the five paradigms described earlier, and highlights the chronological development of the different techniques and their influence map. 

\noindent$\bullet$ We provide a brief survey of the SotA graph-based vector search methods. We describe their design principles and pinpoint their strengths and limitations.

\noindent$\bullet$ We conduct an exhaustive experimental evaluation of twelve SotA methods using synthetic and real-world datasets from neuroscience, computer vision and seismology, with sizes reaching up to 1 billion vectors. We provide a rich discussion of the results. 
We confirm some conventional wisdom, e.g., methods that construct graphs through incremental insertion exhibit superior query performance and scalability compared to other existing methods~\cite{elpis,graph-survey-vldb}. We also explain results that differ from the literature, in particular with respect to the most recent study on the same topic~\cite{graph-survey-vldb}. For instance, in the recent study, SPTAG-BKT~\cite{SPTAG4} and HNSW~\cite{hnsw} rank low in terms of query and indexing efficiency, respectively. 
However, our experiments show that SPTAG-BKT is among the most efficient techniques in query answering on small-sized datasets (1M-25GB), and that many methods lag behind HNSW in terms of indexing time, including on small datasets. 
In addition, contrary to previous works~\cite{elpis}, we demonstrate that Vamana~\cite{vamana} is also competitive with ELPIS and HNSW~\cite{elpis,hnsw}.

\noindent$\bullet$ We share new insights about vector search that have never been published before.
For instance, (i) we show that not all ND techniques are equally effective, and we identify the one that is most effective for large dataset sizes; and (ii) we illustrate how different SS strategies affect query answering and index building (some methods perform a beam search during graph construction).

\noindent$\bullet$ Based on the results of the comprehensive experimental study, we pinpoint some promising research directions in this field: 
(i) novel graph structures that can scale to large data collections should be proposed for methods based on NP and ND strategies;
(ii) more effective and data-adaptive seed selection strategies should be developed to further improve indexing and search performance on large-scale datasets; 
(iii) a deeper theoretical understanding of ND approaches could pave the way for the development of enhanced methods that can adapt to data distributions; and 
(iv) devising techniques (e.g., clustering, neighborhood diversification) tailored to DC-based methods has the potential to improve their performance. 

\noindent{\bf Scope}: The goal of this study is to conduct an experimental evaluation of the SotA graph-based methods designed for in-memory dense vector search. A theoretical analysis of the indexing and search algorithms is an open research question (no existing graph-based vector search method provides this analysis), and is out-of-scope. The extension to the out-of-core scenario and the theoretical analysis are both part of our future work. 

%% file: src/preliminaries.tex
\section{Preliminaries}
\label{sec:preliminaries}

\label{subsec:preliminaries-definitions}
The $ng$-approximate vector search problem is typically modeled as an $ng$-approximate $k$-NN search problem in high-dimensional vector space. 
Data points are represented as $d$-dimensional vectors in ${R}^d$, and the \textit{dissimilarity} between the points is measured using the Euclidean distance \textit{dist}. We consider a dataset collection 
$S$ = $\{V_{C_1}, V_{C_2}, \ldots, V_{C_n}\}$
 of $n$ $d$-dimensional points and a query vector $V_Q$.

\begin{defn} \label{def:knnquery}
    Given a positive integer $k$, an \textbf{exact k-NN query} over a dataset $S$ retrieves the set of vectors $A$ = $\{ \{V_{C_1}, \ldots, V_{C_k}\} \subseteq S$ \text{|} 
    $\forall V_C \in A $
    and 
    $\forall V_{C'} \notin A$, 
    $\text{dist}\left(V_Q, V_C\right) \leq \text{dist}\left(V_Q, V_{C'}\right) \}$
    ~\cite{lernaeanhydra2}.
\end{defn}

\begin{defn} \label{def:appmatch} (An equivalent definition has appeared in~\cite{lernaeanhydra2}.)
	Given a positive integer $k$, an \textbf{ng-approximate k-NN query} over a dataset ${S}$ retrieves the set ${A}$ = $ \{V_{C_1},...,V_{C_k}\} \subseteq {S}$ in a heuristic manner, i.e., there are no theoretical guarantees about the quality of the answers. 
\end{defn}

We use the terms vector search, similarity search, $ng$-approximate search
, and approximate search 
interchangeably.

\subsection{Approximate Vector Search Methods}
Vector search has been heavily studied  for over fifty years. 
Proposed approaches are either based on scans, trees, graphs, hashing, inverted indexes, or on hybrid designs combining two or more of these data structures. 
Vector search methods often rely on summarization to reduce the complexity of the search. In this section, we provide an overview of the main summarization techniques exploited by state-of-the-art approaches, then describe the fundamental concepts of each family, discussing its strengths and limitations.

\noindent{\bf{Summarization Techniques}} \textit{Random projections} 
project the original high-dimensional vector into a lower dimensional space using a random matrix. The pairwise distances are guaranteed to be nearly preserved with high probability if the dimension of the projected space is large enough~\cite{conf/map/johnson84}. \textit{Quantization} compresses a vector of infinite values to a finite set of codewords that constitute the codebook. \emph{Scalar} quantization maps each vector dimension independently, whereas \emph{vector} quantization maps the whole vector at once exploiting inter-dimension correlations~\cite{journal/tit/gray1998}.  \emph{Product} quantization~\cite{gist} divides the vector and operates a vector quantizer on the subvectors. {\it Optimized Product Quantization} (OPQ) optimizes the space decomposition of a product quantizer by decorrelating the dimensions. 
The \textit{Karhunen-Lo\`{e}ve Transform (KLT)}~\cite{karhunen1947ueber,loeve1948functions} performs a linear transformation that adapts to a given signal to decorrelate the dimensions, then applies scalar quantization. Recent work has shown that some summarizations that were developed for data series are equally efficient for generic vectors~\cite{lernaeanhydra2,hercules}. 
In particular, 
\textit{Piecewise Aggregate Approximation} (PAA)~\cite{journal/kais/Keogh2001} and {\it Adaptive Piecewise Constant Approximation} (APCA)~\cite{journal/acds/Chakrabarti2002} segment a vector into equal or variable length segments, respectively, and summarize each segment with its mean value. 
{\it Extended APCA} (EAPCA)~\cite{conf/vldb/Wang2013} improves upon APCA by using both the mean and standard deviation to summarize each segment. {\it Symbolic Aggregate Approximation} (SAX)~\cite{conf/dmkd/LinKLC03} transforms a vector using PAA and discretizes the values using a scalar quantizer. 
{\it Deep Embedding Approximation} (DEA) learns summarizations using the SEAnet deep neural network~\cite{seanetjournal}.

\noindent{\bf{Tree-based indexes}} organize the data using a tree structure and have been the method of choice for exact vector search for data series and generic high-dimensional vectors~\cite{conf/sigmod/Guttman1984,conf/icmd/Beckmann1990,journal/edbt/Schafer2012,conf/vldb/Wang2013,zoumpatianos2016ads,coconutjournal,ulissejournal,twinsubsequences,dpisaxjournal,localgeolocatetedsimilarity,parisplus,messijournal,sing,phd-workshop-karima,seanetjournal,dumpy,oddysey,fresh}. 
Some works have proposed using tree indexes to support approximate search with~\cite{mtree-pac,lernaeanhydra2} or without guarantees~\cite{flann,hdindex,lernaeanhydra2,dumpy}. 
The tree-based methods that support ng-approximate search either build multiple randomized K-D Trees that are searched in parallel 
~\cite{flann}, segment the space into smaller dimensions indexed by an RDB tree~\cite{hdindex}, or use heuristics to select candidates from some leaf node(s)~\cite{lernaeanhydra2,dumpy}. 
The ng-approximate search accuracy/efficiency tradeoff is tuned with user-defined parameters specific to each method, e.g., specifying the maximum number of leaves to visit during search.

\noindent{\bf{Scan-based approaches}} have been designed primarily for exact search. Sequential scans compare the query to each single candidate in a dataset. 
Skip-sequential scans summarize the high-dimensional data in a filter file that fits in memory and only search the original data, typically stored out-of-core, if a candidate cannot be pruned using the filter file. 
Some scan-based methods can also support ANN search with~\cite{lernaeanhydra2} or without guarantees~\cite{lernaeanhydra2,vafile,va+file}. These methods tune the ng-approximate search accuracy/efficiency tradeoff with a user-defined parameter that specifies the maximum number of vectors to be visited. 

\noindent{\bf{Inverted indexes}} typically compress the original high-dimensional vector data using quantization to reduce the space overhead and improve efficiency. 
Popular methods such as the Inverted Multi-Index (IMI)~\cite{journal/pami/babenko15}, and the IVF-PQ~\cite{gist} organize the data by associating lists of data points (a.k.a. posting lists) with their proximity to a codeword. 
These codewords are representative points, e.g. the medoid/centroid of the posting list, derived after clustering the original data, collectively forming a codebook. The
ng-approximate search algorithm accesses the codebook to retrieve points linked to the closest codeword(s).
The accuracy/efficiency tradeoff is tuned with a user-defined parameter that specifies the maximum number of posting lists to be explored during the search. 
This tradeoff is also affected by the index building parameters such as the codebook size and the quantization bit budget. 

\noindent{\bf{Hash-based approaches}} belong primarily to the family of locality-sensitive hashing (LSH)~\cite{conf/stoc/indyk1998,lsh-survey}. 
They are designed to support $\delta$-$\epsilon$-approximate vector search~\cite{lernaeanhydra2} by exploiting hash functions that group with high-probability similar data points into the same bucket and non-similar data points into different buckets. 
Numerous LSH variants exist, each with its unique strengths. 
Among these, the SRS~\cite{srs} and QALSH~\cite{qalsh} methods stand out. SRS provides efficient search results with a relatively small index size, while QALSH enhances bucketing precision by using the incoming query as a reference point. 
The $\delta$-$\epsilon$-approximate search accuracy/efficiency tradeoff is tuned with user-defined parameters that specify the approximation error and probability thresholds. 
This tradeoff is also affected by the index building parameters, e.g., the number of hash tables and the number of random projections.

\noindent{\bf{Graph-Based approaches}} support $ng$-approximate vector search. They often employ a proximity graph structure~\cite{gabriel69,toussaint02,schvaneveldt1989network}, where each vertex represents a data point, and edges connect similar vertices. Two vertices are linked if the data points they represent are close in their respective spaces, usually determined by a common distance measure such as the Euclidean distance~\cite{edelsbrunner87}. When searching for a specific query point, the process typically begins from a set of initial points or seeds. These seeds can be chosen randomly or based on certain criteria.  The search uses one of these seeds as an entry node and the others as initial candidate answers.
It then progresses by visiting the neighboring vertices of the current node in a best-first, greedy manner, concluding when no better matches are identified~\cite{reddy77bm}. State-of-the-art graph-based methods, including KGraph~\cite{kgraph}, EFANNA~\cite{efanna}, HNSW~\cite{hnsw}, DPG~\cite{dpg}, NSG~\cite{nsg}, SPTAG~\cite{tptree}, SSG~\cite{nssg}, Vamana~\cite{vamana}, HCNNG~\cite{hcnng}, ELPIS~\cite{elpis} and others~\cite{nsw14,ieh,fanng} 
use the same search algorithm~\cite{reddy77bm} but differ in how they construct the graph and select seed points.

\noindent{\bf{Hybrid approaches}} combine ideas from the different families described earlier. IEH~\cite{ieh} retrieves initial neighbors to build the graph using hash-based methods, such as LSH~\cite{iehlsh} and ITQ~\cite{iehitq}. EFANNA~\cite{efanna} builds an initial graph using randomized K-D Trees where nodes are connected to the neighbors returned by K-D Trees, and exploits these trees during search to find initial candidates. 
SPTAG~\cite{SPTAG4} divides the dataset using multiple Trinary-Projection Trees (TP Trees)~\cite{tptree} before merging multiple graphs constructed for each of the subsets, and constructs multiple K-D Trees or BKTrees to retrieve initial seeds. 
HCNNG~\cite{hcnng} divides the dataset using random hierarchical clusterings, and builds multiple Minimum Spanning Tree (MST) structures on the different subsets. 
The graph is built by merging the multiple MSTs. 
HCNNG also exploits K-D Trees to retrieve seeds to initiate the search. 
ELPIS~\cite{elpis} builds a Hercules tree~\cite{hercules} to divide the dataset into subsets, and constructs an HNSW~\cite{hnsw} graph on each subset. 
During search, ELPIS prunes subsets using the EAPCA~\cite{conf/vldb/Wang2013} lower-bounding distance. 

\noindent{\bf Summary.} Tree-based approaches typically demonstrate efficient indexing performance, boasting low index construction time and memory usage~\cite{lernaeanhydra2,elpis,tptree}. 
Many of these methods support various search flavors, from exact to approximate, with some offering guarantees. However, tree-based methods often fall short in search efficiency, especially when faced with challenging query workloads~\cite{lernaeanhydra2}. 
Methods based on inverted indexes are less scalable in terms of indexing time and footprint than tree-based indexes but can answer queries faster. 
However, achieving high query accuracy typically requires a significant indexing time and space overhead~\cite{lernaeanhydra2}. 
Hash-based methods provide theoretical guarantees on query accuracy and are the only SotA approximate methods that support theoretical guarantees on query performance. 
However, building the index requires a significant overhead both in space and time, often requiring different indexes to be pre-built to support different guarantees during search. 
Graph-based approaches, while expensive in terms of indexing time and memory usage, and lacking guarantees on query acccuracy, showcase impressive empirical query accuracy and efficiency~\cite{lernaeanhydra2, elpis, lin19, dpg}.

%% file: src/survey.tex
\section{Graph-Based Vector Search}
\label{sec:survey}
We now present an overview of the main SotA graph-based $ng$-approximate vector search methods. 
We outline the base data structures and algorithms in this field, and identify five main paradigms exploited by the SotA approaches. 
We propose a new taxonomy that categorizes these approaches along the five paradigms, highlighting also their chronological development and influence map. 

\subsection{A Primer}

A proximity graph is a graph $G\left({V},{E}\right)$ in which two vertices $V_i$ and $V_j$ are connected by an edge if and only if they satisfy particular geometric requirements, namely the \textit{neighborhood criterion}~\cite{shamos1975closest}.
A proximity graph can be constructed using different distances such as the dot product~\cite{mipsg}, nevertheless the Euclidean distance remains the most popular one~\cite{edelsbrunner87}. One of the earliest proximity graphs in the literature is the Delaunay Graph (\textit{DG}). It is a planar dual graph for the Voronoi Diagram~\cite{vd95}, where each vertex is the center of its own voronoi cell, and two vertices are linked if and only if their corresponding voronoi cells share at least one edge. A DG 
satisfies the Delaunay Triangulation :$\forall q,p,r \in {V}, \left(q,p\right), \left(q,r\right), \left(r,p\right) \in {E}$
if the circumcircle of the triangle $q, p, r$ is empty~\cite{aurenhammer2013voronoi}.
A beam search~\cite{reddy77bm} (Algorithm~\ref{alg:beamsearch}) on a \textit{DG} can find the exact nearest neighbors~\cite{dobkin1990delaunay} when the dataset has a high dimensionality or the beam search uses a large beam width.

However, using a DG in high dimensions is impractical, because the graph becomes almost fully connected as the dimensionality $d$ grows~\cite{dobkin1990delaunay}. 
Thus, SotA methods build alternative graph structures and use beam search to support efficient query answering
~\cite{gabriel69,matula80,toussaint02}. 

\begin{algorithm}[tb]
{\small
\caption{Beam Search (G, $V_Q$, \textit{s}, \textit{k}, \textit{L})}\label{alg:beamsearch}
\KwIn{Graph \textit{G}, query vector $V_Q$, initial seeds s, result size \textit{k}, beam width $\textit{L}\geq \textit{k}$ }
\KwOut{\textit{k} approximate nearest neighbors to $V_Q$}
 initialize candidate set $\textit{C} \longleftarrow {\textit{s}}$\; \label{alg:beam:line:init}
 initialize visited list $\textit{R} \longleftarrow \emptyset$\; \label{alg:beam:line:init2}
 \While{$\textit{C}\backslash\textit{R} \neq \emptyset$ }{
  let $p^{*} \longleftarrow \operatorname*{argmin}_{V_i \in \textit{C}\backslash\textit{R}} dist\left(V_Q, V_i\right)$\;\label{alg:beam:line:best}

  update $\textit{C} \longleftarrow \textit{C} \cup N_{out}\left(p^*\right)  $\;\label{alg:beam:line:updatecand}
  update $\textit{R} \longleftarrow  \textit{R}\cup p^{*}$\;\label{alg:beam:line:updaterslt}
  \If{$\vert \textit{C} \vert > \textit{L}$}{
   update $\textit{C}$ to retain closest \textit{L} points to $V_Q$\;
   }
 }
 return the $\textit{k}$ candidates in  $\textit{C}$ closest to $V_Q$\;\label{alg:beam:line:return}
 } 
\end{algorithm}

\subsection{Main Paradigms}

We provide a brief overview of the five main paradigms exploited by SotA methods.
Then, we describe in more detail the two  paradigms that have the greatest impact on query performance (as will be demonstrated in Section~\ref{sec:experiments}).

\noindent{\bf{Seed Selection (SS)}} chooses initial nodes to visit during search. It is also used during index building by approaches that exploit a beam search during the construction of the index to decide which edges to build. Some methods simply select one or more seed(s) randomly, while others use special data structures, e.g., a K-D Tree.

\noindent{\bf{Neighborhood Propagation (NP)}} refines a pre-existing graph following a user-defined number of iterations, a.k.a. NNDescent~\cite{nndescent}. 
During each iteration, the potential neighbors of a given node are sourced both from its immediate neighbors and the neighbors of its neighbors. Then, the node only keeps the $m$ closest neighbors, where $m$ is a user-parameter. 
The pre-existing graph could be a random graph 
or some other type of graph.

\noindent{\bf{Incremental Insertion (II)}} refers to building a graph by inserting one vertex at a time. 
Each vertex is connected using bi-directional edges to its nearest neighbors and some distant vertices. The neighbors are selected using a beam search on the already inserted portion of the graph. At the end of graph construction, some vertices retain early connections which act as long-range links. 
This approach was first proposed in the object-based peer-to-peer overlay network VoroNet~\cite{voronet}, with the idea of adding long-range links being inspired from Kleinberg's small-world model~\cite{kleinberg2000, kleinberg2002}, with the difference that the latter selects the long-range links randomly.

\noindent{\bf{Neighborhood Diversification (ND)}} was first introduced by the Relative Neighborhood Graph (RNG)~\cite{rng}. It aims to sparsify the graph by pruning unnecessary edges while preserving connectivity. For each node, ND exploits the geometrical properties of the graph to remove edges to neighbors that lead to redundant regions or directions
This indirectly causes the creation of long-range links allowing nodes to maintain diversified neighborhood lists, which reduces the number of comparisons during search.

\noindent{\bf{Divide-and-Conquer (DC)}} is a strategy that splits a dataset into multiple, possibly overlapping, partitions, then builds a separate graph on each partition. 
Some approaches such as SPTAG~\cite{SPTAG4} and HCNNG~\cite{hcnng} combine the individual graphs into one large graph, on which a beam search is performed, while ELPIS~\cite{elpis} maintains the graphs separate and searches them in parallel.


\subsection{Seed Selection}
\label{sec:ss}
While SotA graph-based vector search methods adopt diverse strategies for constructing the graph, they virtually all use beam search for query answering (Algorithm~\ref{alg:beamsearch}).
because it usually retrieves good answers if the graph is well-connected. However, choosing the right nodes to visit first has an impact on how quickly good answers are found. The longer the graph traversal, the higher the number of visited nodes, and the slower the search.
Several methods build one or more index(es), in addition to the graph, on top of a sample of data points. These additional indexes are used during query answering to find the entry points in the graph.
However, to the best of our knowledge, none of these methods have provided sufficient theoretical or empirical evidence to support their choices for seed selection. In this paper, we conduct an in-depth study of the different seed selection techniques proposed in the literature:

\noindent (1) \textbf{Stacked-NSW (SN)}: Inspired by skip lists~\cite{skiplist}, HNSW~\cite{hnsw} constructs hierarchical multi-resolution graphs~\cite{nsw14} for seed selection. Each level contains a diversified NSW graph built using a sample of nodes from the level below, with the lowest layer sampling from the entire dataset. Stacked NSW is constructed by assigning each node a maximum level L. Nodes with L>0 are incrementally inserted into the graphs from layer L down to 1. 
The maximum level L is: 
\begin{equation}
\hspace{3cm}
L = \frac{-\ln \left(\xi\right)}{\ln\left(M\div2\right)} 
 \hspace{2cm} (\text{Eq. 1 \cite{hnsw} }) \tag*{}
\end{equation}
where $\xi$ is a uniformly random number between 0 and 1, 
and $M$ is the maximum out-degree in the graphs, controlling the probability distribution of a node's maximum layer. 
A high M decreases the number of nodes represented in hierarchical layers as well as the number of layers, while a lower M allocates more nodes to the hierarchical layers, thus generating more hierarchical layers. 
During query answering, it starts a greedy search from a fixed entry point at the top layer, descending layer by layer, each time starting from the closest node found in the previous layer, until reaching the bottom layer. The node selected in the bottom layer and the nodes connected to it serve as seed nodes.

\noindent (2) \textbf{K-D Trees (KD)}: Utilized by EFANNA~\cite{efanna}, SPTAG-KDT~\cite{SPTAG2}, and HCNNG~\cite{hcnng}, this technique involves constructing single or multiple K-D Tree(s)~\cite{kdtree} on a dataset sample. During search, a set of seed points is retrieved by running a depth-first search traversal (DFS) on the K-D Tree structure(s), warming up the set of candidate answers. The node closest to the query is selected as an entry node.

\noindent (3) \textbf{LSH}: Utilized by IEH~\cite{ieh}, this strategy constructs an LSH index
on a sample of the dataset and uses it during search to return a set of seed points, using one as an entry node. 

\noindent (4) \textbf{Medoid (MD)}: 
This strategy fixes the medoid node as entry point during query answering and uses its neighbors as seeds.

\noindent (5) \textbf{Single Fixed Random Entry Point (SF)}: A random node is selected and fixed as the entry point for all searches. This node and the nodes connected to it are used as seeds.

\noindent (6) \textbf{K-Sampled Random Seeds (KS)}: 
For each query, \( k \) random nodes are selected to warm up the set of candidate answers. This approach is supported in DPG~\cite{dpg}, NSG~\cite{nsg}, and Vamana~\cite{vamana} which choose the medoid as the entry point and enhance the list of initial seeds with the random nodes. 

\noindent (7) \textbf{Balanced K-means Trees (KM)}: Utilized by SPTAG-BKT~\cite{SPTAG2}, this technique constructs Balanced K-means Trees (BKT) ~\cite{bkmtree} on a dataset sample. During search, seed points are retrieved via depth-first search (DFS) on the BKT structure(s).

\subsection{Neighborhood Diversification}
\label{sec:nd}
 The goal of Neighborhood Diversification (ND) is to create a sparse graph structure, where nodes are connected both to their close neighbors and to some further nodes. This is because traversing a graph in which nodes are only connected to their close neighbors incurs many unnecessary comparisons before reaching the promising regions. This method first appeared in the Relative Neighborhood Graph (RNG)~\citep{rng,toussaint02}, which builds an undirected graph from scratch or by pruning an existing Delaunay Graph, then removes the longest edge in each triangle of three connected points.
 This approach, and other variations that exploit different geometric properties of the graph, have been adapted for directed graphs by several graph-based ng-approximate vector search methods ~\citep{hnsw,dpg,nsg,nssg,vamana,SPTAG4,elpis}. We identify three main ND strategies used by these methods: Relative Neighborhood Diversification (RND), Relaxed Relative Neighborhood Diversification (RRND) and Maximum-Oriented Neighborhood Diversification (MOND).
Note that the ND strategy is different from Kleinberg's small world network graph~\cite{kleinberg2000, kleinberg2002}. The long-range links in the latter are achieved through a random selection of nodes, while in ND they result indirectly from pruning the close neighbors of each node.
Assume: 

\begin{itemize}
  \setlength\itemindent{1em}  
  \item \(X_q\) is the query node, i.e. the node to be inserted in the graph.
  \item \(R_q\), the list of current closest neighbors to \(X_q\).
  \item \(C_q\), the list of candidate neighbors to \(X_q\) not yet in \(R_q\).
  \item \(X_j\) is a candidate from \(C_q\) considered for inclusion in \(R_q\).
  \item \(X_i\) is a node already in the set \(R_q\).
  \item \(\text{dist} \left(X_i, X_j\right)\) is the Euclidean distance between \(X_i\) and \(X_j\).
\end{itemize}
\begin{defn}[RND]\label{def:rnd}
The node \(X_j\) is added to \(R_q\) if and only if the following condition holds:
	\[
	\forall X_i \in R_q, \, \text{dist}\left(X_q, X_j\right) < \text{dist}\left(X_i, X_j\right) \quad (\text{Eq. 2})
	\]
\end{defn}

\begin{defn}[RRND]\label{def:rrnd}
For a relaxation factor \(\alpha \geq 1\), the node \(X_j\) is added to \(R_q\) if and only if the following condition holds:
	\[
	\forall X_i \in R_q, \, \text{dist}\left(X_q, X_j\right) < \alpha \cdot \text{dist}\left(X_i, X_j\right) \quad (\text{Eq. 3})
	\]
\end{defn}

\begin{defn}[MOND]\label{def:mond}
For an angle \(\theta \geq 60^\circ\), the node \(X_j\) is added to \(R_q\) if and only if the following condition holds:
 	\[
	\forall X_i \in R_q, \, \angle\left(X_j X_q X_i\right) > \theta \quad (\text{Eq. 4})
	\]
\end{defn}

Figure ~\ref{fig:ND:example} shows the results after applying different ND approaches on the candidate neighborhood list \( C_q = \{ X_1, X_2, X_3, X_4 \} \) for node \( X_q \).
RND (Figure~\ref{fig:ND:RND}) was first used in the context of vector search by HNSW~\cite{hnsw}. Other methods adopted RND including NSG~\cite{nsg}, SPTAG~\cite{SPTAG4} and ELPIS~\cite{elpis}.  Per Definition~\ref{def:rnd}, RND operates over the candidate neighborhood list by adding $X_j$ to $R_q$ if its distance to $X_q$ is smaller than the distance between $X_j$ and any neighbors $X_i$ of $X_q$. 
In the example of Figure~\ref{fig:ND:RND}, $X_q$ is going to connect to $X_1$ since it is the closest neighbor ($\sigma = dist\left(X_q,X_1\right)$). Both $X_2$ and $X_3$ are pruned from the $X_q$ neighborhood list as they are closer to $X_1$ than $X_q$, and $X_4$ will be added to $X_q$ eventually since dist($X_q$,$X_4$) $\leq$ dist($X_1$,$X_4$).
In contrast, RRND (Figure~\ref{fig:ND:RRND}), proposed by Vamana~\cite{vamana}, introduces a relaxation factor $\alpha$ (with $\alpha \geq 1.5$) where $X_j$ is added to $R_q$ if its distance to $X_q$ is smaller than $\alpha$ times its distance to $X_i$, a neighbor of $X_q$ in $R_q$, relaxing the property to prune less candidate neighbors (hence $X_2$ will not be pruned in this case, but $X_3$ is pruned due to its proximity to $X_2$). 
When $\alpha = 1$, RRND is reduced to RND.
Finally, MOND (Figure~\ref{fig:ND:MOND}) was proposed by DPG~\cite{dpg} and used in NSSG~\cite{nssg} as well. 
It aims at maximizing angles in the graph topology (cf. Definition~\ref{def:mond}), guiding the selection process via the angle threshold $\theta$ (e.g., $\theta = 60^\circ$). MOND prunes the candidate neighbors of a node to favor edges pointing in different directions ($X_2$ is pruned since $\angle X_1X_qX_2$ < $60^\circ$, while $X_q$ connects with $X_3$ since  $\angle X_1X_qX_3$ > $60^\circ$ and with $X_4$ since  $\angle X_1X_qX_4$ < $60^\circ$ and  $\angle X_3X_qX_4$ < $60^\circ$). 
Note that any nodes pruned by RRND and MOND will eventually be pruned by RND, but not vice versa. 
Refer to~\cite{url/GASS} for a detailed proof.

\newcommand{\subfigwidth}{0.27\columnwidth} 
\newcommand{\dottedheight}{3.5cm} 
\newcommand{\vdottedline}[1]{
	\begin{tikzpicture}
	\draw[dotted] (0,0) -- (0,#1);
	\end{tikzpicture}
}
\begin{figure}[tb]
	\begin{subfigure}[b]{\subfigwidth}
		\centering
		\captionsetup{justification=centering}
		\includegraphics[width=\textwidth]{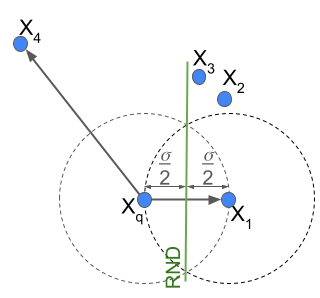}
		\caption{RND}
		\label{fig:ND:RND}
	\end{subfigure}
	\vdottedline{\dottedheight}
	\begin{subfigure}[b]{\subfigwidth}
		\centering
		\captionsetup{justification=centering}
		\includegraphics[width=\textwidth]{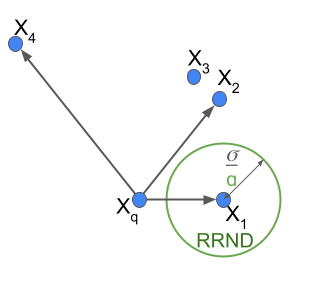}
		\caption{RRND}
		\label{fig:ND:RRND}
	\end{subfigure}
	\vdottedline{\dottedheight}
	\begin{subfigure}[b]{\subfigwidth}
		\centering
		\captionsetup{justification=centering}
		\includegraphics[width=\textwidth]{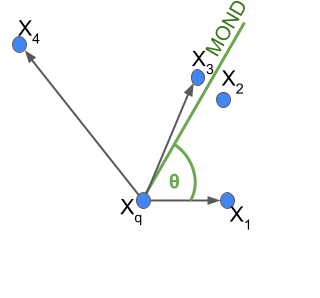}
		\caption{MOND}
		\label{fig:ND:MOND}
	\end{subfigure}
    \caption{Neighborhood diversification approaches}
    \label{fig:ND:example}
\end{figure}

\subsection{A Taxonomy}
\begin{figure}[tb] 
		\captionsetup{justification=centering}
		\includegraphics[width=0.9\columnwidth]{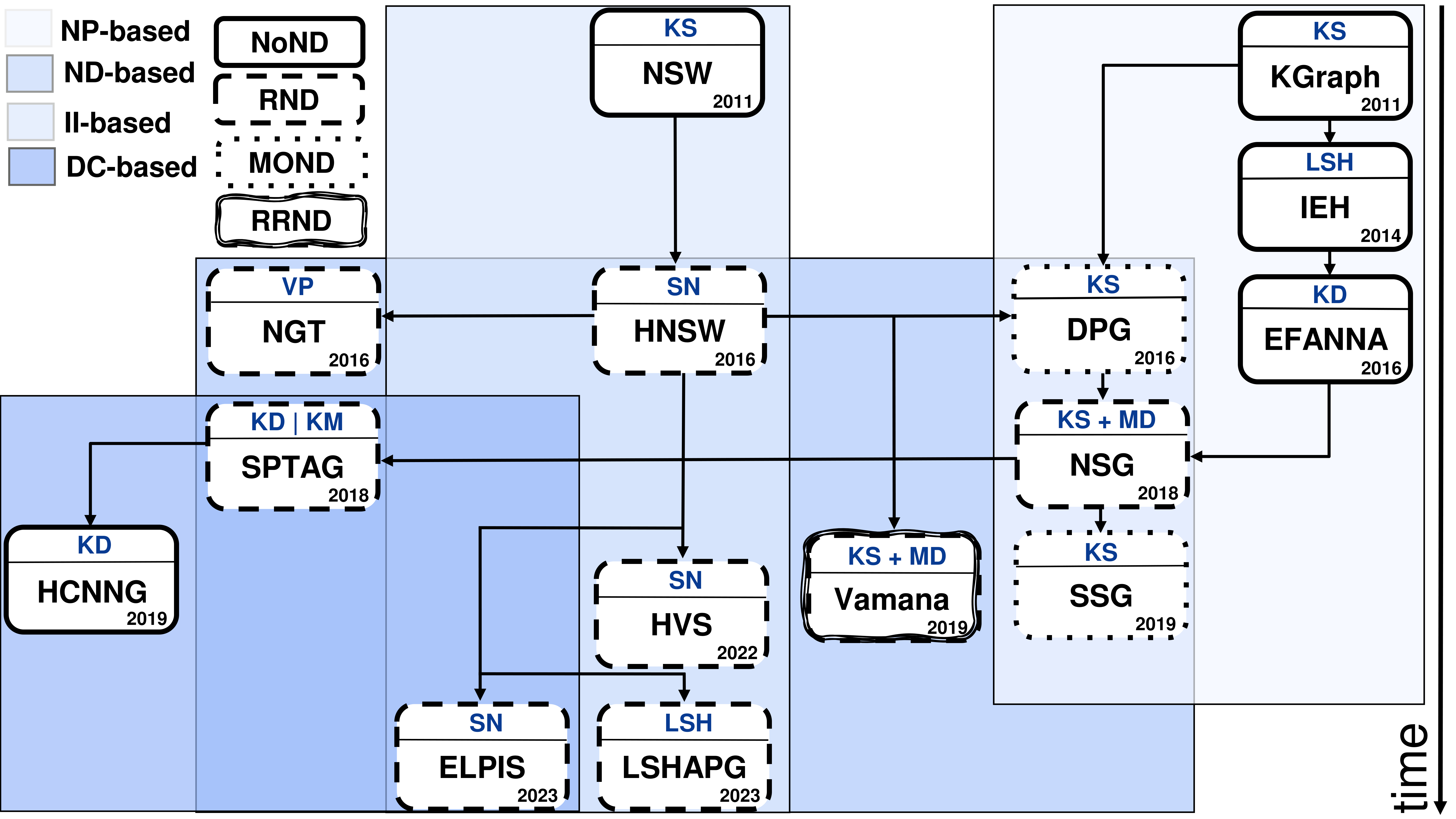}
        \caption{Graph-based ANN indexing paradigms}
		\label{fig:roadmap}
	
 \end{figure}

Figure~\ref{fig:roadmap} depicts the SotA graph-based approaches, classified based on the five design paradigms: SS, NP, II, ND, and DC.
The taxonomy also reflects the chronological development of the methods. 
Directed arrows 
indicate the influence of one method on another. Within the ND category, distinctions are made between different strategies, i.e., No Neighborhood Diversification (NoND), RND, RRND, and MOND (cf. Section~\ref{sec:nd}). 
We identify the SS strategy of each method: KS, KD, SN, MD, LSH, and KM (SF is not used by any SotA method, but we consider it as an alternative strategy). 
Additionally, some methods use more than one strategy (e.g. NSG and VAMANA use KS and MD), or offer the flexibility to use different strategies (e.g., SPTAG can use either KD or KM). 
Note that a method can exploit one or more paradigms; e.g., HNSW uses incremental node insertion and prunes each node's neighbors using the RND approach, thereby being classified as both II and ND. 
KGraph~\cite{kgraph} was the first to use NP to approximate the exact k-NN graph (k-NNG) (with quadratic complexity), and influenced numerous subsequent methods, including IEH~\cite{ieh} and EFANNA~\cite{efanna}. In parallel, NSW~\cite{nsw11} introduced the II strategy for graph construction.
HNSW~\cite{hnsw} and DPG~\cite{dpg} leveraged ND to enhance NSW and KGraph~\cite{kgraph}, respectively.
The good performance of HNSW and DPG encouraged more methods to adopt the ND paradigm, including NGT~\cite{ngt_library}, NSG~\cite{nsg} and SSG~\cite{nssg}, which apply ND on the NP-based graph EFANNA~\cite{efanna}. SPTAG~\cite{SPTAG4} combined DC with ND. 
Vamana~\cite{vamana} adopts NSG's idea of constructing the graph through beam search 
 and ND. 
 However, Vamana constructs its graph by refining an initial base random graph in two rounds of pruning, using RRND and RND. Inspired by HNSW, Vamana and NGT proposed variants 
 that support incremental graph building~\cite{diskanncode, ngt_library}, but we classify them as ND-based per the ideas proposed in the original papers. HCNNG~\cite{hcnng} was influenced by SPTAG~\cite{SPTAG4} and adopted a DC approach for constructing the graph without adopting ND. ELPIS~\cite{elpis} also adopted a DC strategy but leveraged both II and ND. 
HVS~\cite{hvs} and LSHAPG~\cite{lshapg} both propose new seed selection structures for HNSW, with the latter additionally adopting a new probabilistic rooting approach. Note that earlier approaches, except from NSW, were mainly NP-based; however, recent studies have focused on devising methods that leverage the ND, II, and DC paradigms because they lead to superior performance 
(cf. Section~\ref{sec:experiments}).

\subsection{State-of-the-Art Approaches}
\noindent{\bf KGraph}~\cite{kgraph} reduces the construction cost of an exact k-NNG, which has a quadratic worst-case complexity. It constructs an approximate k-NNG by refining a random initial graph with an empirical cost of \( O\left(n^{1.14}\right) \)~\cite{nndescent}. 
This refinement process, also known as NNDescent~\cite{nndescent} (Neighborhood Propagation), aims at improving the approximation of the \( k \)-NN graph by assuming that the neighbors of a vertex \( u \) are more likely to be neighbors of each other. 
The process iterates over all graph vertices \( u \in {V} \): 
for each vertex \( u \) and pair \( (x,y) \) of its neighbors, it adds \( x \) to the neighbors of \( y \) and vice-versa, keeping the closest \( k \) neighbors of \( u \).

 \noindent{\bf Navigable Small World (NSW)}~\cite{nsw11,nsw14} is an approximation of a Delaunay graph which guarantees the small world property~\cite{watts98}, i.e. the number of hops $L$ between two randomly chosen vertices grows to the logarithm of graph size $n$ such that $L \propto Log\left(n\right)$.
An {\it NSW} graph is based on the VoroNet graph~\cite{voronet}, an extention of Kleinberg's variant of Watts-Strogatz's small world model graph~\cite{kleinberg2000,kleinberg2002},  The VoroNet graph is built incrementally by inserting a randomly picked vertex to the graph and connecting it to 2d+1 
neighbors selected using a beam search on the existing vertices in the graph.
Once this process completes, the first built edges would serve as long-range edges to quickly converge toward nearest neighbors~\cite{voronet}. The resulting graph was proved to guarantee the small world network property~\cite{voronet,beaumont07}. 

 \noindent{\bf Iterative Expanding Hashing (IEH)}~\cite{ieh} follows the same process as KGraph to construct an approximate k-NNG; however, it refines an initial graph where the candidates for each node are generated using a hashing function.
Two extensions of IEH have been proposed to better leverage advanced hashing methods for generating initial candidates: IEH-LSH~\cite{iehlsh} and IEH-ITQ~\cite{iehitq}. All these methods use NNDescent to finalize the graph connections.

 \noindent{\bf EFANNA}~\cite{efanna} selects seeds similarly to KGraph~\cite{kgraph} and IEH~\cite{ieh} and refines candidates using NNdescent. It builds an approximate $k$-NNG by selecting initial neighbors of each node using randomized truncated K-D Trees \cite{dasgupta2008random} and refining the graph using NNDescent~\cite{nndescent}. 
During search, EFANNA uses the pre-built trees to select seeds, then
runs a beam search on the graph index.
\noindent{\bf Hierarchical Navigable Small World (HNSW)}~\cite{hnsw} improves the scalability of NSW~\cite{nsw11,nsw14}  by proposing RND to sparsify the graph and a hierarchical seed selection strategy (SN) 
to shorten the search path during index building and query answering. Each hierarchical layer includes all nodes in the layer above it, with the bottom (a.k.a. {\it base}) layer containing all points of the dataset \( {S} \), 
HNSW builds an NSW graph incrementally. However, HNSW diverges from NSW in that it refines the candidate nearest neighbors, identified through beam search on the nodes already in that layer using RND. 
During query answering, HNSW utilizes SN to quickly find an entry point in the base layer to start the beam search.


\noindent{\bf Diversified Proximity Graph (DPG)}~\cite{dpg} 
extends KGraph~\cite{kgraph} by diversifying the neighborhoods of its nodes through edge orientation, a technique we refer to as Maximum-Oriented Neighborhood Diversification (MOND) in Section 3.4.
MOND’s main objective is to maximize the angles between neighboring nodes, contributing to a sparsed graph structure. This process is iteratively applied to all nodes. After that, the directed graph is transformed into an undirected one, enhancing its connectivity. Nevertheless, note that DPG's publicly available implementation~\cite{dpgrepo} utilizes RND rather than MOND for neighborhood diversification.

\noindent{\bf NGT}~\cite{ngt_library} is an approximate nearest neighbor (ANN) search library developed by Yahoo Japan. It offers two construction methods: one extends KNN graphs with reverse edges, forming bi-directed KNN graphs~\cite{ngtpanng1}, while the other incrementally builds graphs similar to HNSW with a range-based search strategy~\cite{ngtpanng2}.  In this study, we consider the former~\cite{ngtpanng1}. Additionally, the library includes methods that employ quantization for highly efficient search.
NGT maintains efficiency by pruning neighbors via RND and using Vantage-Point Trees~\cite{vptree} to select seed nodes for accurate query results.

\noindent{\bf Navigating Spreading-out Graph (NSG)}~\cite{nsg}, similarly to DPG, builds an approximate k-NNG first. But, unlike DPG, it builds an EFANNA graph rather than a KGraph. It then diversifies the graph using RND. 
At the end, NSG creates a depth-first search tree to verify the connectivity of the graph. If there is a vertex that is disconnected from the tree, NSG connects it to the nearest node in the tree to ensure graph connectivity.

\noindent{\bf SPTAG}~\cite{SPTAG4} is a library for approximate vector search proposed by Microsoft. 
SPTAG follows a DC approach and is based on multiple existing works.  
It selects small dataset samples on which it builds either K-D Trees~\cite{kdtree} or Balanced K-means Trees~\cite{bkmtree}. These strutures will be used for seed selection during query answering. Then it clusters the full dataset using multiple hierarchical random divisions of TP Trees~\cite{tptree}, builds an exact k-NN graph on each cluster (i.e., leaf) and refines each graph using ND. The graphs are merged into one large graph index for query processing. 

\noindent{\bf Vamana}~\cite{vamana} is similar to NSG in considering the set of visited nodes when building long-range edges within the graph. However, instead of using EFANNA~\cite{efanna}, Vamana uses a randomly generated graph with node degree~$\geq~log\left(n\right)$ to ensure the initial graph connectivity~\cite{erconnect}. 
Then, for each node, Vamana runs a beam search on the graph structure to get the visited node list $R$, which will be refined in the first round using RRND. After adding bi-directional edges to selected neighbors, the neighbors that exceed the maximum allowed out-degree will refine their neighborhood list following an RND process. Then, Vamana repeats the same refinement process a second time to improve the graph quality, this time using RRND with $\alpha \geq 1$ to increase the connectivity within the graph.

\noindent{\bf SSG}~\cite{nssg} integrates the MOND approach from DPG~\cite{dpg} and closely follows the steps of NSG~\cite{nsg} and DPG~\cite{dpg} in index building from a foundational graph. Instead of performing a search for each node to acquire candidates, SSG~\cite{nssg} employs a breadth-first search on each node to assemble candidate neighbors through local expansion on a base graph (EFANNA). When the maximum size for the candidate neighbors is achieved, SSG reduces the neighbors in the list by enforcing the MOND diversification strategy, pruning the candidate nodes forming an angle smaller than a user-defined parameter $\theta$ with the already existing neighbors of the concerned node. After iteratively applying this method to all nodes, SSG~\cite{nssg} enhances connectivity by constructing multiple DFS trees from various random points, in contrast to NSG's~\cite{nsg} singular DFS approach.

\noindent{\bf Hierarchical Clustering-based Nearest Neighbor Graph (HCNNG)}~\cite{hcnng} was inspired by SPTAG. It employs hierarchical clustering to randomly divide the dataset into multiple subsets. This subdivision process is executed several times, resulting in a collection of intersecting subsets. On each subset, HCNNG constructs a Minimum Spanning Tree (MST) graph. 
Following this, the vertices and edges from all the MSTs are merged to form a single, connected graph. 
To facilitate the search process, HCNNG constructs multiple K-D Trees~\cite{kdtree}, to identifying entry points during query search. 

\noindent{\bf HVS}~\cite{hvs} extends HNSW's base layer by refining the construction of hierarchical layers. Instead of random selection, nodes are assigned to layers based on local density to better capture data distribution. Each layer forms a Voronoi diagram 
and uses multi-level quantization, increasing dimensionality by a factor of 2 in each lower layer.
Search at the base layer is similar to that of HNSW.

\noindent{\bf LSHAPG}~\cite{lshapg} combines HNSW graphs with multiple hash tables based on the LSB-Tree structure~\cite{lsb} to enhance search efficiency. It leverages $L$ hash tables to retrieve seeds for beam search on the base layer, unlike HNSW, which selects a single seed through SN. LSHAPG also utilizes these hash tables for probabilistic rooting during search, pruning neighbors based on the projected distance 
before evaluating and pruning the raw vectors.

\noindent{\bf ELPIS}~\cite{elpis} is a DC-based approach that splits the dataset into subsets using the Hercules EAPCA tree~\cite{hercules}, where each leaf corresponds to a different subset, then builds in parallel a graph-based index for each leaf using HNSW~\cite{hnsw}. During search, ELPIS first selects heuristically an initial leaf 
and executes a beam search on its respective graph. 
The retrieved set of answers feed the search priority queues for the other leaves. 
Only a subset of leaves is selected based on the answers and the lower-bounding distances of the query to the EAPCA summarization of each leaf.  Then, ELPIS initiates multiple concurrent beam searches on the graph structures of the candidate leaves. 
Finally, ELPIS aggregates all results from candidate clusters and returns the top-k answers. 

%% file: src/experiments.tex
\section{Experimental Evaluation}
\label{sec:experiments}
We experimentally evaluate twelve state-of-the-art graph-based vector search methods, 
based on the two key paradigms described in Section~\ref{sec:survey}, i.e., SS and ND. 
To single out the effect of each strategy, we first implement a basic II-based method, where nodes are inserted incrementally and each node $i$ acquires its list of candidate neighbors through a beam search on the current partial graph of already inserted nodes.  Then, we implement each strategy independently on the resulting graph. 
Finally, we assess the indexing and query-answering performance of these methods on a variety of real and 
synthetic datasets.
All artifacts are available in~\cite{url/GASS}.
\subsection{Framework}

\noindent{\bf Setup.} 
Methods were compiled with GCC 8.2.0 under Ubuntu Linux 20.04 (Rocky Linux 8.5 on HPC) using default compilation flags and optimization level 3. Experiments were conducted on an Intel Xeon Platinum 8276 server with 4 sockets, each with 28 cores and 1 thread per core.
The CPU cache configuration is: 32KB L1d, 32KB L1i, 1,024KB L2, and 39,424KB L3 cache. The server includes a 1.5TB RAM via 24x 64GiB DDR4 DIMMs.

\noindent{\bf Algorithms.} We cover the following methods: HNSW~\cite{hnsw}, NSG~\cite{nsg}, Vamana~\cite{vamana}, DPG~\cite{dpg}, EFANNA~\cite{efanna}, HCNNG~\cite{hcnng}, KGraph~\cite{kgraph}, NGT~\cite{ngt_library}, DPG~\cite{dpg}, and two versions of SPTAG~\cite{SPTAG4} (SPTAG-BKT and SPTAG-KDT, using BKT and K-D Trees, respectively). We also include ELPIS~\cite{elpis} and LSHAPG~\cite{lshapg}, new techniques not evaluated in the latest survey~\cite{graph-survey-vldb}. IEH~\cite{ieh} and FANNG~\cite{fanng} are excluded due to suboptimal performance~\cite{graph-survey-vldb, nsg}, and HVS~\cite{hvs} due to difficulties running the official implementation~\cite{hvsgithub}. We use the most efficient publicly available C/C++ implementations for each algorithm, leveraging multithreading and SIMD vectorization to optimize performance.  We also carefully inspected all code bases and, as is common in the literature ~\cite{graph-survey-vldb, diskanncode, nsgcode,ssgcode,ngtcode,sptagcode}, disabled the optimizations that would lead to an unfair evaluation such as cache pre-warming in Vamana and L2-normalized Euclidean distance in NSG, EFANNA, and Vamana. Since all methods except ELPIS and HNSW use a single linear buffer as a priority queue, we modified the original implementations of these two algorithms (which used two max-heap priority queues)~\cite{url/Elpis,url/hnsw}. The modifications to each code base are documented in~\cite{url/GASS}.

\noindent{\bf Datasets.} 
We use seven real-world datasets from various domains, including deep network embeddings, computer vision, neuroscience, and seismology:  
(i) \emph{Deep}~\cite{url/data/deep1b} contains 1 billion 96-dimensional vectors extracted from the final layers of a convolutional neural network;  
(ii) \emph{Sift}~\cite{conf/icassp/jegou2011,url/data/sift} consists of 1 billion 128-dimensional SIFT vectors representing image feature descriptors;  
(iii) \emph{SALD}~\cite{url/data/eeg} provides neuroscience MRI data with 200 million 128-dimensional data series;  
(iv) \emph{Seismic}~\cite{url/data/seismic} contains 100 million 256-dimensional time series representing earthquake recordings from seismic stations worldwide;  
(v) \emph{Text-to-Image}~\cite{url/data/text2image} offers 1 billion 200-dimensional image embeddings from Se-ResNext-101 along with 50 million DSSM-embedded text queries for cross-modal retrieval under domain shifts;  
(vi) \emph{GIST}~\cite{gist} contains 1 million 960-dimensional vectors, using GIST descriptors~\cite{gistdesc} to capture spatial structure and color layout of images;  and
(vii) \emph{ImageNet1M}, a new dataset that we generated from the original ImageNet~\cite{imagenet}, producing embeddings of 1 million original vectors using the ResNet50 model~\cite{resnet}, with PCA applied to reduce dimensionality to 256. We select subsets of different sizes from the Sift, Deep, SALD and Seismic datasets, and we refer to each subset with the name of the dataset followed by the subset size in GBs (e.g., Deep25GB). 
We refer to the 1-million and 1-billion datasets with the 1M and 1B prefixes, respectively. 
To evaluate the methods on datasets with different distributions, we generate three random 25GB datasets RandPow0, 
RandPow5 and RandPow50, each with 256 dimensions, following the power law distribution~\cite{powerlaw} using three power law exponents: 0 (uniform~\cite{url/power-law}), 5 and 50 (very skewed).
The power law distribution models many real world phenomena (including in economics, physics, 
social networks, etc.). 
It is a relationship of type $Y= kX^a$, where $Y$ and $X$ are variables of interest, $a$ is the power law exponent and $k$ is a constant. 
The skewness of a dataset distribution increases with $a$. 
When $a = 0$, the dataset is evenly distributed.

\begin{figure}[htb]
    \centering
    \begin{subfigure}{0.39\columnwidth}
    \centering
			\captionsetup{justification=centering}	
        \includegraphics[width=\textwidth]{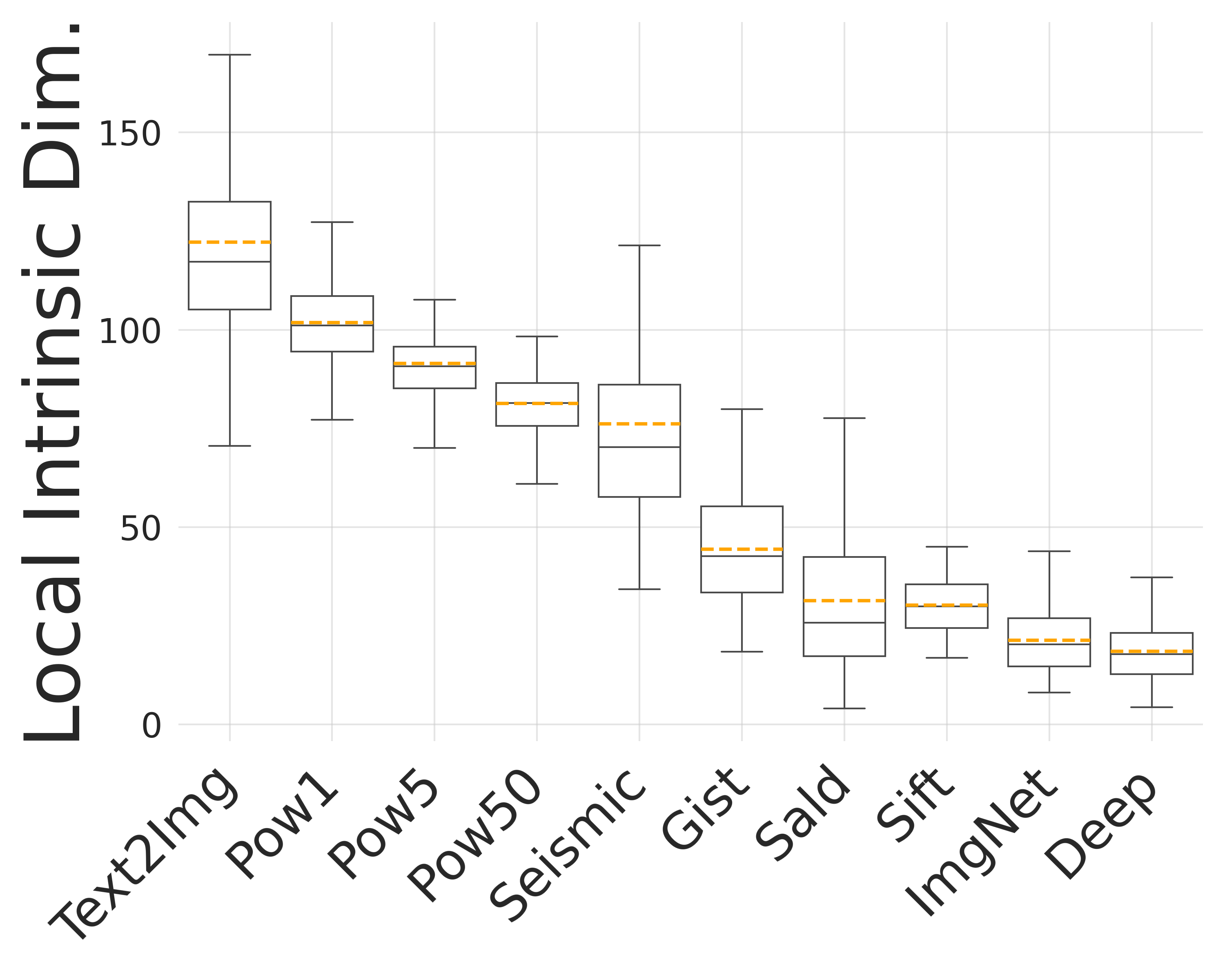}
        \caption{\karima{Local Intrinsic Dimensionality (LID): low values indicate easy search}}
        \label{fig:datacomp:lid}
    \end{subfigure}
    \hspace{1cm}
    \begin{subfigure}{0.39\columnwidth}
    \centering
			\captionsetup{justification=centering}	
        \includegraphics[width=\textwidth]{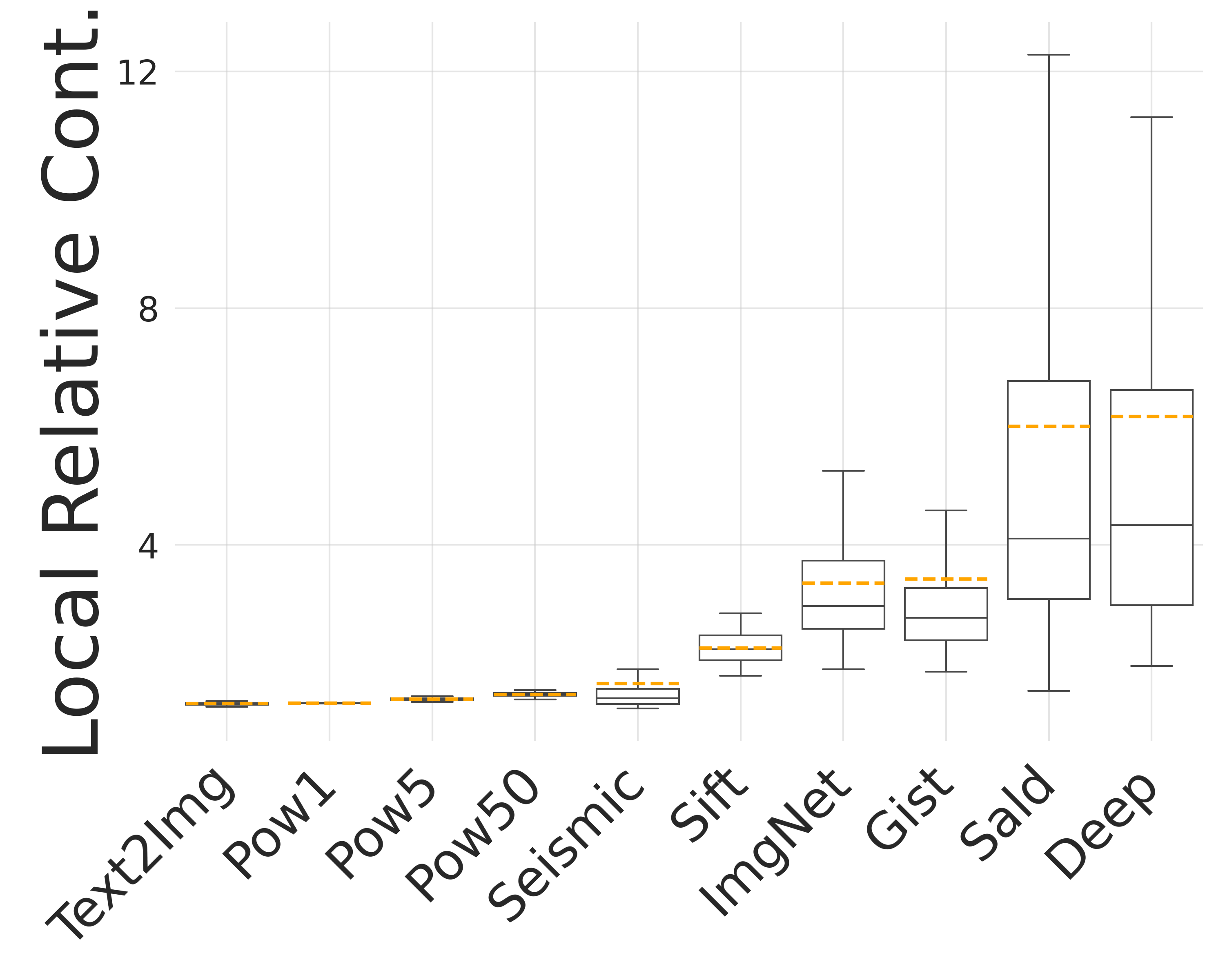}
        \caption{\karima{Local Relative Contrast (LRC): high values indicate easy search}}
        \label{fig:datacomp:rc}
    \end{subfigure}
    \caption{\karima{Dataset Complexity}}
    \label{fig:datacomp}
\end{figure}

\noindent{\bf Dataset Complexity.} 
The complexity of the datasets in our experimental evaluation is assessed using Local Intrinsic Dimensionality (LID)~\cite{lid15,DBLP:journals/is/AumullerC21} and Local Relative Contrast (LRC)~\cite{rc,DBLP:journals/is/AumullerC21}.  
The LID and LRC for a query point \( x \) are  defined as follows: 
\begin{displaymath}
\hspace{1.25cm}
\text{LID}(x) = -\left(\frac{1}{k} \sum_{i=1}^{k} \log \frac{\text{dist}_i(x)}{\text{dist}_{k}(x)}\right)^{-1} \hspace{0.7cm}(\text{Eq. 5}) \tag*{}
\end{displaymath}

\begin{equation}
\hspace{1.3cm}
\text{LRC}(x) = \frac{\text{dist}_{\text{mean}}(x)}{\text{dist}_k(x)}
\hspace{2.8cm} (\text{Eq. 6}) \tag*{}
\end{equation}
where $dist_i(x)$ is the Euclidean distance between point $x$ and its $i$-th true nearest neighbor, $k$ is the number of nearest neighbors, and $\text{dist}_{mean}(x)$ is the average distance of $x$ to all other points in the dataset.
LID represents the intrinsic dimensionality of the data
(and can be significantly lower than the original dimensionality of the data space): the lower the LID, the easier the search is.
LRC is an intuitive measure of separability of the nearest neighbor of a query from the rest of the points in the dataset: the higher the LRC, the easier the search is. Figure~\ref{fig:datacomp} shows that the LID and LRC results are consistent. The graphs were produced using a subset of 1M points randomly sampled from each dataset, and $k=100$. 
Note that the orange horizontal lines in Figures~\ref{fig:datacomp:lid} and~\ref{fig:datacomp:rc} denote the mean LID and LRC values of each dataset, respectively. 
The Pow0, Pow5, Pow50, Seismic and \karima{Text2Img}~datasets have the highest LID and lowest LRC values, indicating that they are hard datasets for vector search~\cite{DBLP:journals/is/AumullerC21}. 
\karima{Sift, Deep and ImageNet}, on the other hand, are the easiest datasets in our workload, with the lowest LID and highest LRC values.

\noindent{\bf Queries.}  Query sets include 100 vectors processed sequentially, not in batches, mimicking a real-world scenario where queries are unpredictable~\cite{itsawreport,DBLP:conf/edbt/GogolouTPB19,conf/sigmod/gogolou20}. \karima{Results with 1 million queries are extrapolated from 100 query sets. For Deep, Sift, GIST, and Text-to-Image, queries are randomly sampled from available query workloads. For SALD, ImageNet, and Seismic, 100 queries are randomly selected from the datasets and excluded from the index-building phase.}
For hardness experiments, we use Deep query vectors of varying complexity, denoted as a percentage ranging from 1\% to 10\%. These vectors were obtained by adding Gaussian noise (\(\mu = 0\), \(\sigma^2\) ranging from 0.01 to 0.1) to randomly choose dataset vectors, with the percentage reflecting the \(\sigma^2\) value~\cite{johannesjoural2018}.
The 100-query workloads for the power law distribution datasets are generated following the same distributions using different seeds. 
Unless otherwise stated, all experiments were conducted with 10-NN queries per the standard in the community~\cite{neurips-2021-ann-competition,url/Elpis,aumuller2017ann}.
We use 100-NN queries to evaluate dataset complexity because a higher k improves the estimation of LID and LRC (Eqs. 5-6), and to assess the different SS strategies because the higher the k the more overhead for seed selection.

\noindent\textbf{Measures.} We measure 
the {\it wall clock time} and {\it distance calculations} for both indexing and query answering. We also measure the accuracy of each k-NN query using {\it Recall} which quantifies the fraction of the true nearest neighbors that the query $S_Q$ successfully returns.  

\karima{\noindent {\bf Procedure.} We tune each method to achieve the best trade-offs in accuracy/efficiency.  
Then, we carry experiments in two steps: indexing building and query answering, with caches cleared before each step and kept warm during the same query workload. Methods were allowed at most 48 hours to build a single index. During timed experiments, the server was used exclusively to ensure accurate measurements. 
 For each query workload, we ran the experiment six times; we excluded the two best and worst, and reported the mean of the remaining performances. For reproducibility, all parametrization details are provided in~\cite{url/GASS}.}

\subsection{Neighborhood Diversification}
\label{subsec:experiments-ND}
We now evaluate the ND strategies covered in Section~\ref{sec:survey}, i.e., RND, RRND, and MOND against a baseline without ND (NoND). 
We apply each strategy individually to an II-based graph, where each node is inserted sequentially and linked with a pruned list of neighbors, determined via a beam search with maximum out-degree $R=60$ and beam width $L=800$. Bi-directional edges are added to neighbors, and the neighborhood list is pruned to size $R$ using the same ND strategy.
Graphs are built on Deep and Sift (25GB, 100GB and 1B). 
For RRND and MOND, we run experiments with different values of $\alpha$ ($1-2$) and $\theta$ ($50^\circ-80^\circ$), respectively, and use $\alpha = 1.3$ and $\theta = 60^\circ$ because they lead to the best performance. 
Then, we execute workloads with 100 queries against each dataset, and measure the accuracy/efficiency tradeoff using the recall and the number of distance calculations incurred during the search. The results in Figure~\ref{fig:ND:search:real} indicate that both RND and MOND consistently outperform, followed by RRND. NoND is the worst performer overall. As the dataset size increases, the performance gap between NoND and ND methods widens, particularly at high Recall (Figures~\ref{fig:ND:sift1b}, \ref{fig:ND:deep1b}). 
This is due to the higher number of hops needed to find the answers and the density of the neighborhoods in the NoND nodes since no pruning was applied. 
These results indicate the key role played by the ND paradigm in improving query-answering performance and the superiority of the RND and MOND strategies.
\newcommand{\sffive}{0.27\columnwidth}
\begin{figure}[tb]
	\captionsetup{justification=centering}
	\centering	
		\begin{subfigure}{\columnwidth}
			\centering
			\captionsetup{justification=centering}	
			\includegraphics[width=0.35\columnwidth]{../img-png/Experiments/RNG/legend.png}
			\label{fig:ND:legend}
		\end{subfigure}\\
		\begin{subfigure}{\sffive}
			\centering
			\captionsetup{justification=centering}	
			\includegraphics[width=\textwidth]{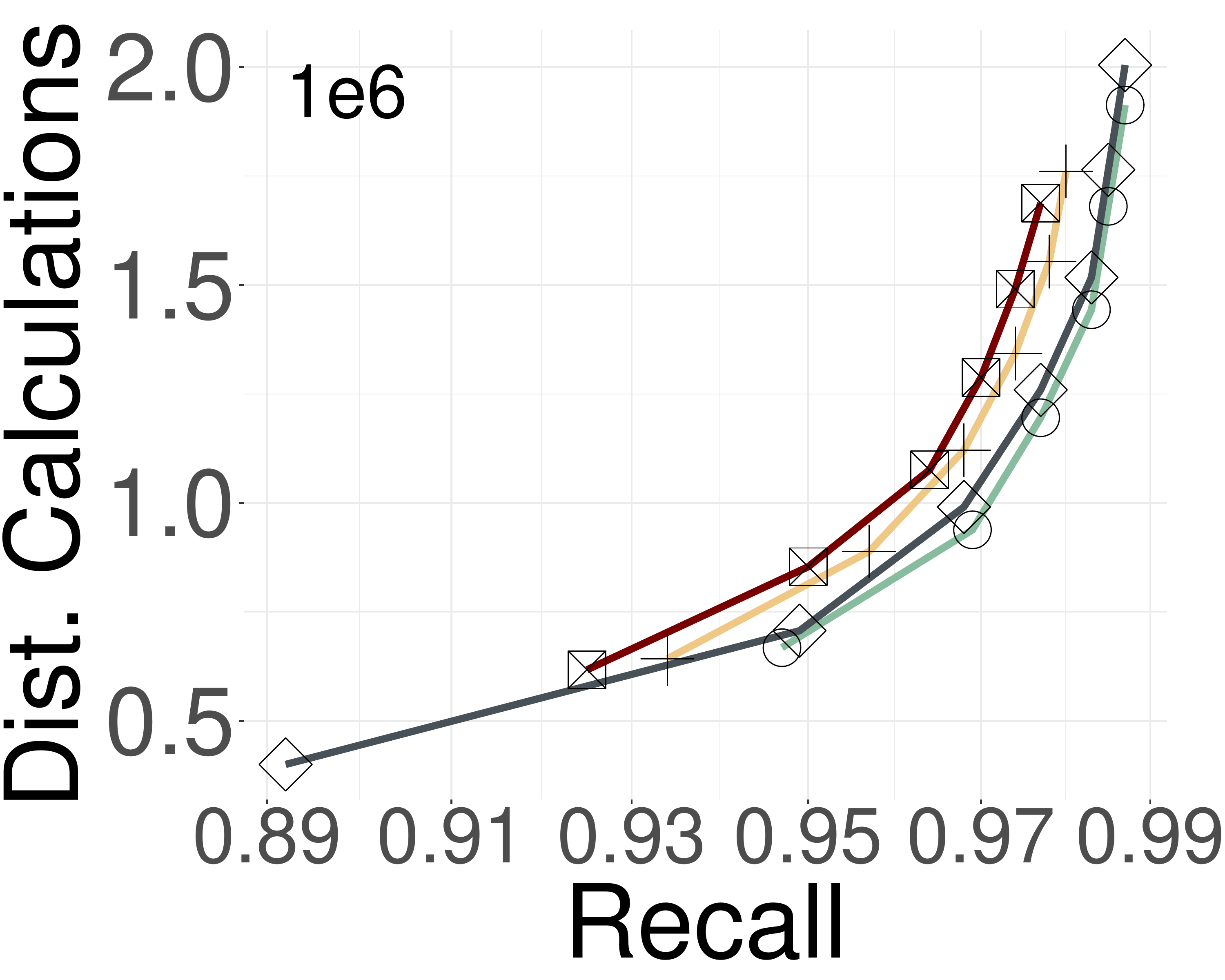}
		\caption{{DEEP25GB}}
		\label{fig:ND:deep25GB}	
		\end{subfigure}	
		\begin{subfigure}{\sffive}
			\centering
			\captionsetup{justification=centering}	
			\includegraphics[width=\textwidth]{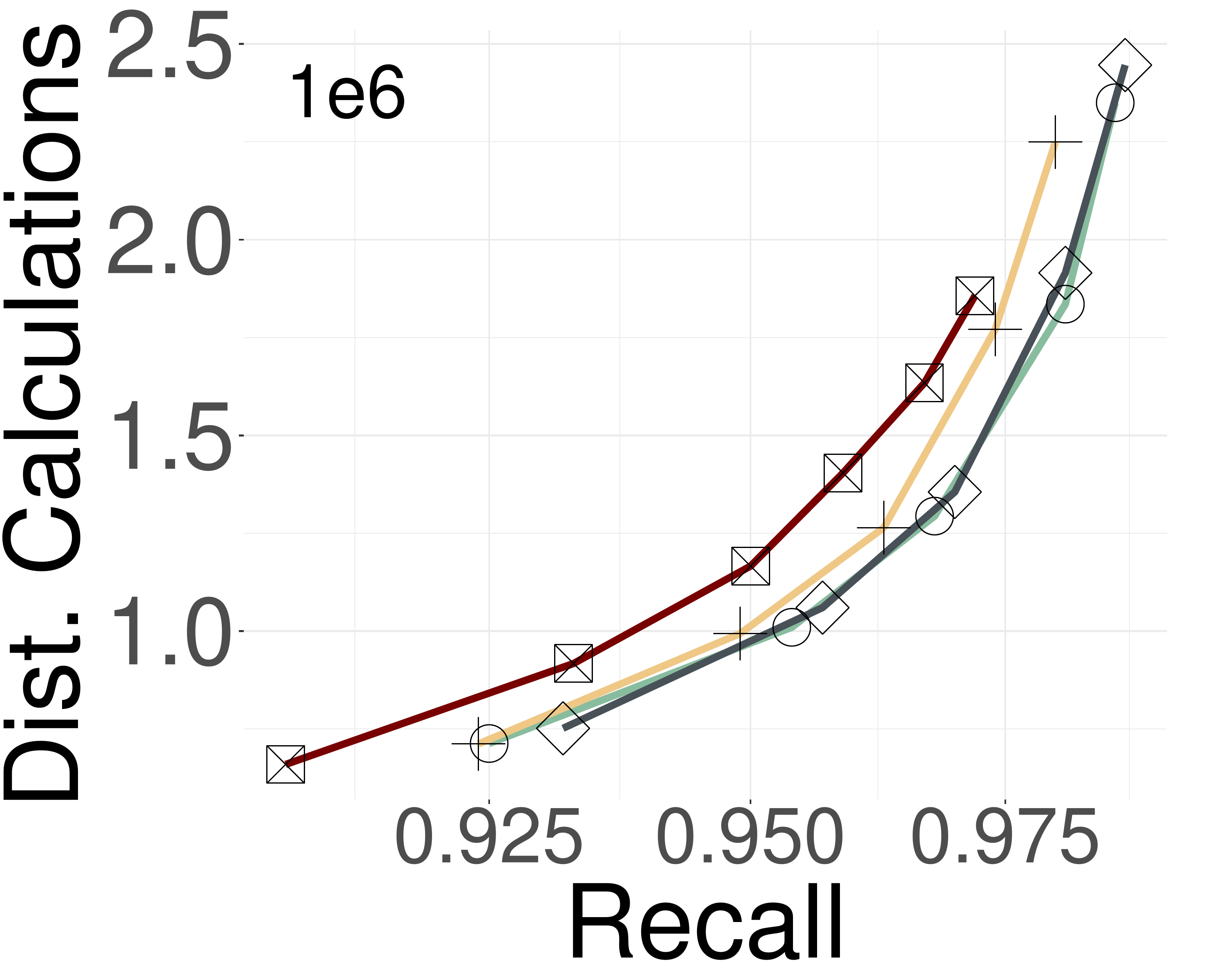}
		\caption{{DEEP100GB}}
		\label{fig:ND:deep100GB}
		\end{subfigure}	
		\begin{subfigure}{\sffive}
			\centering
			\captionsetup{justification=centering}	
			\includegraphics[width=\textwidth]{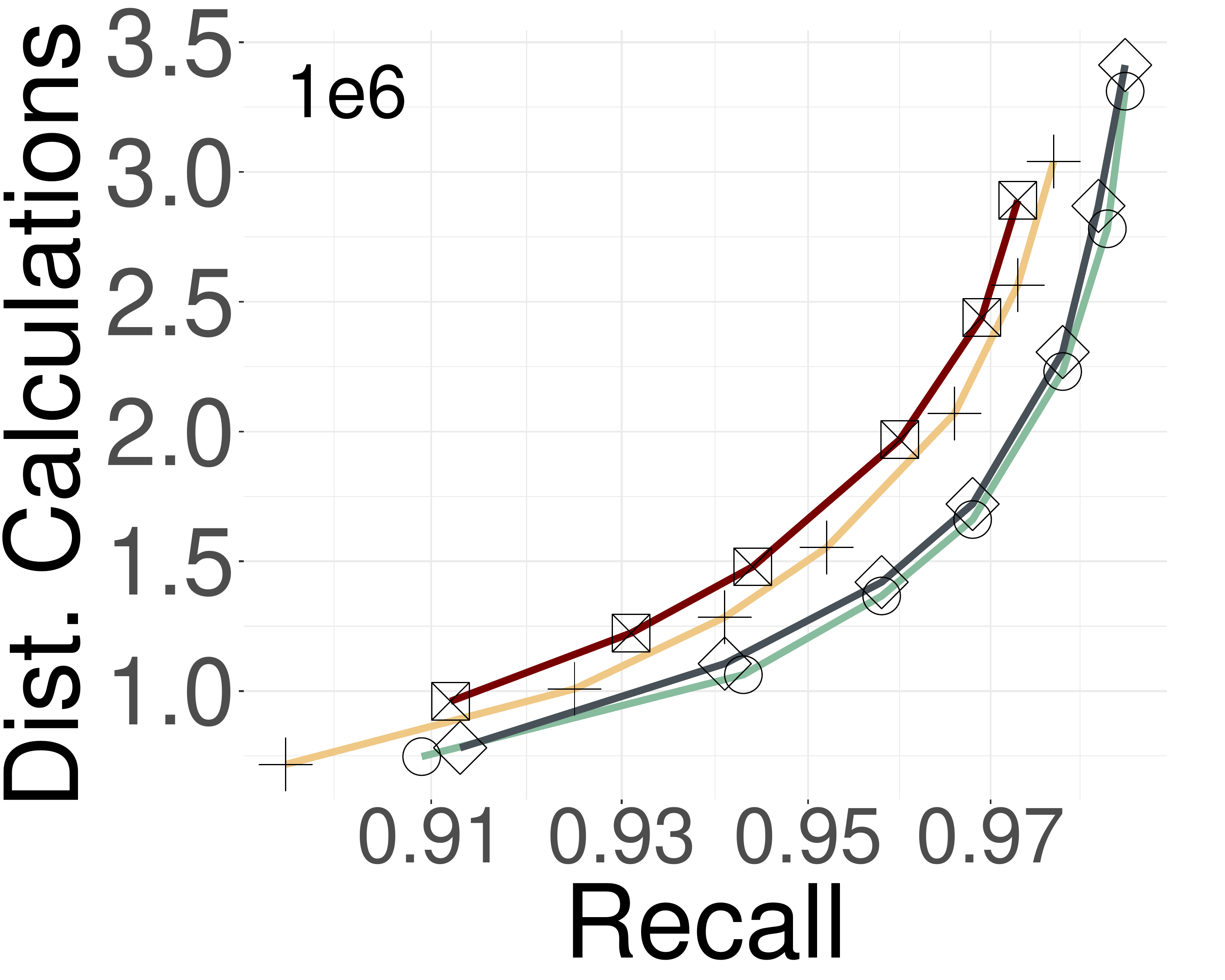}
		\caption{{DEEP1B}}
		\label{fig:ND:deep1b}	
  \end{subfigure}	
  \\
		\begin{subfigure}{\sffive}
			\centering
			\captionsetup{justification=centering}	
			\includegraphics[width=\textwidth]{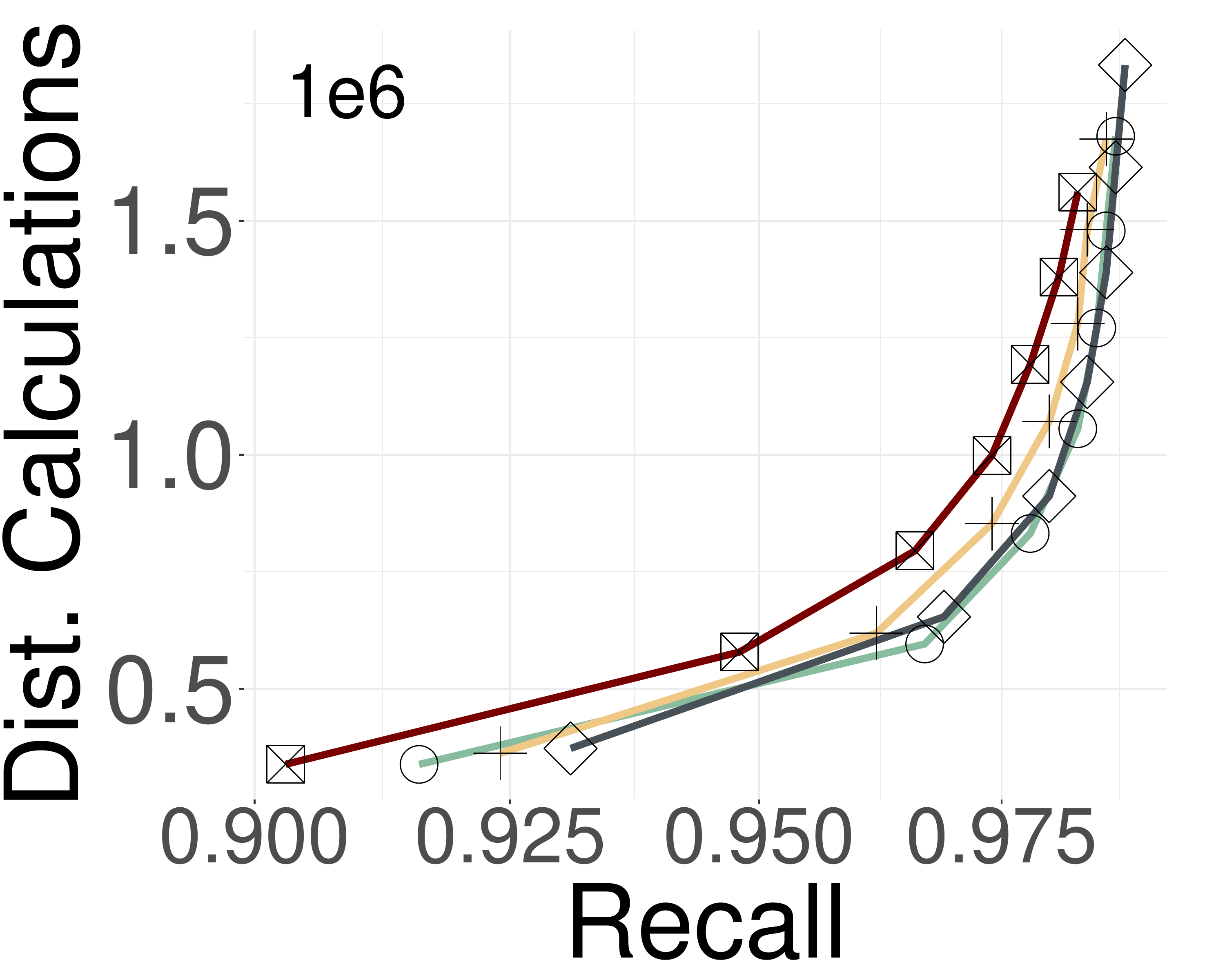}
   \caption{{SIFT25GB}}
		\label{fig:ND:sift25GB}
		\end{subfigure}	
		\begin{subfigure}{\sffive}
			\centering
			\captionsetup{justification=centering}	
			\includegraphics[width=\textwidth]{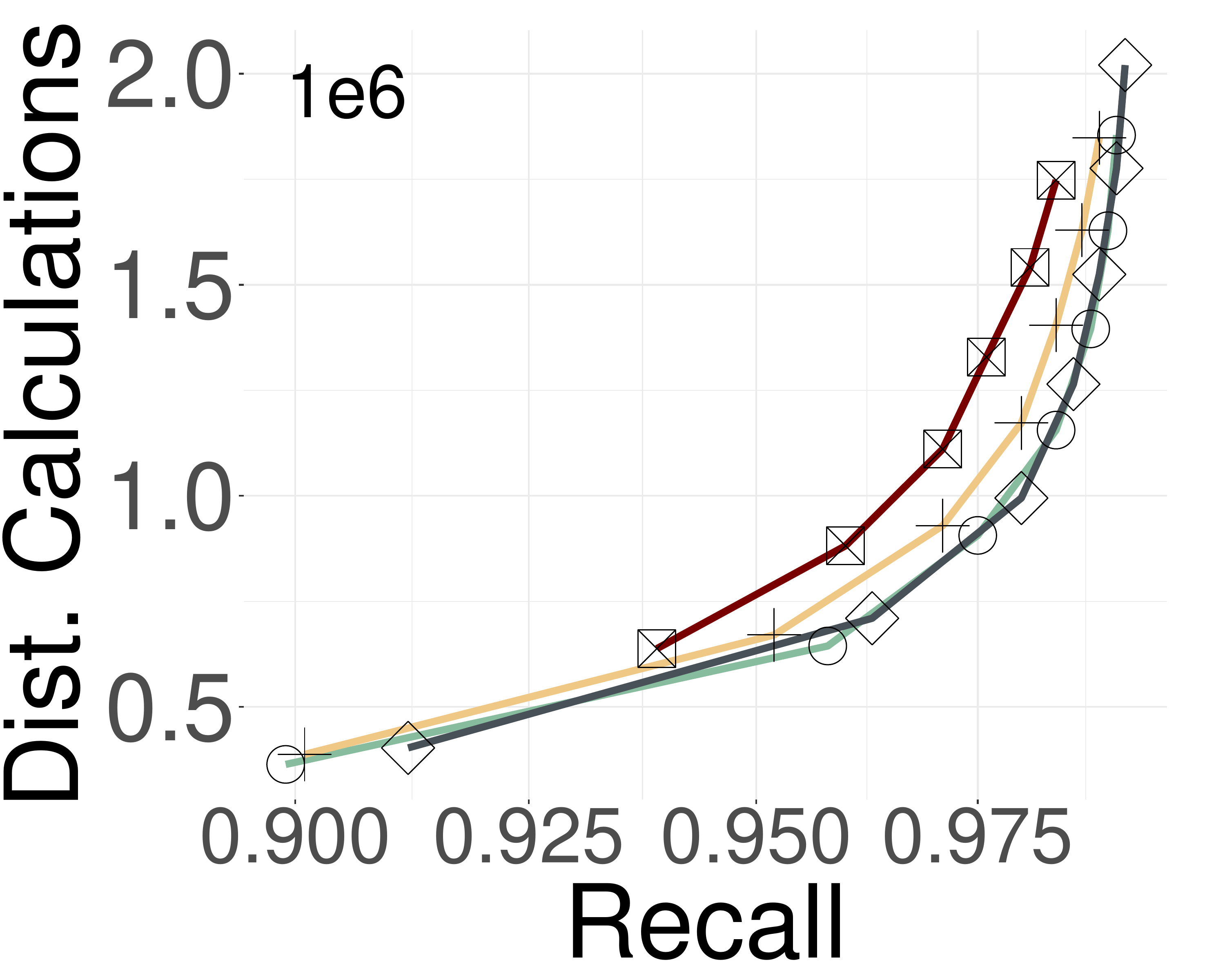}
             \caption{{SIFT100GB}}
		      \label{fig:ND:sift100GB}
		\end{subfigure}	
		\begin{subfigure}{\sffive}
			\centering
			\captionsetup{justification=centering}	
			\includegraphics[width=\textwidth]{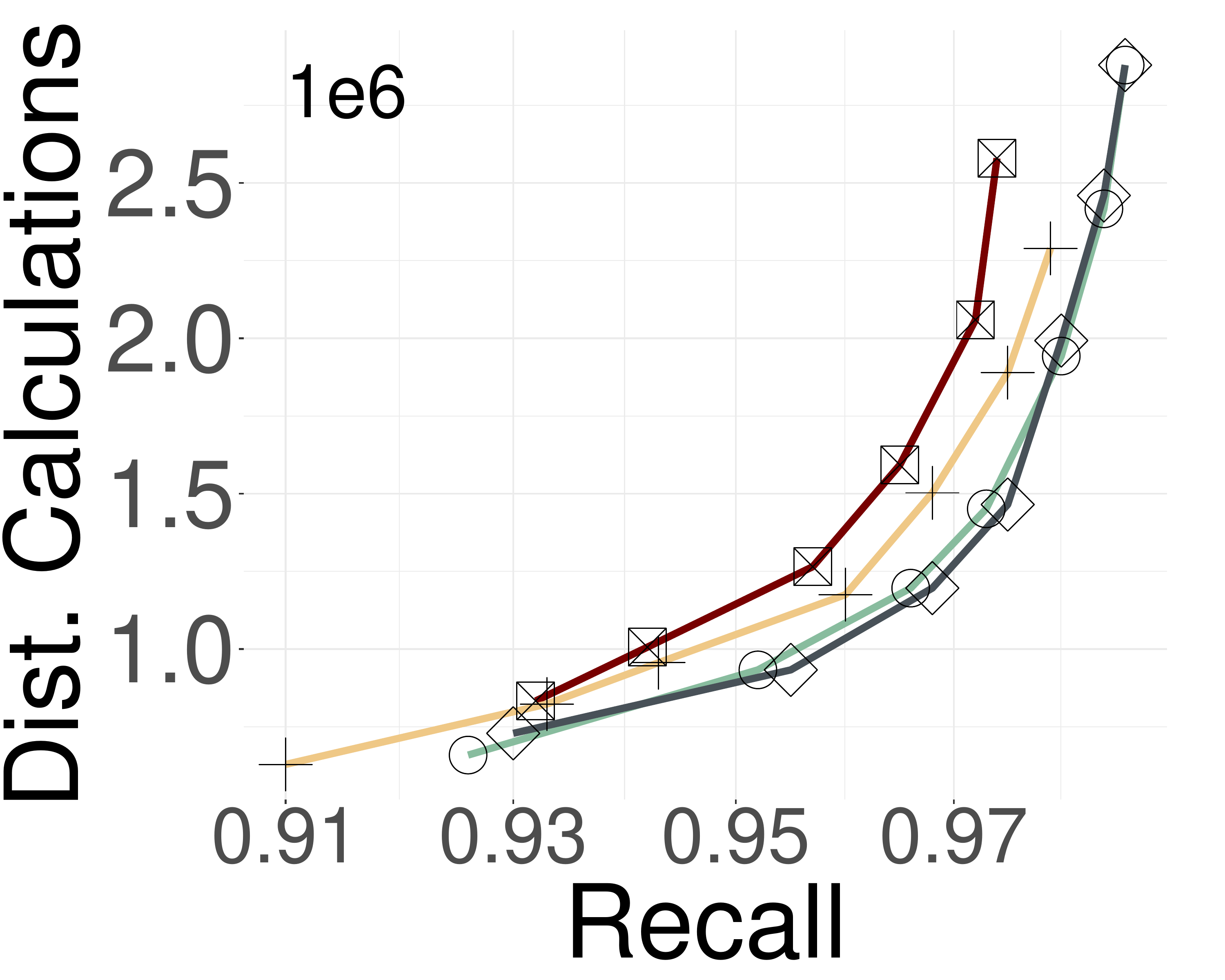}
   \caption{{SIFT1B}}
		\label{fig:ND:sift1b}
		\end{subfigure}	
		\caption{{ND methods performance on real-world datasets}}
	
\label{fig:ND:search:real}
 \end{figure}

In Table~\ref{tab:pruning_ratio}, we report the pruning ratios of the three neighborhood ND methods on the Deep and Sift 25GB datasets. 
 The pruning ratio quantifies the percentage reduction in the size of the candidate neighbor list before and after the diversification step. Higher pruning ratios indicate more aggressive pruning, which directly affects the graph size and memory usage. RND achieves the highest pruning ratios, MOND provides moderate pruning, and RRND exhibits the least pruning. 
 As a result, RND leads to smaller graph sizes and reduced memory requirements, while RRND creates larger graphs with higher memory usage.
\begin{table}[tb]
{\normalsize
\centering
\begin{tabular}{@{}lccc@{}}
\toprule
         & \textbf{RND} & \textbf{MOND} & \textbf{RRND} \\ 
\midrule
\textbf{Deep} & 20\%        & 2\%          & 0.6\%         \\ 
\textbf{Sift} & 25\%        & 4\%          & 0.7\%         \\ 
\bottomrule
\end{tabular}
} 
\caption{Pruning ratios of ND methods on Deep and Sift datasets.}
\label{tab:pruning_ratio}
\end{table}

\subsection{Seed Selection}
In these experiments, we focus on the four most common SS strategies for the beam search algorithm: SN~\cite{hnsw,elpis}, MD~\cite{nsg,vamana}, KS~\cite{kgraph,nsw11,dpg,vamana,nssg}, and KD~\cite{efanna,SPTAG4,hcnng} (KM and LSH were excluded because they are not among the commonly used seed selection strategies in graph-based methods). We consider the baseline method SF which has not been used in the literature before. 
In these experiments, we focus on the four most common SS strategies for the beam search algorithm: SN~\cite{hnsw,elpis}, MD~\cite{nsg,vamana}, KS~\cite{kgraph,nsw11,dpg,vamana,nssg}, and KD~\cite{efanna,SPTAG4,hcnng} (KM and LSH were excluded because they are not among the commonly used seed selection strategies in graph-based methods). We consider the baseline method SF which has not been used in the literature before. 
These strategies are compared using the same insertion-based graph structure that exploits RND pruning since this is the best baseline from the results in Section~\ref{subsec:experiments-ND}.  
We run 100 queries for each strategy on the Deep and Sift datasets with sizes 25GB, 100GB, and 1B. We extrapolate the results to 1M queries and report the number of distance calculations to achieve a 0.99 accuracy in Figure~\ref{fig:ss:search}. We observe that SN and KS are the most efficient strategies across all scenarios, while SF and MD are the least efficient overall. The KD strategy is competitive on 25GB and 100GB Deep and Sift datasets but its performance deteriorates on billion-scale datasets. 
KS outperforms SN on dataset sizes 25GB and 100GB; however, this trend reverses with the 1B size dataset. 
The difference in distance calculations between SN and KS on the 25GB and 1B datasets is $\sim$1M and $\sim$10M, respectively. 
As the dataset size increases, it becomes imperative to sample more nodes (beyond the beam width utilized during search in KS) to obtain a representative sample of the dataset, thereby enhancing the likelihood of initiating the search closer to the graph region where the query resides (SN adapts its size logarithmically with the growth of the dataset, leading to a better representation of the dataset). Figure~\ref{fig:ss:search} also illustrates that both MD and SF are among the least performing strategies, with MD performing better than SF on Deep and vice-versa on Sift. This indicates neither MD nor SF are effective and robust seed selection strategies.

We now study the effect of SS strategies on indexing performance. We focus on the two best strategies KS and SN and study their effect on the same baseline based on II and RND~\cite{nsw11,dpg,hnsw,nsg,nssg,vamana,elpis,SPTAG4}. This is because these methods are the most impacted by the SS strategy used, since they perform a beam search, which includes a seed selection step, at the insertion of each node.
We build an index using each strategy on Deep1M and Deep25GB and measure distance calculations. 
We calculate the distance overhead of SN compared to KS, and we estimate the number of additional 100-NN queries that the KS-based graph can answer, with 0.99 recall, before the SN-based graph completes its construction.  
We observe (Table~\ref{tab:ss:idx}) that the SN-based graph requires 182 million and 22.3 billion more distance calculations than the KS-based graph on Deep1M and Deep25GB respectively. Furthermore, the KS-based graph can answer ~45K and 1.17 million queries on Deep1M and Deep25GB respectively before the SN-based graph finishes construction.

\newcommand{\sfsix}{0.27\columnwidth}
\begin{figure}[tb]
	\captionsetup{justification=centering}
	\centering	
 		\begin{subfigure}{0.015\columnwidth}
			\centering
			\captionsetup{justification=centering}	
			\includegraphics[width=\textwidth]{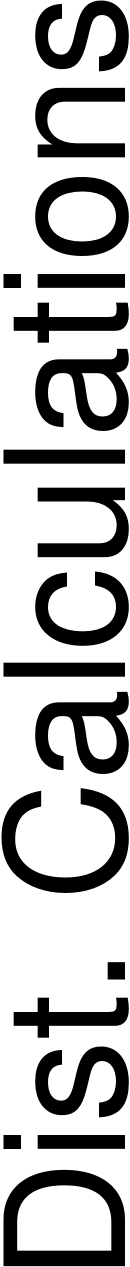}
   \vspace{0.24in}
		\end{subfigure}	
		\begin{subfigure}{\sfsix}
			\centering
			\captionsetup{justification=centering}	
			\includegraphics[width=\textwidth]{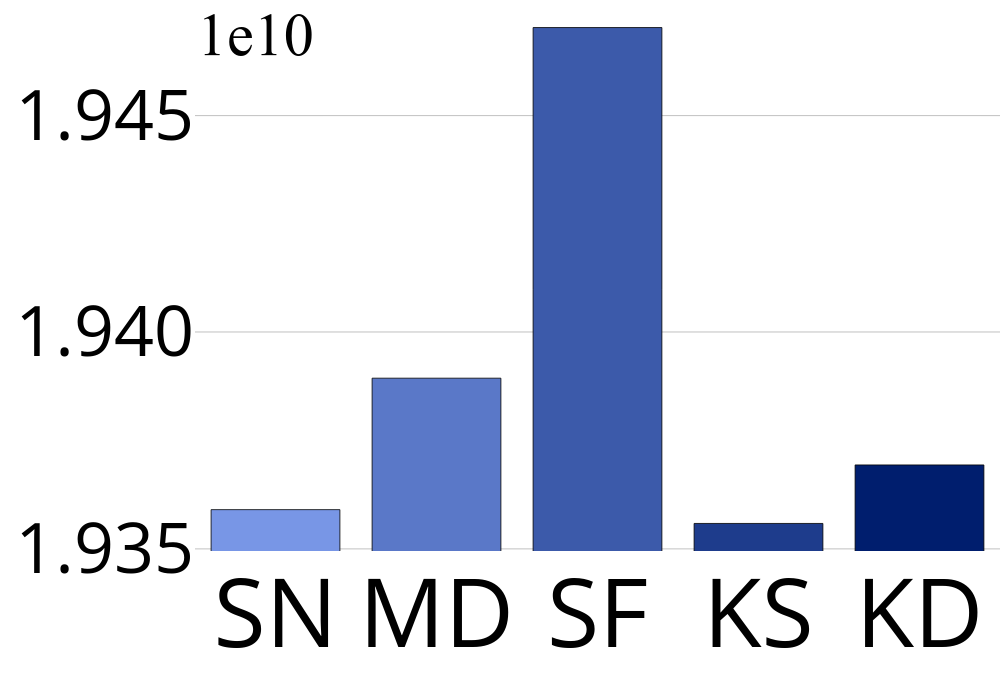}
		\caption{{Deep25GB}}
		\label{fig:ss:deep25gb}
		\end{subfigure}	
		\begin{subfigure}{\sfsix}
			\centering
			\captionsetup{justification=centering}	
			\includegraphics[width=\textwidth]{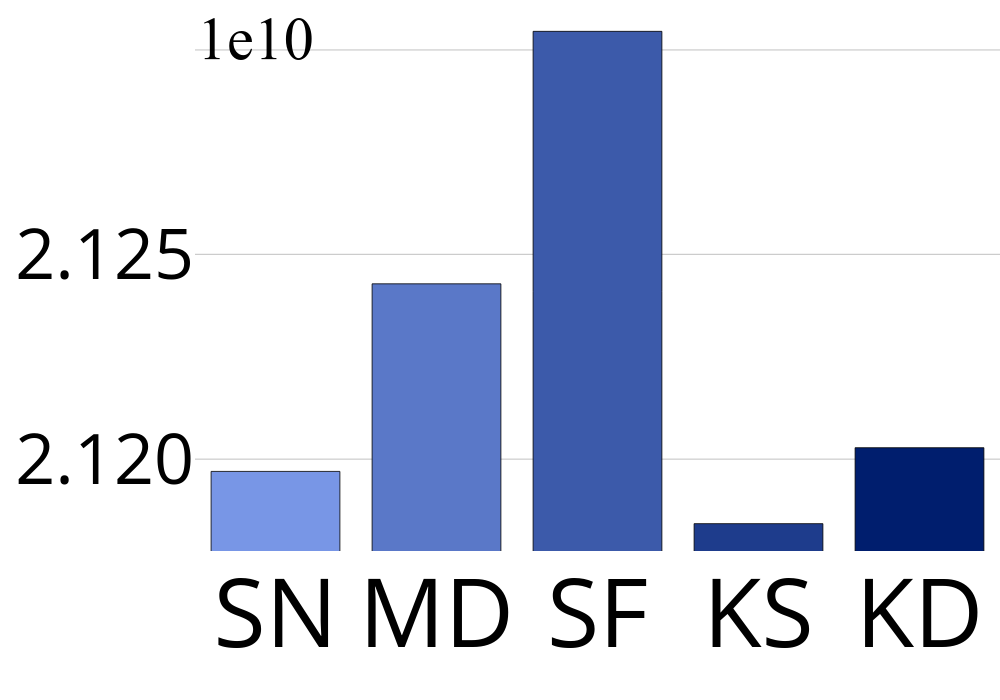}
		\caption{{Deep100GB}}
		\label{fig:ss:deep100gb}
		\end{subfigure}	
		\begin{subfigure}{\sfsix}
			\centering
			\captionsetup{justification=centering}	
			\includegraphics[width=\textwidth]{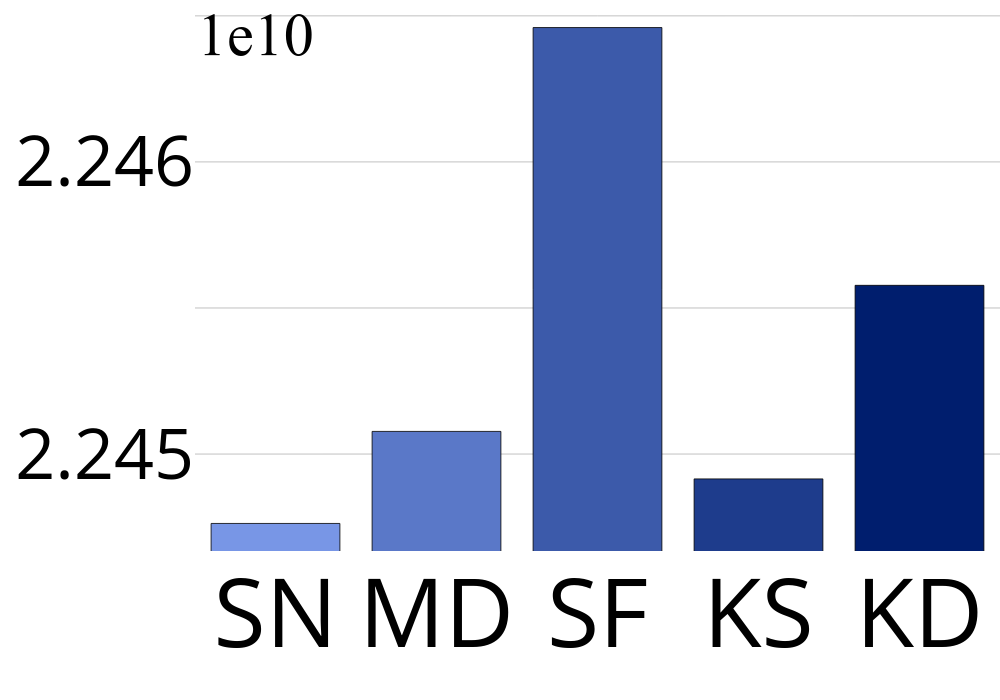}
		\caption{{Deep1B}}
		\label{fig:ss:deep1b}
		\end{subfigure}	
        
 		\begin{subfigure}{0.015\columnwidth}
			\centering
			\captionsetup{justification=centering}	
			\includegraphics[width=\textwidth]{../img-png/Experiments/EP/dc.png}
   \vspace{0.24in}
		\end{subfigure}	
		\begin{subfigure}{\sfsix}
			\centering
			\captionsetup{justification=centering}	
			\includegraphics[width=\textwidth]{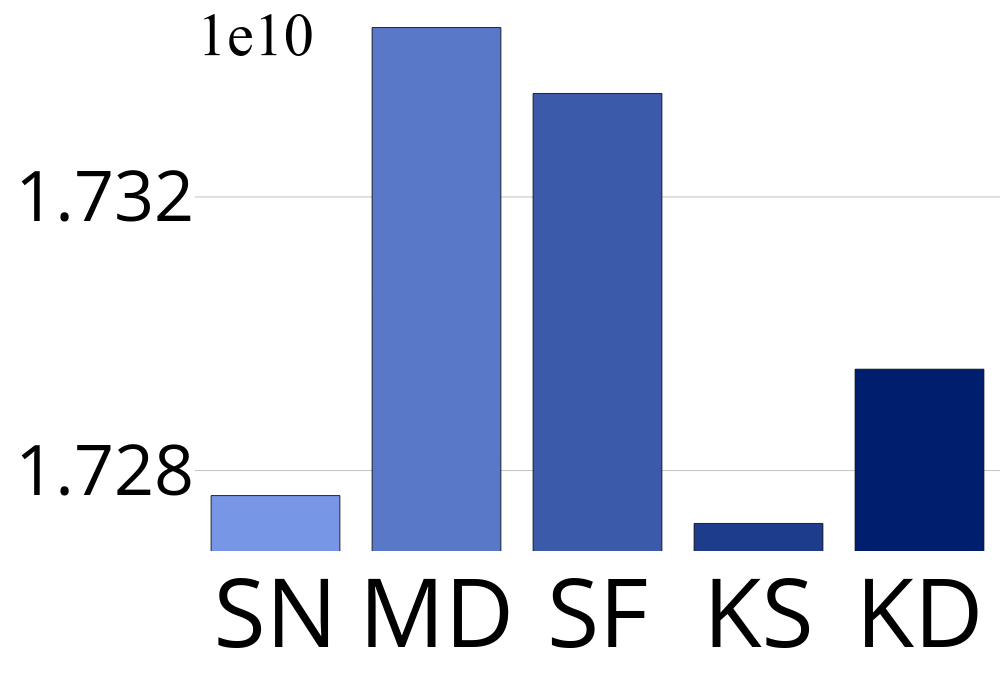}
		\caption{{Sift25GB}}
		\label{fig:ss:sift25gbß}
		\end{subfigure}	
		\begin{subfigure}{\sfsix}
			\centering
			\captionsetup{justification=centering}	
			\includegraphics[width=\textwidth]{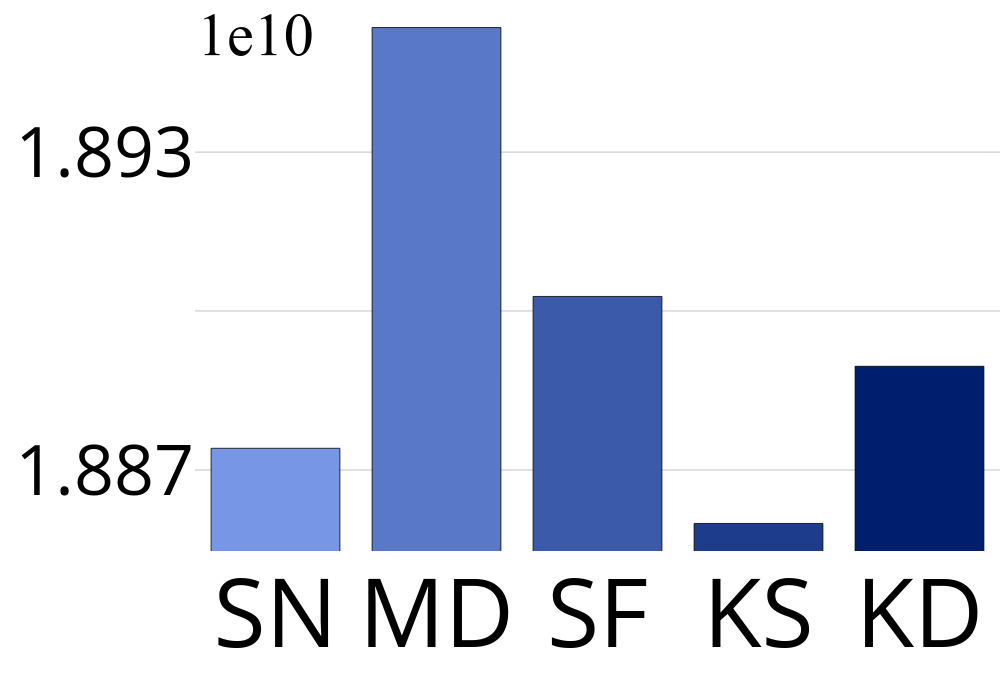}
		\caption{{Sift100GB}}
		\label{fig:ss:sift100gb}
		\end{subfigure}	
		\begin{subfigure}{\sfsix}
			\centering
			\captionsetup{justification=centering}	
			\includegraphics[width=\textwidth]{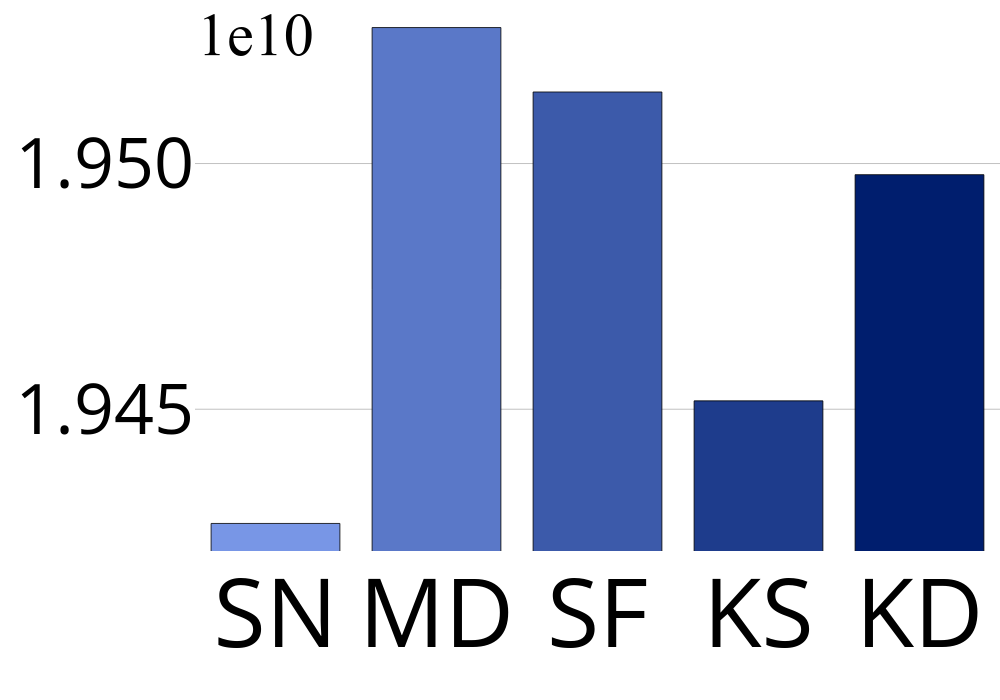}
		\caption{{Sift1B}}
		\label{fig:ss:sift1b}
		\end{subfigure}	
\caption{The impact of SS Methods on query answering}
\label{fig:ss:search}
 \end{figure}

\begin{table}[tb]
\centering
\begin{tabular}{@{}lcc@{}}
\toprule
& \textbf{Deep1M} & \textbf{Deep25GB} \\ 
\midrule
\textbf{Dist. Calculations (SN)} & 4.3 billion & 1.49 trillion \\
\textbf{Dist. Calculations (KS)} & 4.1 billion & 1.46 trillion \\
\midrule
\textbf{Overhead (SN vs. KS)} & 182 million & 22.3 billion \\
\textbf{Additional Queries} & 44,959 & 1,165,870 \\
\bottomrule
\end{tabular}
\caption{The impact of SS methods on Indexing Performance}
\label{tab:ss:idx}
\end{table}

\subsection{Indexing Performance}
We evaluate twelve state-of-the-art vector search methods, varying dataset sizes and reporting total indexing time and memory footprint. For brevity, we present results only for the Deep dataset as trends are consistent across other datasets. Full results are in~\cite{url/GASS}.  We use subsets of Deep ranging from 1 million to 1 billion vectors (equivalent to ~350GB). Indexes are built to allow a 0.99 recall efficiently. \karima{ Initial experiments on 1 million vectors include all methods. Methods that could not scale to larger datasets are excluded from subsequent experiments. Specifically, HCNNG, SPTAG-BKT, NGT, and SPTAG-KDT take over 24 hours to build indexes on 25GB datasets and exceed 48 hours on 100GB datasets.
KGraph, DPG, EFANNA, and LSHAPG delivered unsatisfactory results on 25GB, so they were not included in larger datasets.} Furthermore, KGraph and EFANNA require over 300GB and 1.4TB of RAM for 25GB and 100GB datasets, respectively. As DPG, NSG, and SSG rely on KGraph and EFANNA, they were also excluded from larger datasets.

\begin{figure*}[!htb]
	\captionsetup{justification=centering}
	\centering	
	\begin{minipage}{\textwidth}
		\begin{subfigure}{\textwidth}
			\centering
			\captionsetup{justification=centering}	
			\includegraphics[width=\textwidth]{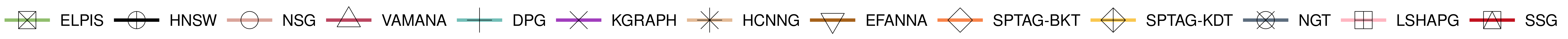}
		\end{subfigure}
	\end{minipage}
	\begin{minipage}{0.19\textwidth}				
			\centering
		\begin{subfigure}{\textwidth}
			\captionsetup{justification=centering}	
			\includegraphics[width=\textwidth]{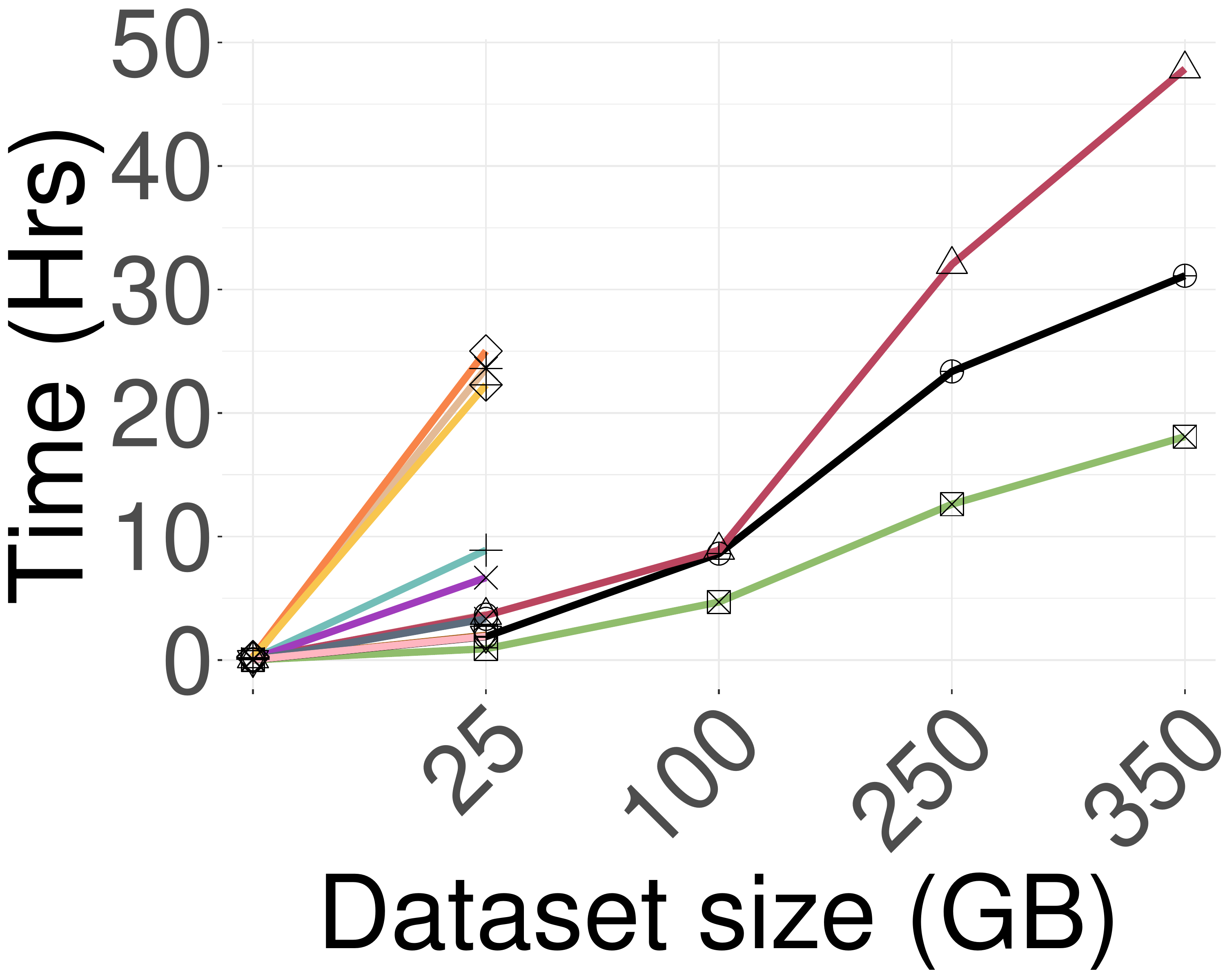}
		\end{subfigure}	
		\caption{\karima{Indexing Time }}
		\label{fig:elpis:idx:time}
	\end{minipage}	
	\begin{minipage}{0.19\textwidth}				
		\begin{subfigure}{\textwidth}
			\centering
			\captionsetup{justification=centering}	
			\includegraphics[width=\textwidth]{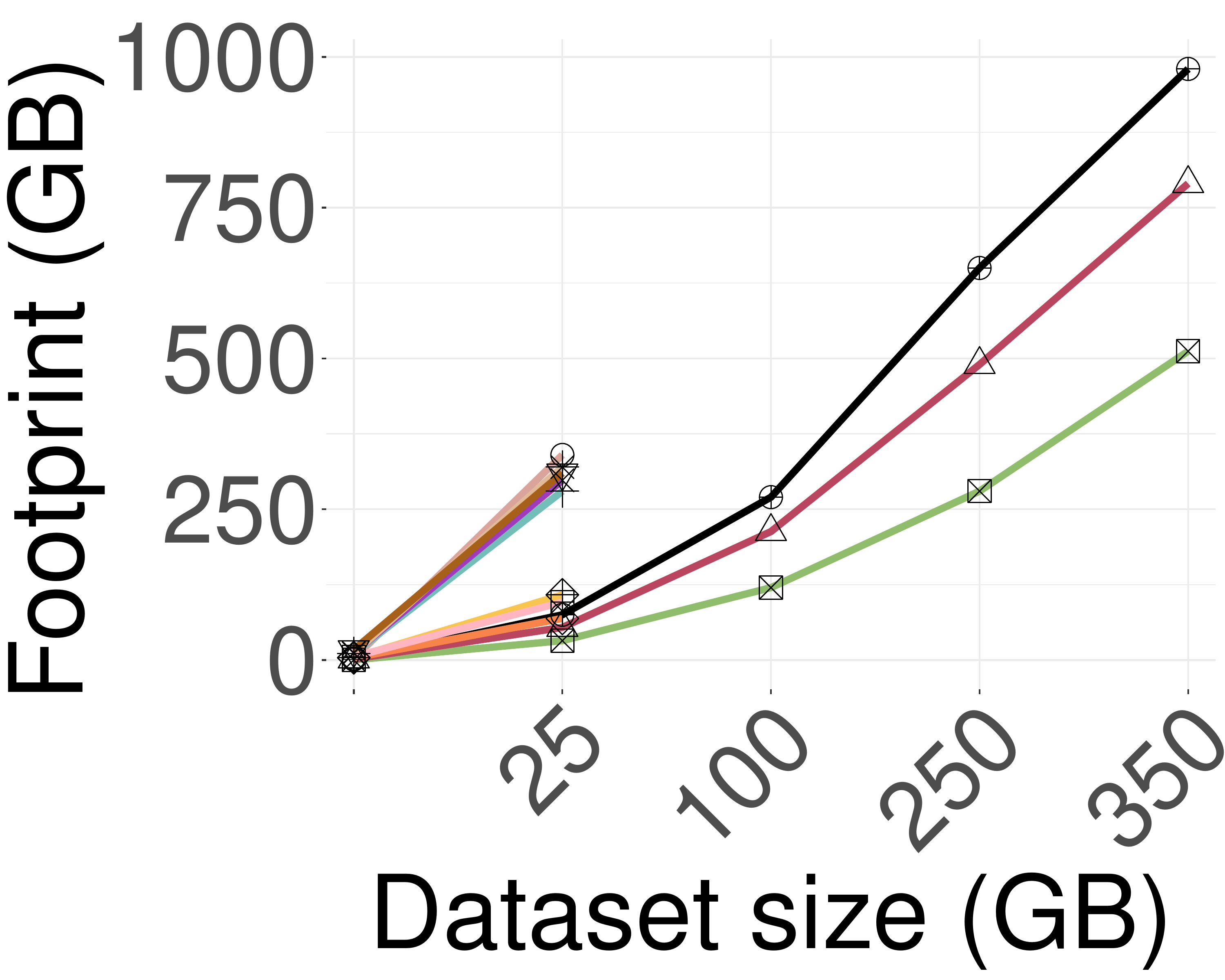}
		\end{subfigure}	
		\caption{\karima{Indexing Memory Footprint}}
		\label{fig:elpis:idx:footprint:memory}
	\end{minipage}		
	\begin{minipage}{0.19\textwidth}				
		\begin{subfigure}{\textwidth}
			\centering
			\captionsetup{justification=centering}	
			\includegraphics[width=\textwidth]{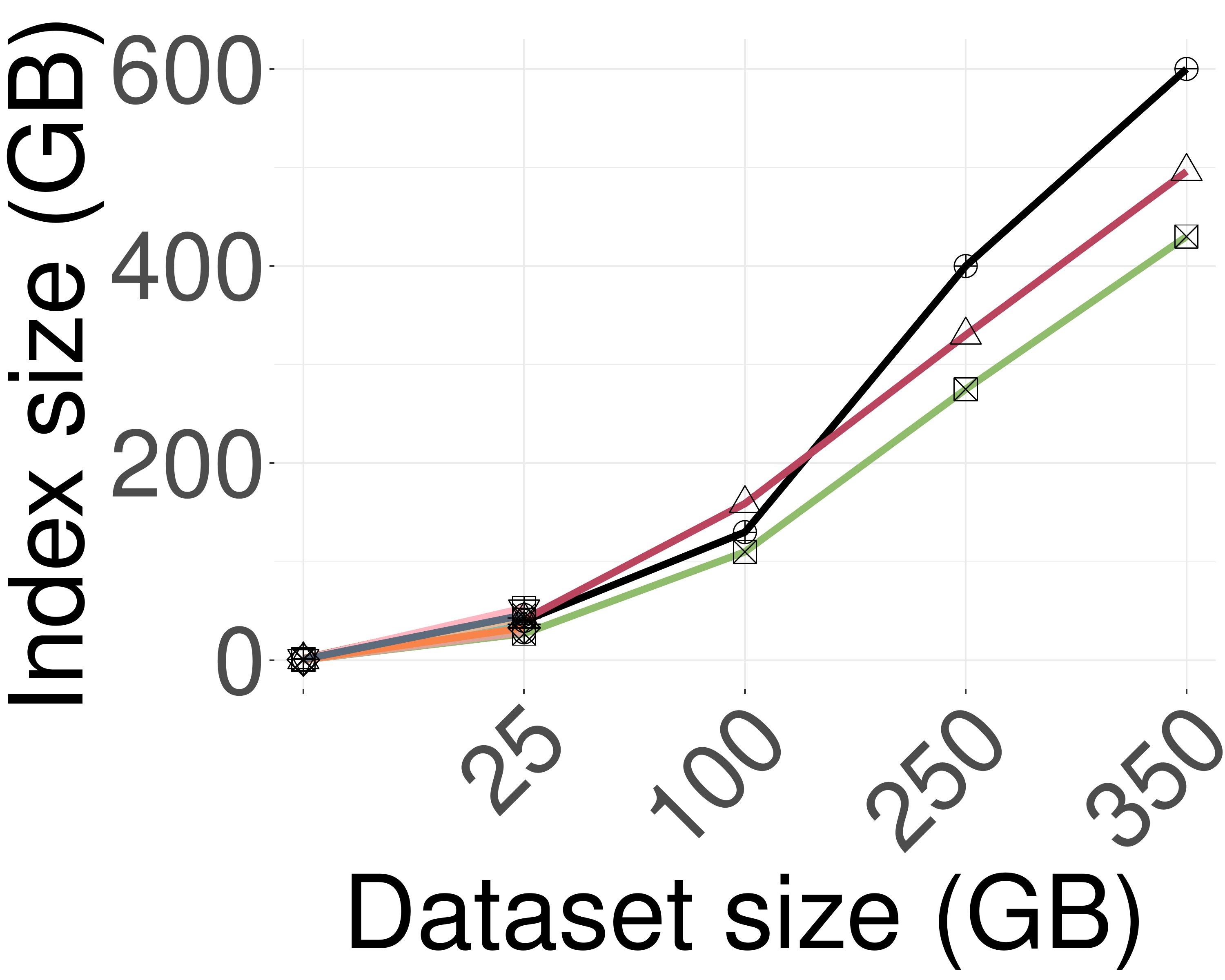}
		\end{subfigure}	
		\caption{\karima{Indexing Disk Footprint}}
		\label{fig:elpis:idx:footprint:disk}
	\end{minipage}		
	\begin{minipage}{0.19\textwidth}				
		\begin{subfigure}{\textwidth}
			\centering
			\captionsetup{justification=centering}	
			\includegraphics[width=\textwidth]{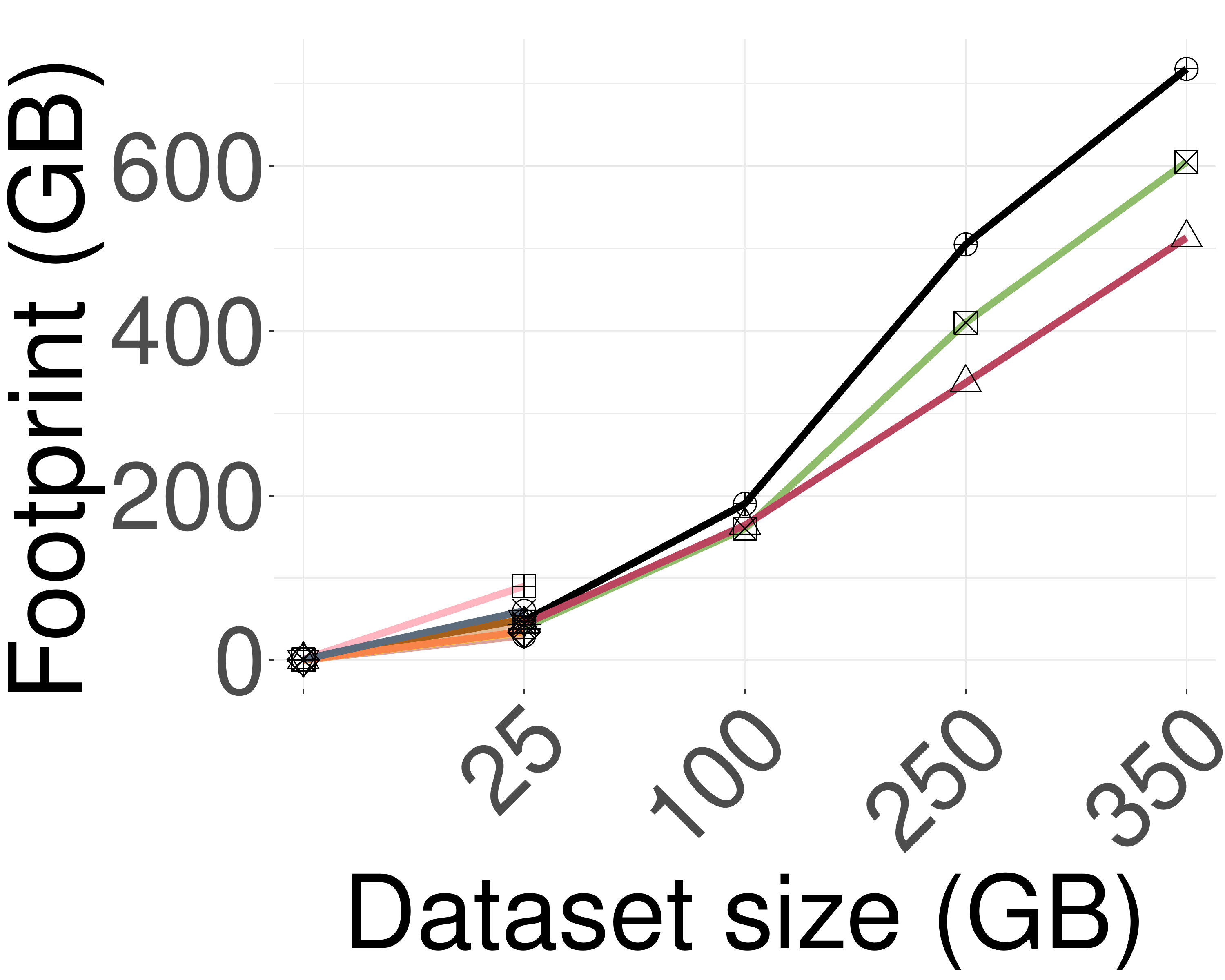}
		\end{subfigure}	
		\caption{\karima{Query Memory Footprint}}
		\label{fig:elpis:query:footprint:memory}
	\end{minipage}	
	\begin{minipage}{0.19\textwidth}
		\centering
		\begin{subfigure}{\textwidth}
			\includegraphics[width=\textwidth]{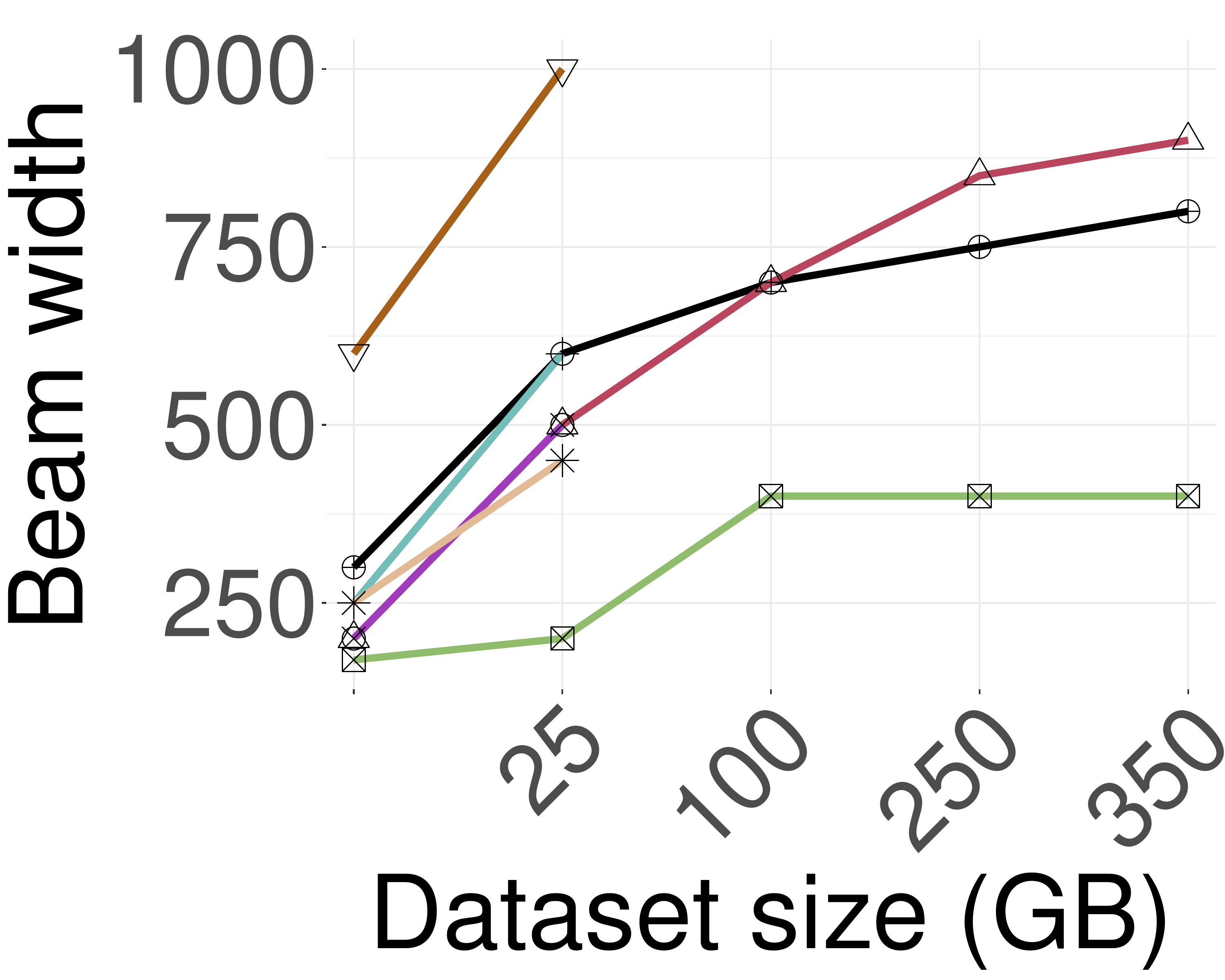}
		\end{subfigure} 
		\caption{\karima{Query Beam Width}}
		\label{fig:elpis:query:beam-width}
	\end{minipage}
\end{figure*}

\noindent{\bf Indexing Time.} Figure~\ref{fig:elpis:idx:time} demonstrates that II-based approaches have the lowest indexing time across dataset sizes. In particular, the II and DC-based approach ELPIS, is 2.7x faster than HNSW and 4x faster than NSG for both 1M and 25GB dataset sizes, \karima{while HNSW is 1.4x faster than NGT}. Note that NSG's indexing time includes both the construction of its base graph, EFANNA, which is time-intensive, and the refinement with NSG. SPTAG-BKT and SPTAG-KDT exhibit high indexing times, requiring over 25 hours to index the Deep25GB dataset-24 times more than ELPIS, the fastest method. This inefficiency in SPTAG arises from its design, which involves constructing multiple TP Trees and graphs, becoming increasingly costly with larger datasets. On datasets with 100GB and 1B vectors ($\approx$350GB), only HNSW, ELPIS, and Vamana scale with acceptable indexing time, with ELPIS being 2 and 2.7 times faster than HNSW and Vamana, respectively.

\noindent\textbf{Indexing Footprint.}
Figure~\ref{fig:elpis:idx:footprint:memory} reports the memory footprint for each index, including the raw data. To perform the evaluation, we record the peak virtual memory usage during construction\footnote{Readings are taken from the proc pseudo filesystem’s Virtual Memory Peak.}. SPTAG-BKT and SPTAG-KDT demonstrate efficient memory utilization (1M and 25GB) despite having the highest indexing time. 
For larger datasets, ELPIS has the lowest indexing memory footprint, occupying up to 40\% less memory than HNSW and 30\% less than Vamana during indexing. This is because ELPIS needs a smaller maximum out-degree and beam width compared to its competitors.
HNSW has a higher indexing memory footprint due to its use of a graph layout optimized for direct access to node edges through a large contiguous block allocation~\cite{url/hnsw}. This layout offers a time advantage over adjacency lists by reducing memory indirections and cache misses. However, it can result in quadratic memory growth when using a large maximum out-degree on large-scale datasets.
In Figure \ref{fig:elpis:idx:footprint:disk}, we compare the size of method indices, including the raw data. The figure shows that certain methods, such as EFANNA, HCNNG, KGraph, and consequently NSG, SSG, and DPG (which use one of these base graphs), exhibit a significantly larger memory footprint relative to their final index size. For instance, HCNNG consumes substantial memory during indexing, requiring over 200GB for Deep25GB (Fig. \ref{fig:elpis:idx:footprint:memory}) due to merging multiple MST from numerous samples generated during hierarchical clusterings. In contrast, its final index size is less than 50GB (Fig. \ref{fig:elpis:idx:footprint:disk}).

\subsection{Search Performance}
\karima{
We now evaluate the search performance of the different methods. All methods were included in the 1M experiments. Some methods were excluded in 25GB plots (KGraph, DPG, SPTAG-KDT, HCNNG, and EFANNA) for the sake of clarity, as their search was significantly slower than the best baselines. Full results are in~\cite{url/GASS}.
Other methods were omitted from the 100GB and 1B dataset sizes due to various limitations. The indexes for SPTAG and NGT could not be built on the larger datasets within 48 hours. EFANNA was excluded due to its high footprint, and likewise for methods based on it such as NSG and SSG.
Finally, KGraph, DPG, and LSHAPG were excluded due to unsatisfactory results on 1M and 25GB.
}


\noindent\textbf{Query Memory Footprint and Beam Width.}
Figure~\ref{fig:elpis:query:footprint:memory} indicates that Vamana, followed by ELPIS, have the lowest memory footprint during search. Even though ELPIS has a smaller index size, it adopts a contiguous memory storing during search, which increases the index footprint when loaded into memory. Besides, Figure~\ref{fig:elpis:query:beam-width} shows that Elpis requires the smallest beam width to reach similar query accuracy. Having a very high beam width indicates that the beam search requires to visit a wider area to and making more distance calculations to retrieves the NN answers.

\begin{figure}[h]
  \centering      
  \begin{minipage}{\columnwidth}
        \includegraphics[width=\columnwidth]{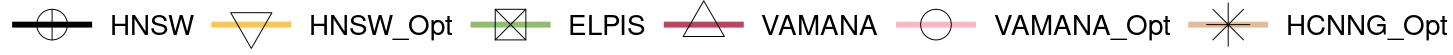}
        \end{minipage}
 \begin{minipage}{0.73\columnwidth}
 \centering
    \begin{minipage}{\textwidth}
        \centering
        \captionsetup{justification=centering}
        \begin{subfigure}{0.32\textwidth}
            \centering
            \captionsetup{justification=centering}
            \includegraphics[width=\textwidth]{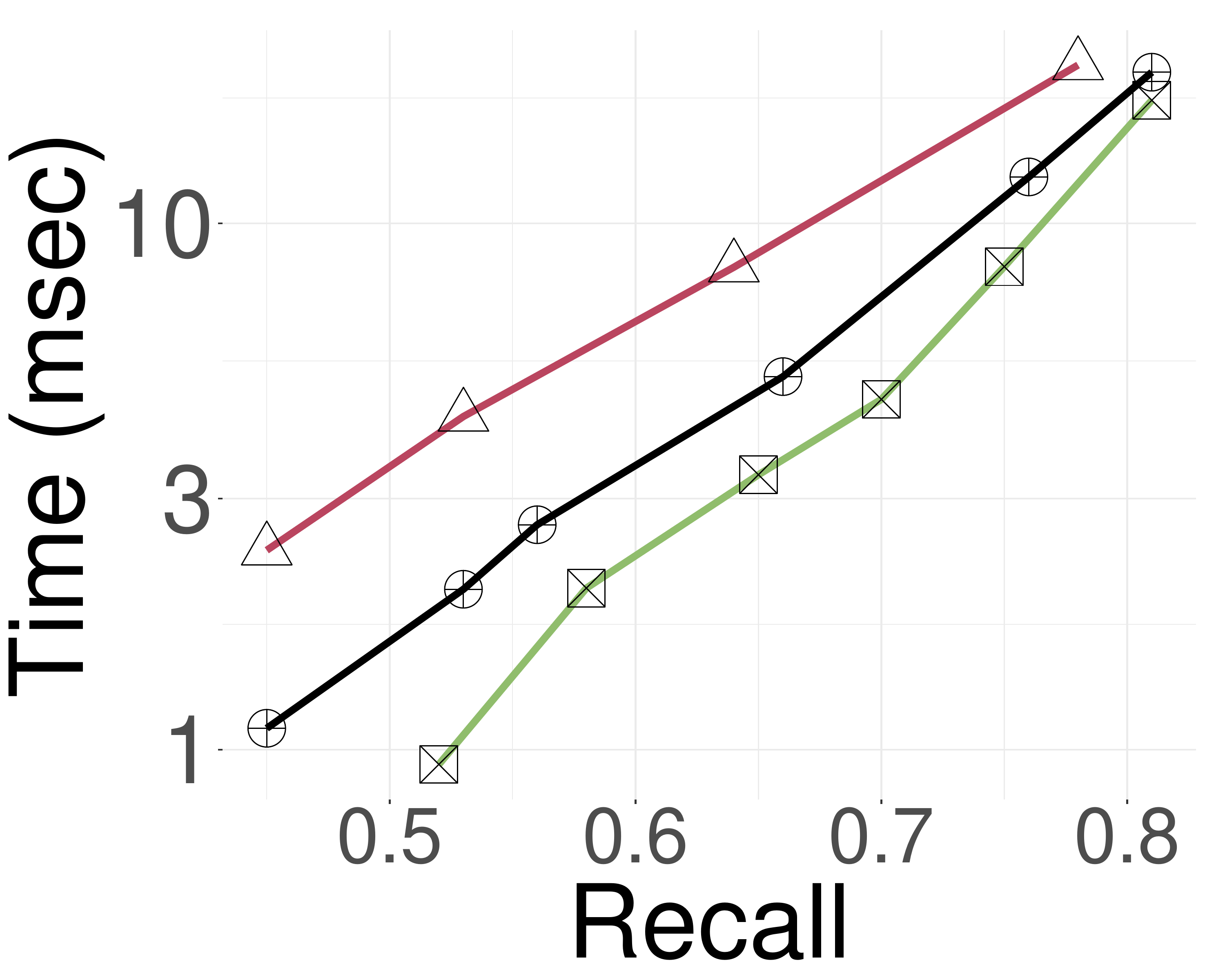}
            \caption{\karima{Text2Image}}
            \label{fig:elpis:query:performance:1B:t2i:10NN}
        \end{subfigure}
        \begin{subfigure}{0.305\textwidth}
            \centering
            \captionsetup{justification=centering}
            \includegraphics[width=\textwidth]{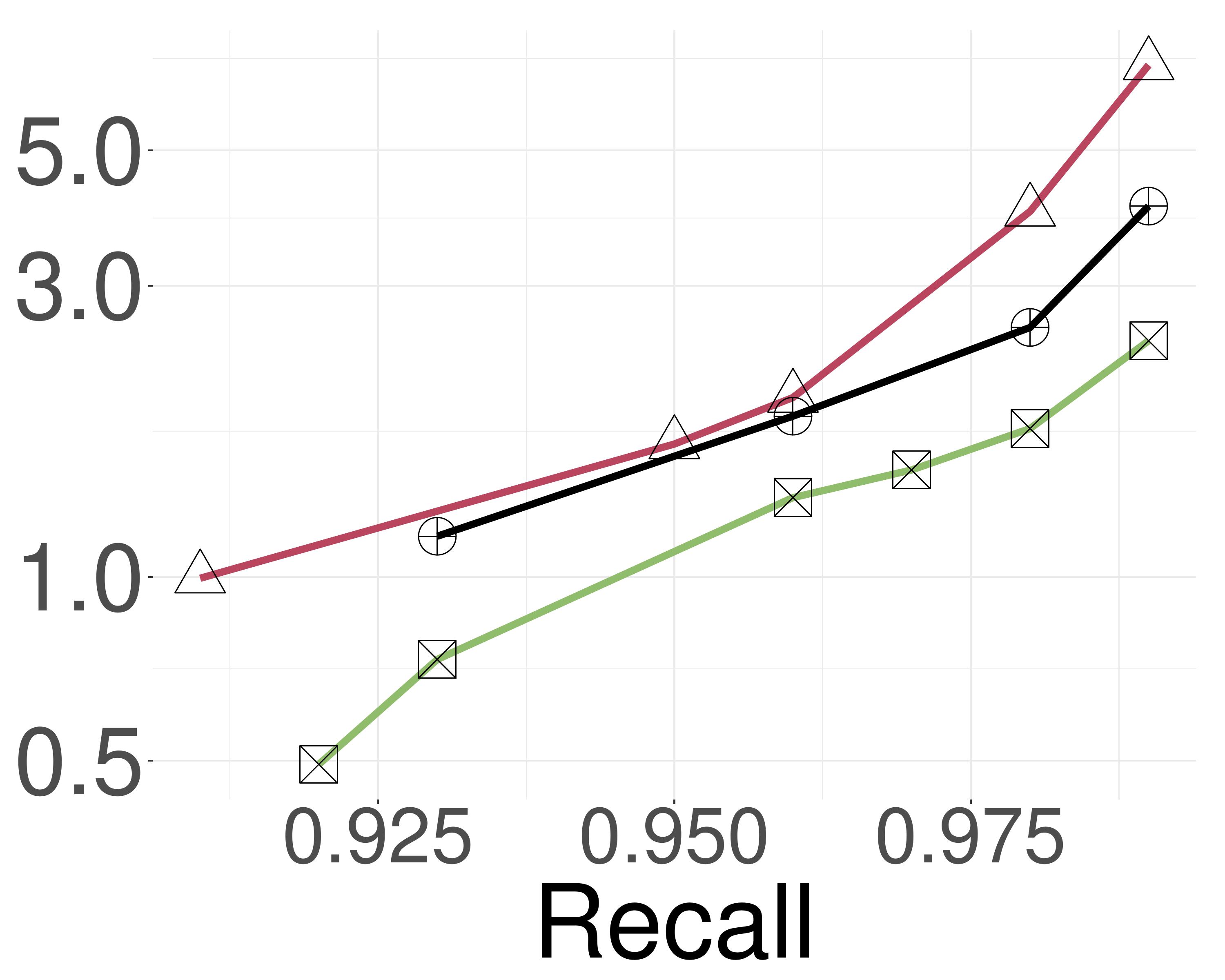}
            \caption{\karima{Deep}}
            \label{fig:elpis:query:performance:1B:deep:10NN}
        \end{subfigure}
        \begin{subfigure}{0.305\textwidth}
            \centering
            \captionsetup{justification=centering}
            \includegraphics[width=\textwidth]{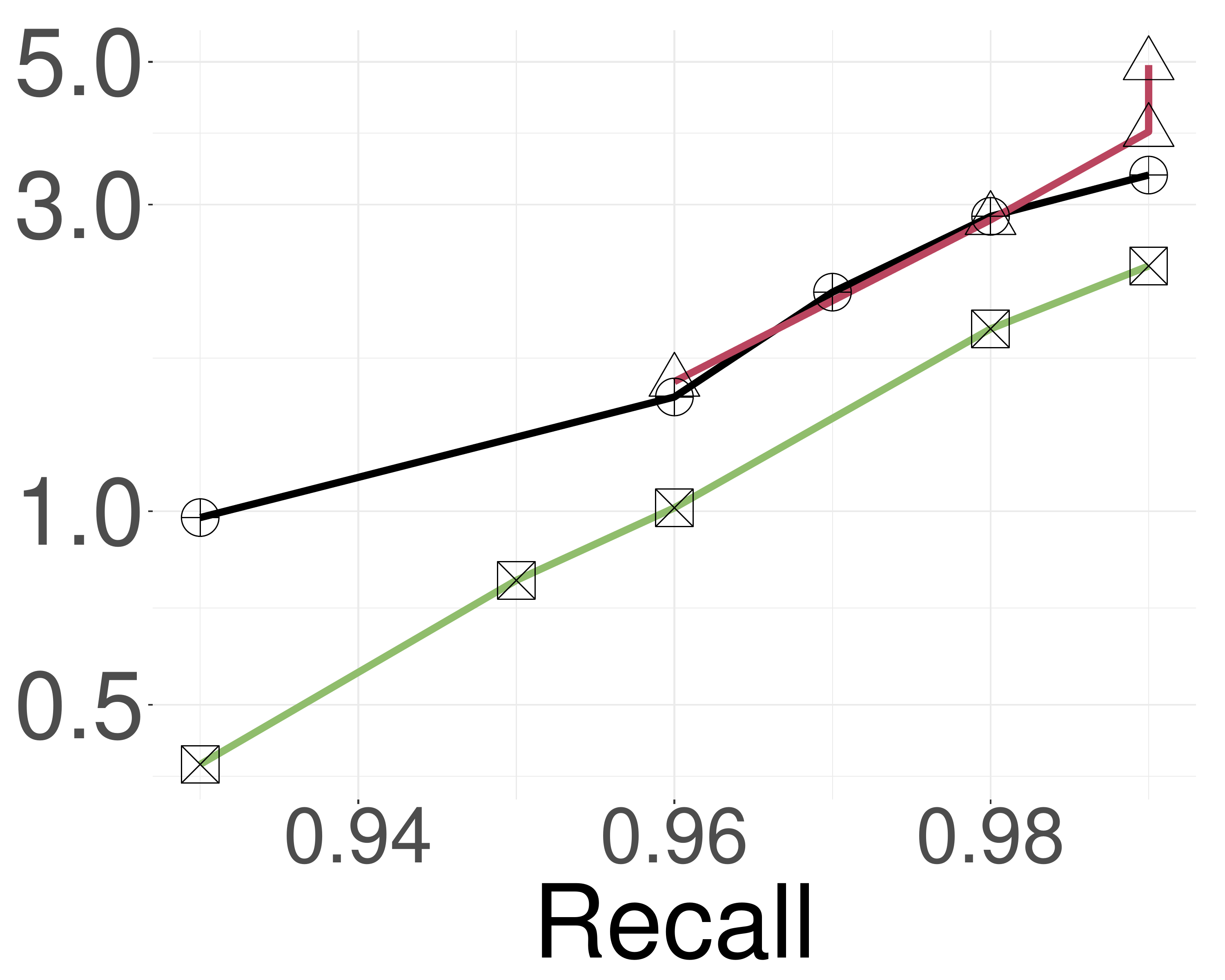}
            \caption{\karima{Sift}}
            \label{fig:elpis:query:performance:1B:sift:10NN}
        \end{subfigure}
    \end{minipage}
    \caption{\karima{1B Datasets}}
    \label{fig:elpis:query:performance:1B}
    \end{minipage}
    \begin{minipage}{0.26\columnwidth}
        \centering
        \captionsetup{justification=centering}
        \includegraphics[width=\textwidth]{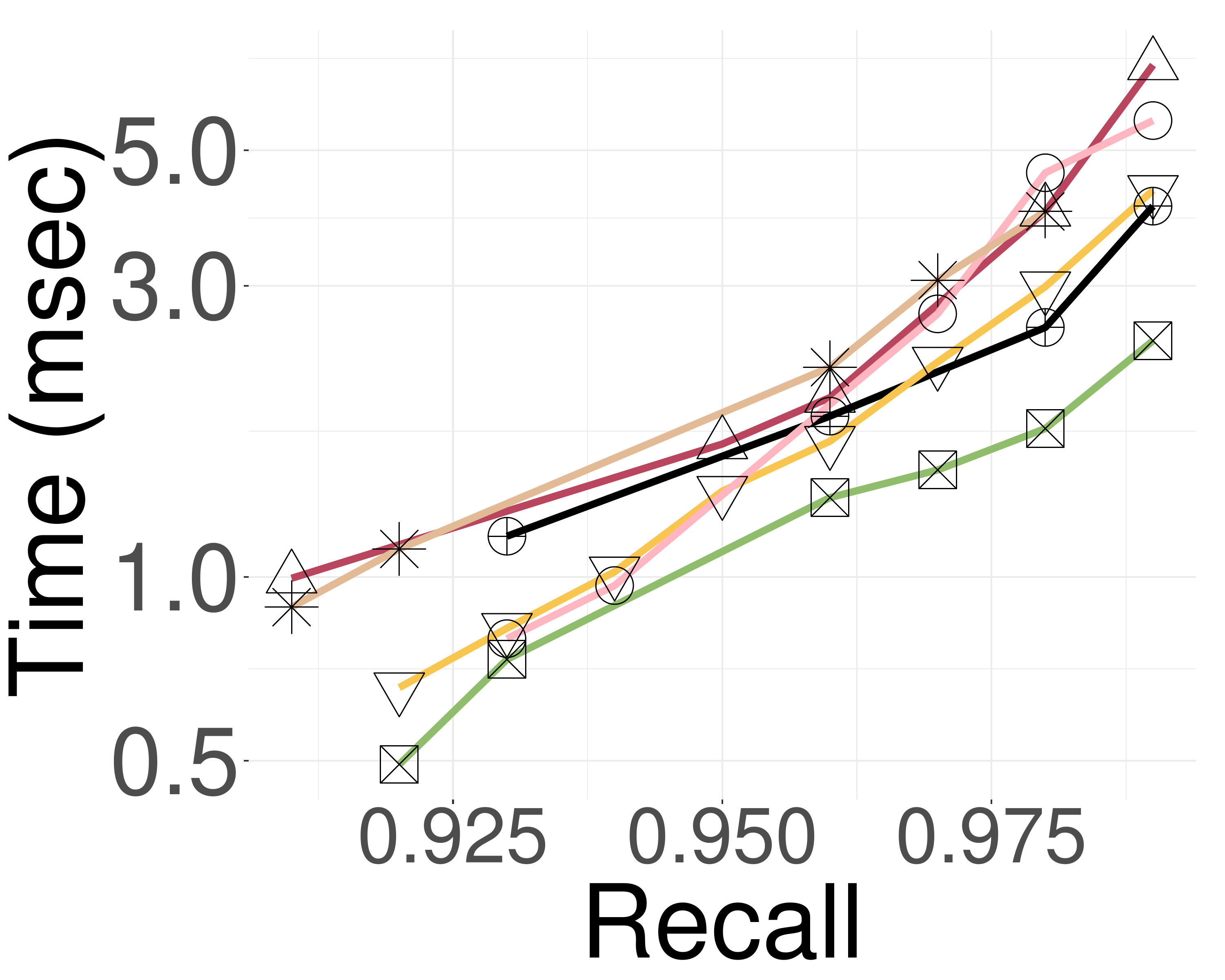}
        \caption{\karima{Optimized Implementations (Deep1B)}}
        \label{fig:optimized_impl}
    \end{minipage}
\end{figure}

\noindent\textbf{Real Datasets.} 
\karima{On datasets with 1M vectors (Figure~\ref{fig:elpis:query:performance:1M}), ELPIS and NSG/SSG perform best on Sift1M, achieving the highest performance for 0.99 and lower recall, respectively. For Seismic1M, HCNNG and ELPIS share the top rank. NGT, SSG, and NSG excel on Deep1M, while HCNNG leads on SALD1M at the highest recall, followed by SPTAG and NSG at lower recall levels. In ImageNet1M, NSG/SSG and HNSW rank as the top performers. 
Across most scenarios, SSG and NSG show similar performance. However, LSHAPG demonstrates limitations, requiring more computation to achieve high accuracy. Its probabilistic rooting prunes promising neighbors, requiring a larger beam width and tighter $L$-bsf lower-bound distance during search.}
\karima{When moving to 25GB datasets (Figure~\ref{fig:elpis:query:performance:25GB}), SSG, NSG, NGT and HCNNG experience a drop in performance, and ELPIS takes the lead with the best overall performance together with SPTAG-BKT for SALD25GB .} 
It is worth noting that none of the methods achieved an accuracy over 0.8 on the Seismic dataset, leading us to report results for these lower recall values. 
The significant indexing footprint of NSG prevented us from extending its evaluation to larger datasets, as constructing the EFANNA graph (which NSG depends on) requires more memory than the available 1.4TB.
For hard query workloads in Figure~\ref{fig:search:query:performance:25GB:hard} we compare the best-performing methods from the two most performing graph paradigms, ND-based and DC-based methods, including HNSW, NSG, ELPIS and SPTAG-BKT. 
SPTAG-BKT achieves the overall best performance for 1\% noise query set, as we increase the noise up to 10\%, SPTAG-BKT's performance deteriorates, which we can relate to SPTAG BKT structures failing to identify good seed points. At the same time, the other competitors gain an advantage, with ELPIS taking the lead. 
When analyzing very large datasets of 1 billion vectors, Figure~\ref{fig:elpis:query:performance:1B} shows the superiority of ELPIS which is up to an order of magnitude faster at achieving 0.95 accuracy, thanks to its design that supports multi-threading for single query answering
This trend is consistent across subsets ranging from 100GB (Figure~\ref{fig:elpis:query:performance:100GB}) to 250GB (detailed results are reported in~\cite{url/GASS}).

\noindent{\bf Data Distributions.} We assess top performers representing different paradigms (EFANNA, Vamana, SSG, HNSW, ELPIS, and SPTAG-BKT) on challenging datasets (Fig. ~\ref{fig:datacomp}). Results (Figs. ~\ref{fig:query:performance:25GB:rand:pow1:10NN} and ~\ref{fig:query:performance:25GB:rand:pow50:10NN}) indicate that ELPIS consistently achieves high accuracy across skewness levels (0 to 50), outperforming other methods. As skewness increases, search becomes easier so most graph-based approaches improve but ELPIS maintains its superiority.

\newcommand{\soneM}{0.27}
\begin{figure}[!htb]
    \centering
    \begin{minipage}{\textwidth}
        \begin{subfigure}{\textwidth}
            \centering
            \captionsetup{justification=centering}	
            \includegraphics[width=\textwidth]{../img-png/Experiments/legendall.png}
        \end{subfigure}
    \end{minipage}
    
    \begin{minipage}{\textwidth}
        \centering
        \captionsetup{justification=centering}
        \captionsetup[subfigure]{justification=centering}
        \begin{subfigure}{0.025\textwidth}
            \raisebox{1.2cm}{\includegraphics[width=\textwidth]{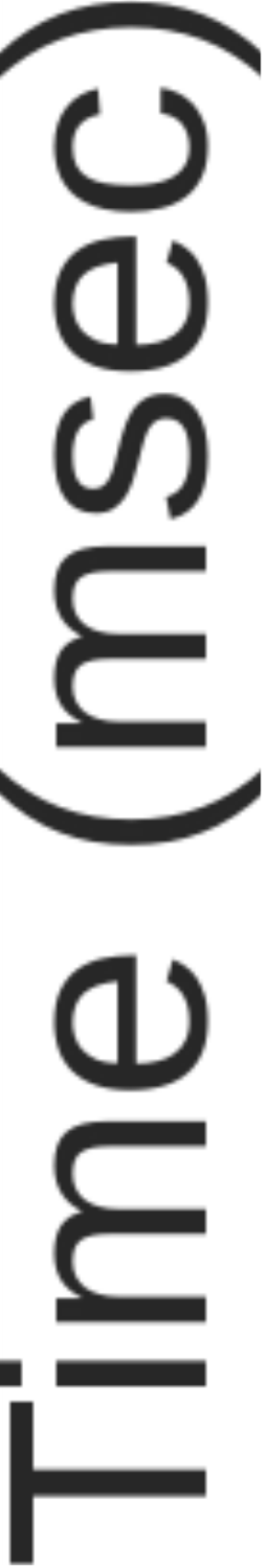}} 
        \end{subfigure}
        \begin{subfigure}{\soneM\textwidth}
            \includegraphics[width=\textwidth]{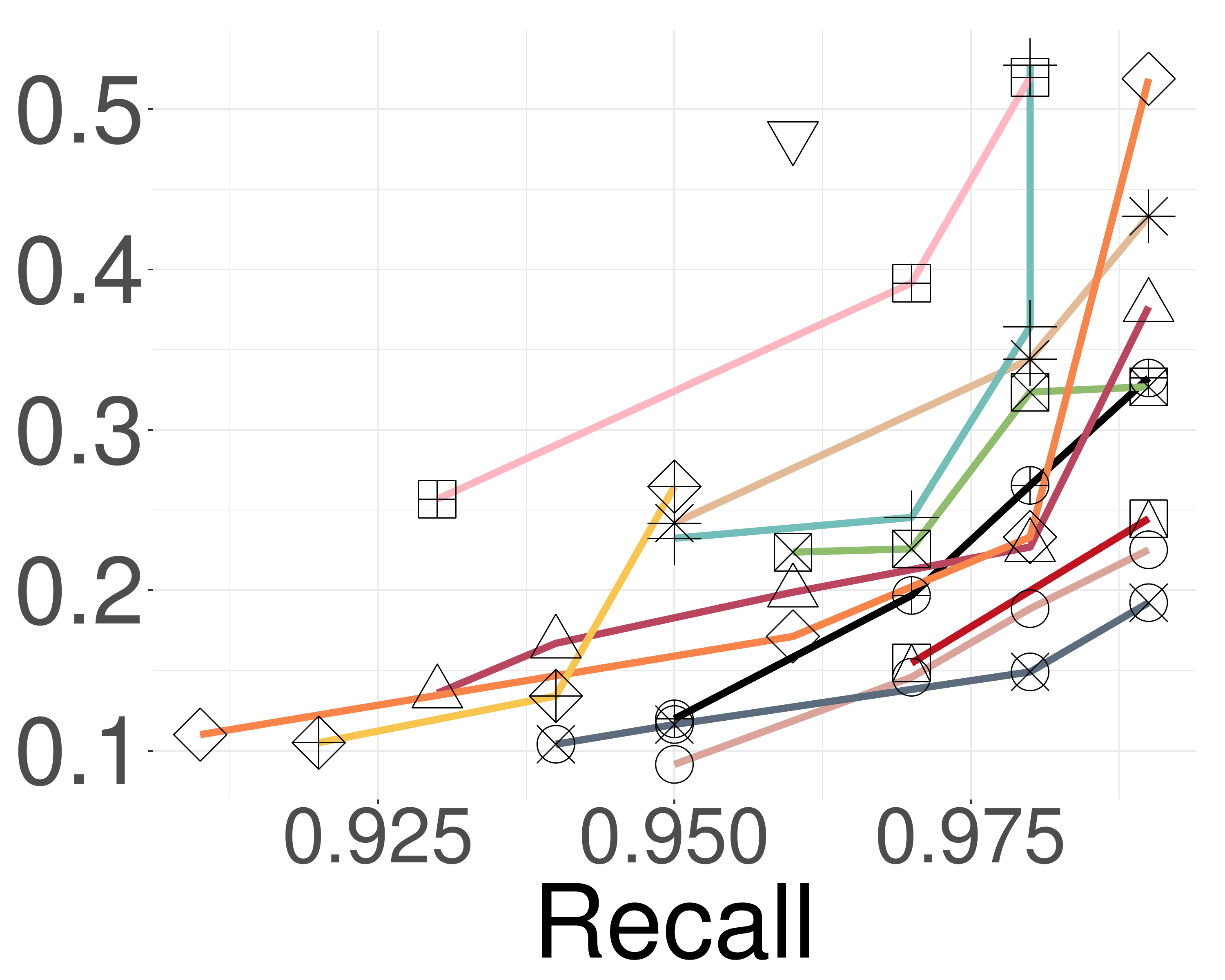}
            \caption{\karima{\textbf{Deep}}} 
            \label{fig:elpis:query:performance:1M:deep:10NN}
        \end{subfigure}
        \begin{subfigure}{\soneM\textwidth}
            \includegraphics[width=\textwidth]{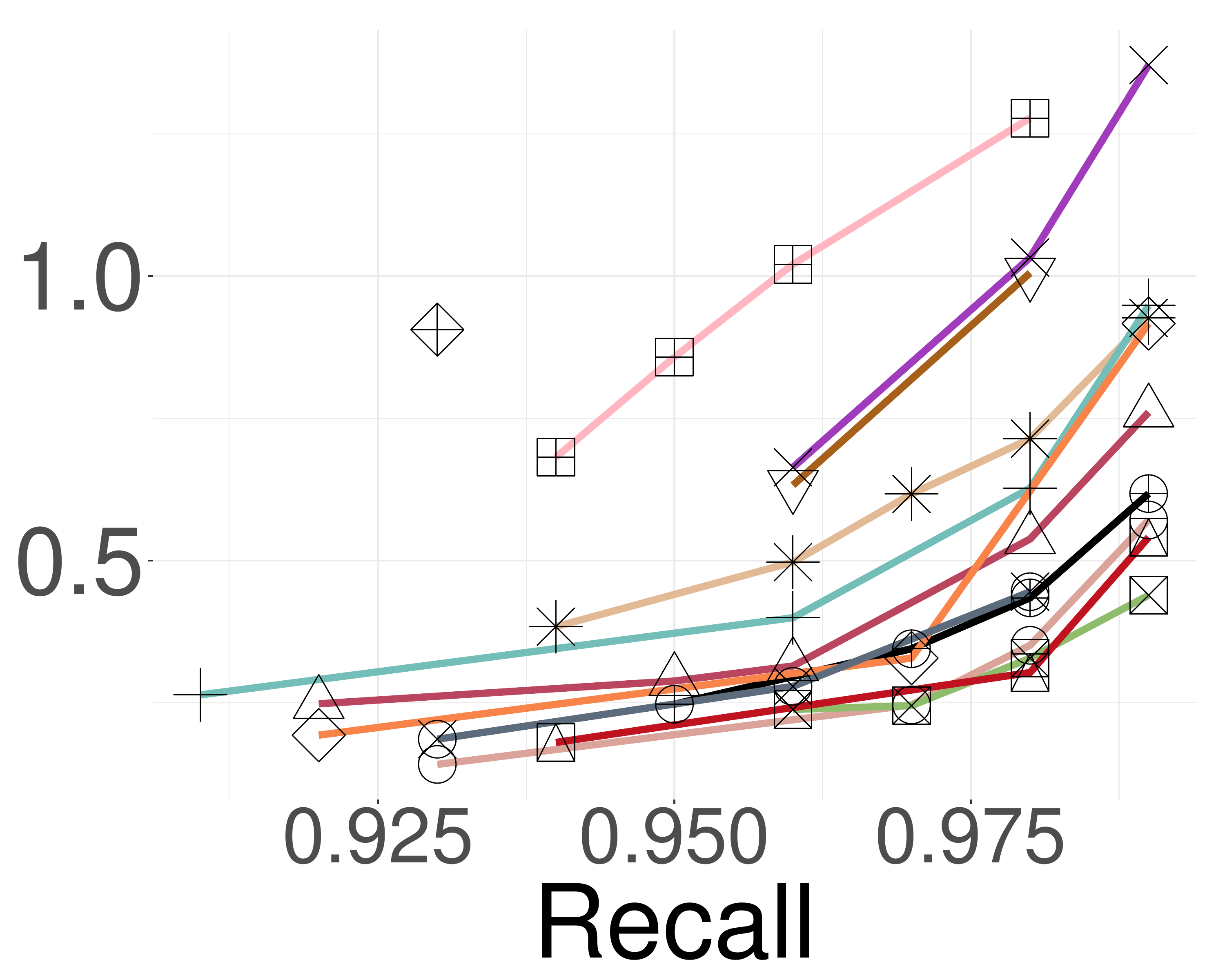}
            \caption{\karima{\textbf{Sift}}} 
        \label{fig:elpis:query:performance:1M:sift:10NN}
        \end{subfigure}
        \begin{subfigure}{\soneM\textwidth}
            \includegraphics[width=\textwidth]{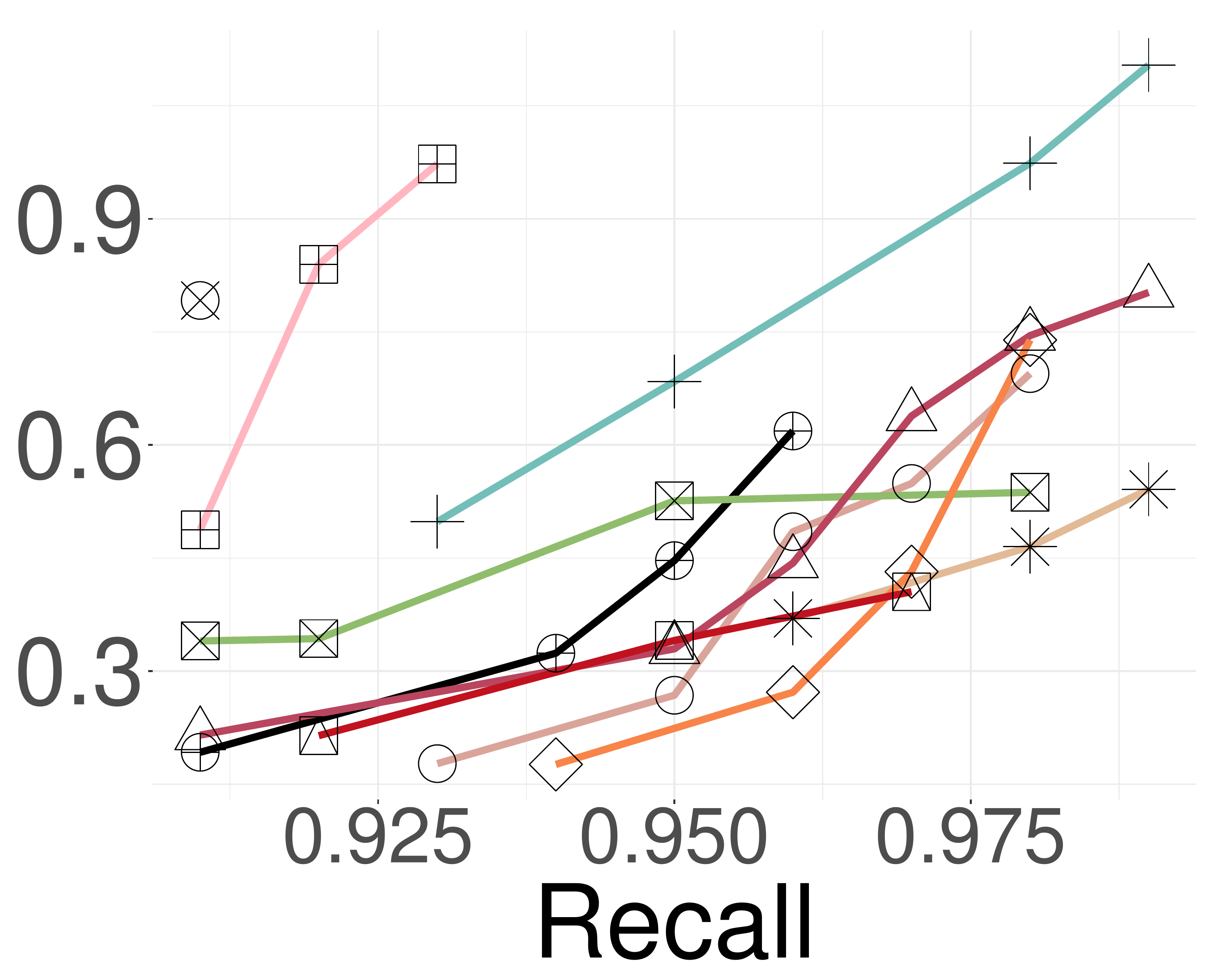}
            \caption{\karima{\textbf{SALD}}} 
            \label{fig:elpis:query:performance:1M:sald:10NN}
        \end{subfigure}
        \\
        \begin{subfigure}{0.025\textwidth}
            \raisebox{1.2cm}{\includegraphics[width=\textwidth]{../img-png/Experiments/time.png}} 
        \end{subfigure}
        \begin{subfigure}{\soneM\textwidth}
            \includegraphics[width=\textwidth]{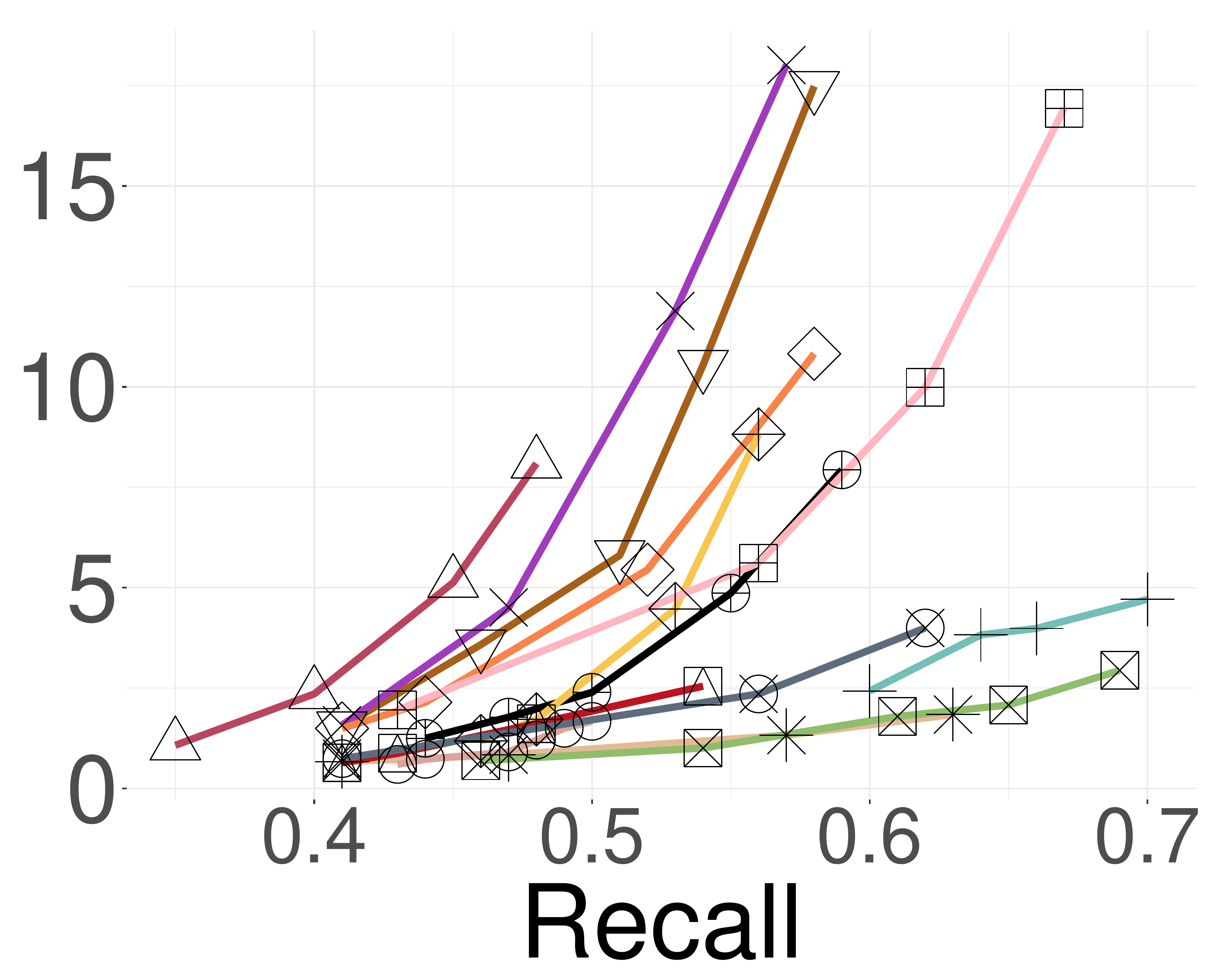}
            \caption{\karima{\textbf{Seismic}}} 
            \label{fig:elpis:query:performance:1M:seismic:10NN}
        \end{subfigure}
        \begin{subfigure}{\soneM\textwidth}
            \includegraphics[width=\textwidth]{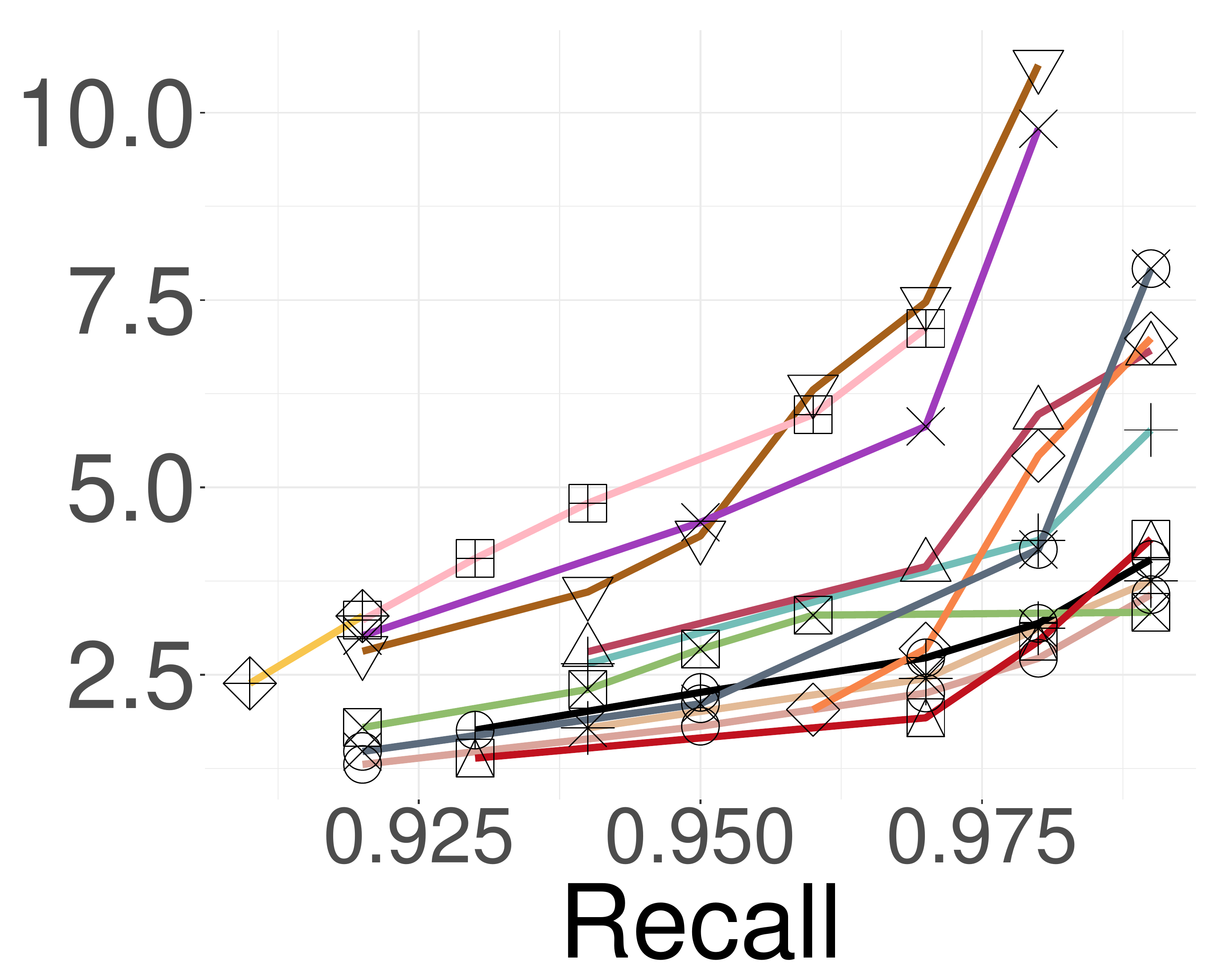}
            \caption{\karima{\textbf{Gist}}} 
            \label{fig:elpis:query:performance:1M:gist:10NN}
        \end{subfigure}
        \begin{subfigure}{\soneM\textwidth}
            \includegraphics[width=\textwidth]{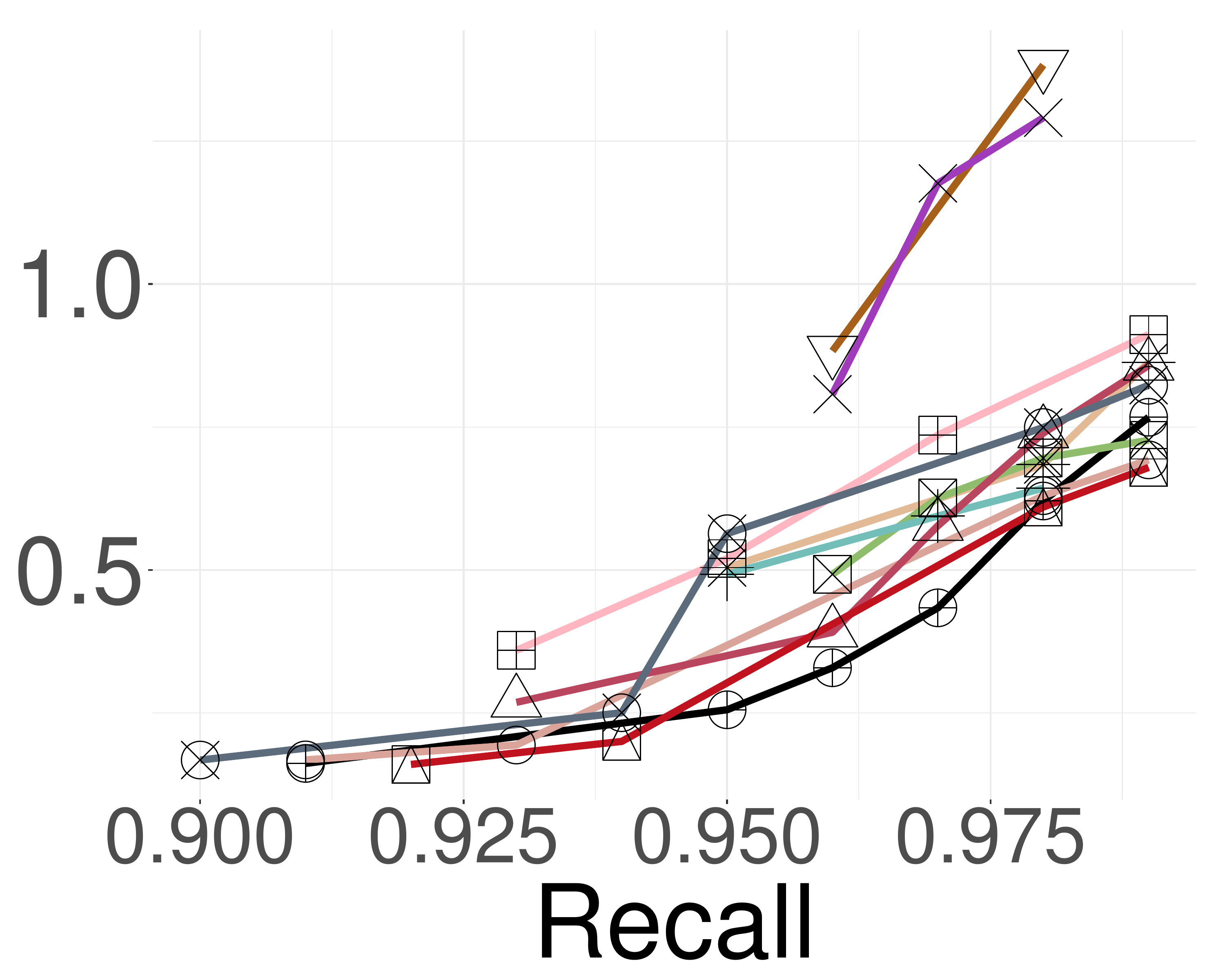}
            \caption{\karima{\textbf{Imagenet}}} 
            \label{fig:elpis:query:performance:1M:imagenet:10NN}
        \end{subfigure}
        
        \caption{\karima{Query performance on 1M vectors}}
        \label{fig:elpis:query:performance:1M}
    \end{minipage}
\end{figure}

\newcommand{\soneMs}{0.19}
\begin{figure}[!htb]
    \centering    
    \begin{minipage}{\textwidth}
        \begin{subfigure}{\textwidth}
            \centering
            \captionsetup{justification=centering}	
            \includegraphics[width=\textwidth]{../img-png/Experiments/legendall.png}
        \end{subfigure}
    \end{minipage}
    
	\begin{minipage}{0.6\textwidth}
		\captionsetup{justification=centering}
		\captionsetup[subfigure]{justification=centering}
        \begin{subfigure}{0.029\textwidth}
            \raisebox{1cm}{\includegraphics[width=\textwidth]{../img-png/Experiments/time.png}} 
        \end{subfigure}
		\begin{subfigure}{0.314\textwidth}
			\includegraphics[width=\textwidth]{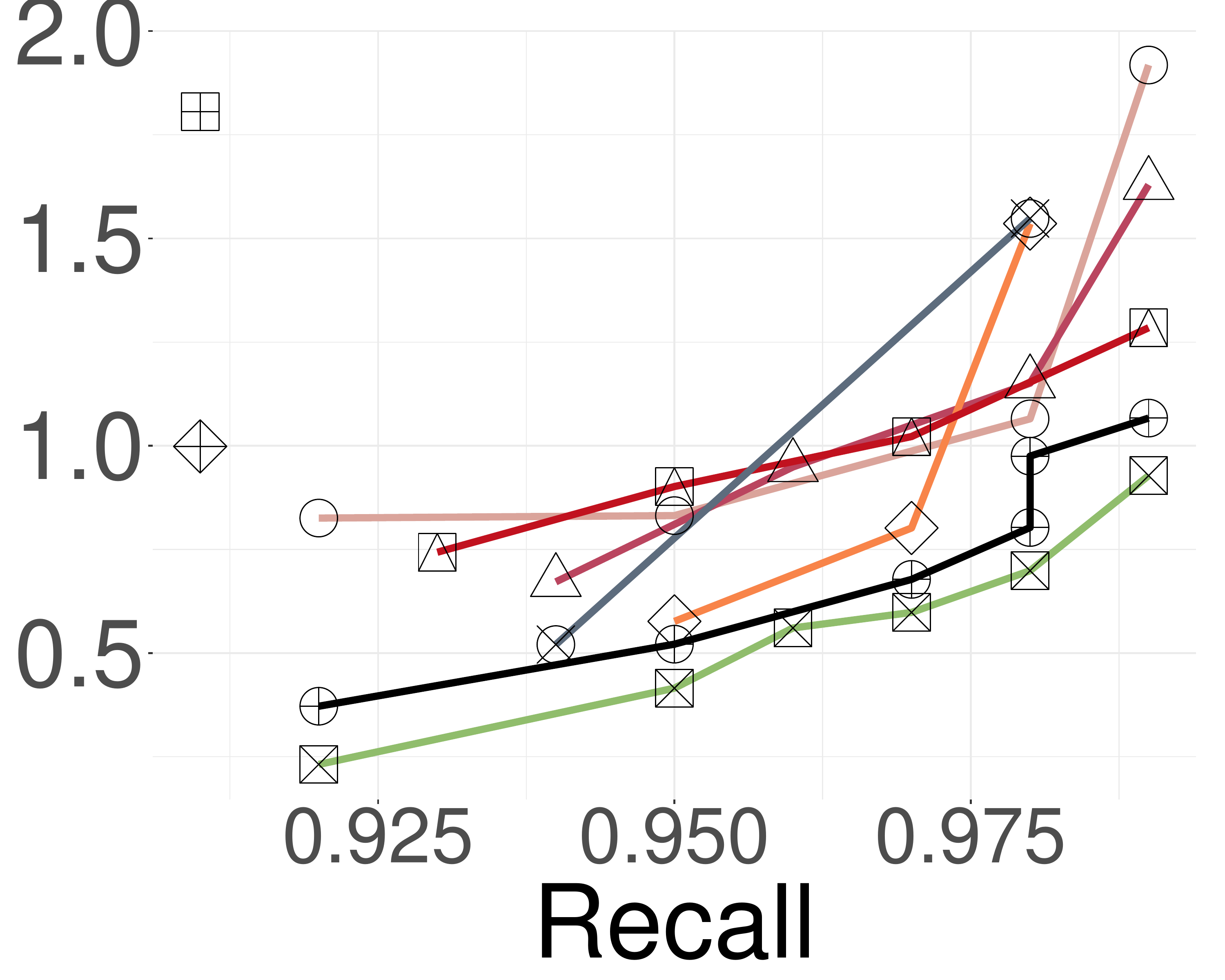}
			\caption{\karima{Deep}}  
			\label{fig:elpis:query:performance:25GB:deep:10NN}
		\end{subfigure}
		\begin{subfigure}{0.314\textwidth}
			\includegraphics[width=\textwidth]{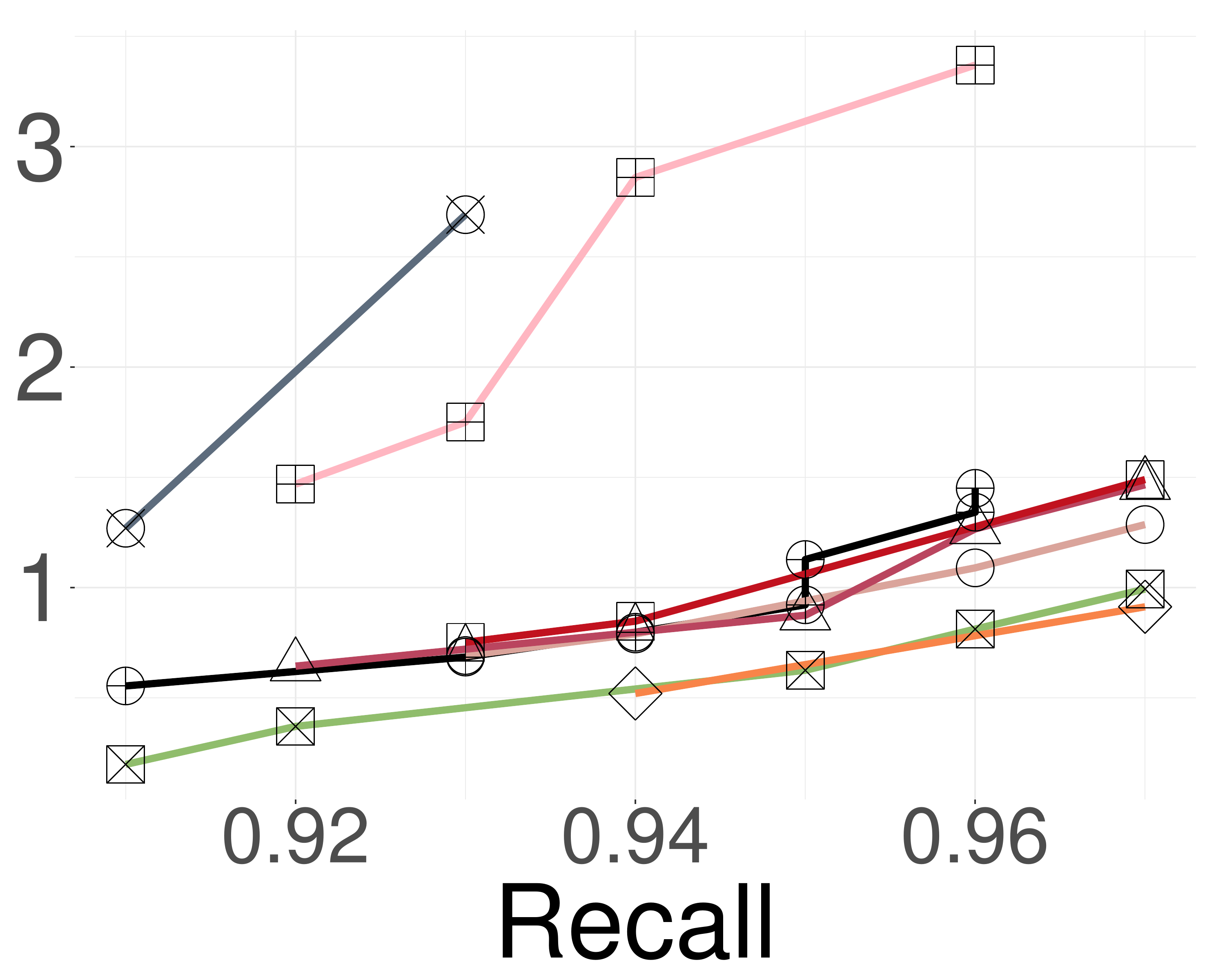}
			\caption{\karima{SALD}}  
			\label{fig:elpis:query:performance:25GB:sald:10NN}
		\end{subfigure}
		\begin{subfigure}{0.314\textwidth}
			\includegraphics[width=\textwidth]{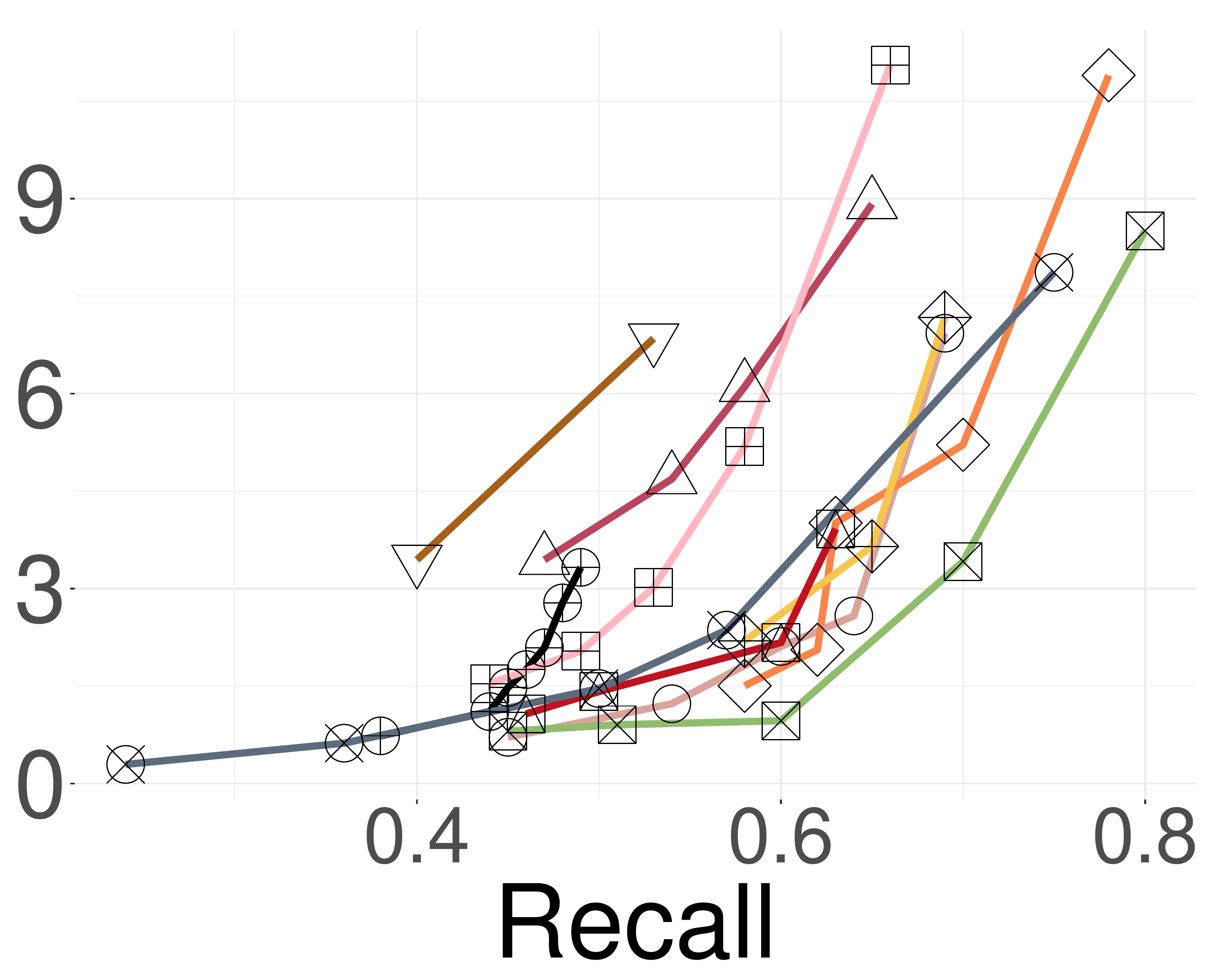}
			\caption{\karima{Seismic}}  
			\label{fig:elpis:query:performance:25GB:seismic:10NN}
		\end{subfigure}	

                \begin{subfigure}{0.029\textwidth}
            \raisebox{1cm}{\includegraphics[width=\textwidth]{../img-png/Experiments/time.png}} 
        \end{subfigure}
		\begin{subfigure}{0.314\textwidth}
			\includegraphics[width=\textwidth]{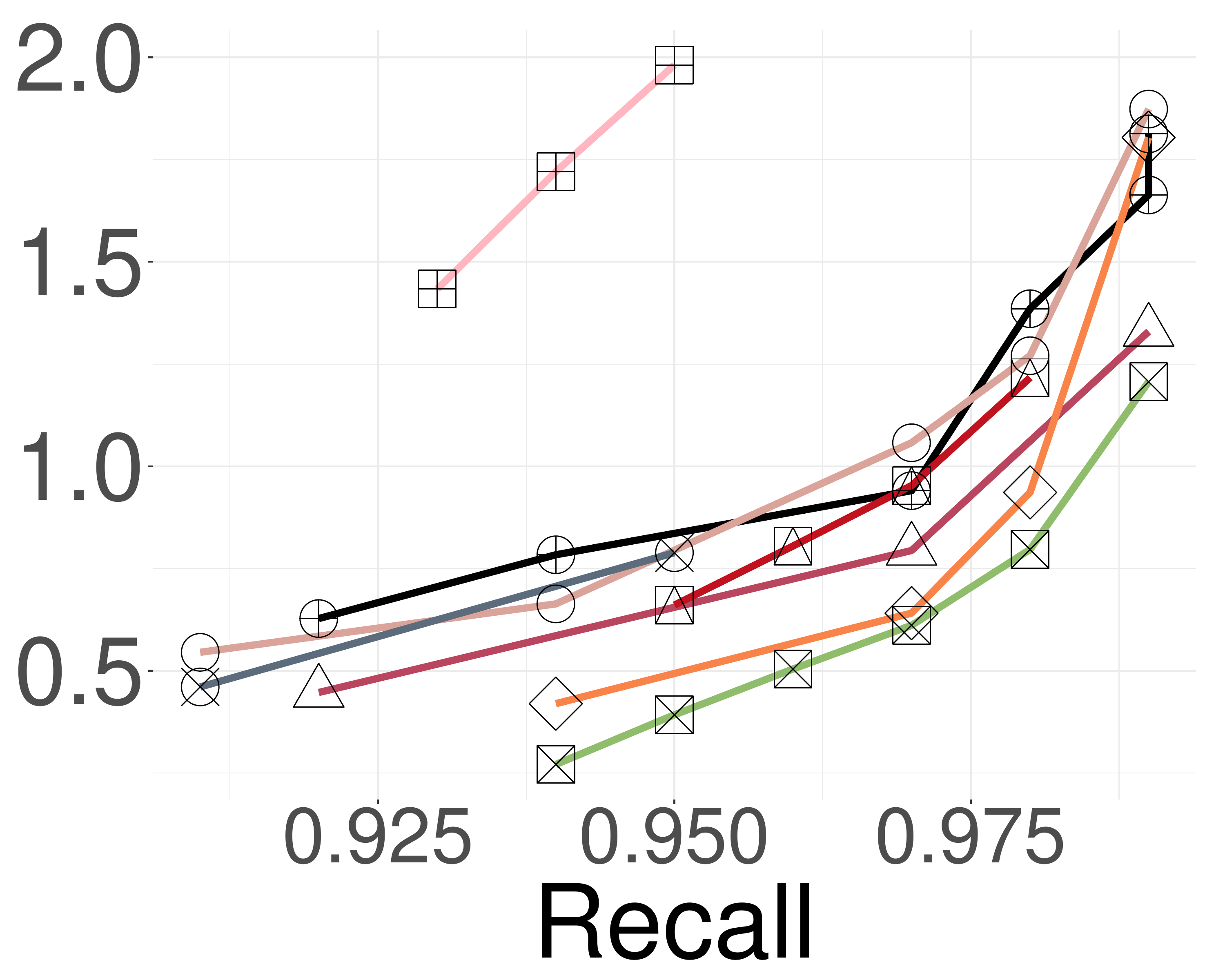}
			\caption{\karima{Sift}}  
			\label{fig:elpis:query:performance:25GB:sift:10NN}
		\end{subfigure}
	    \begin{subfigure}{0.314\textwidth}
		\includegraphics[width=\textwidth]{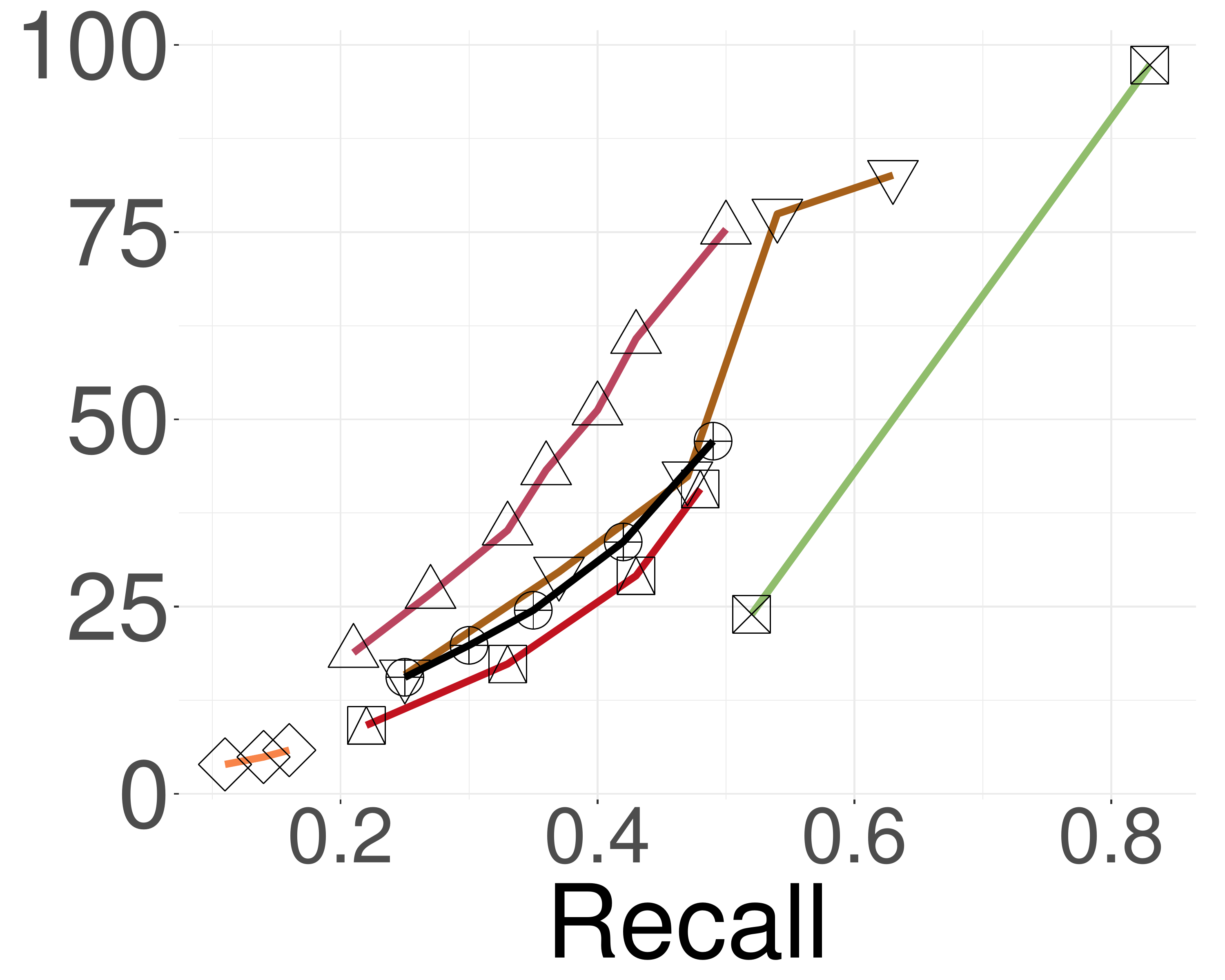}
		\caption{\karima{\textbf{RandPow0}}} 
		\label{fig:query:performance:25GB:rand:pow1:10NN}
	\end{subfigure}
	    \begin{subfigure}{0.314\textwidth}
		\includegraphics[width=\textwidth]{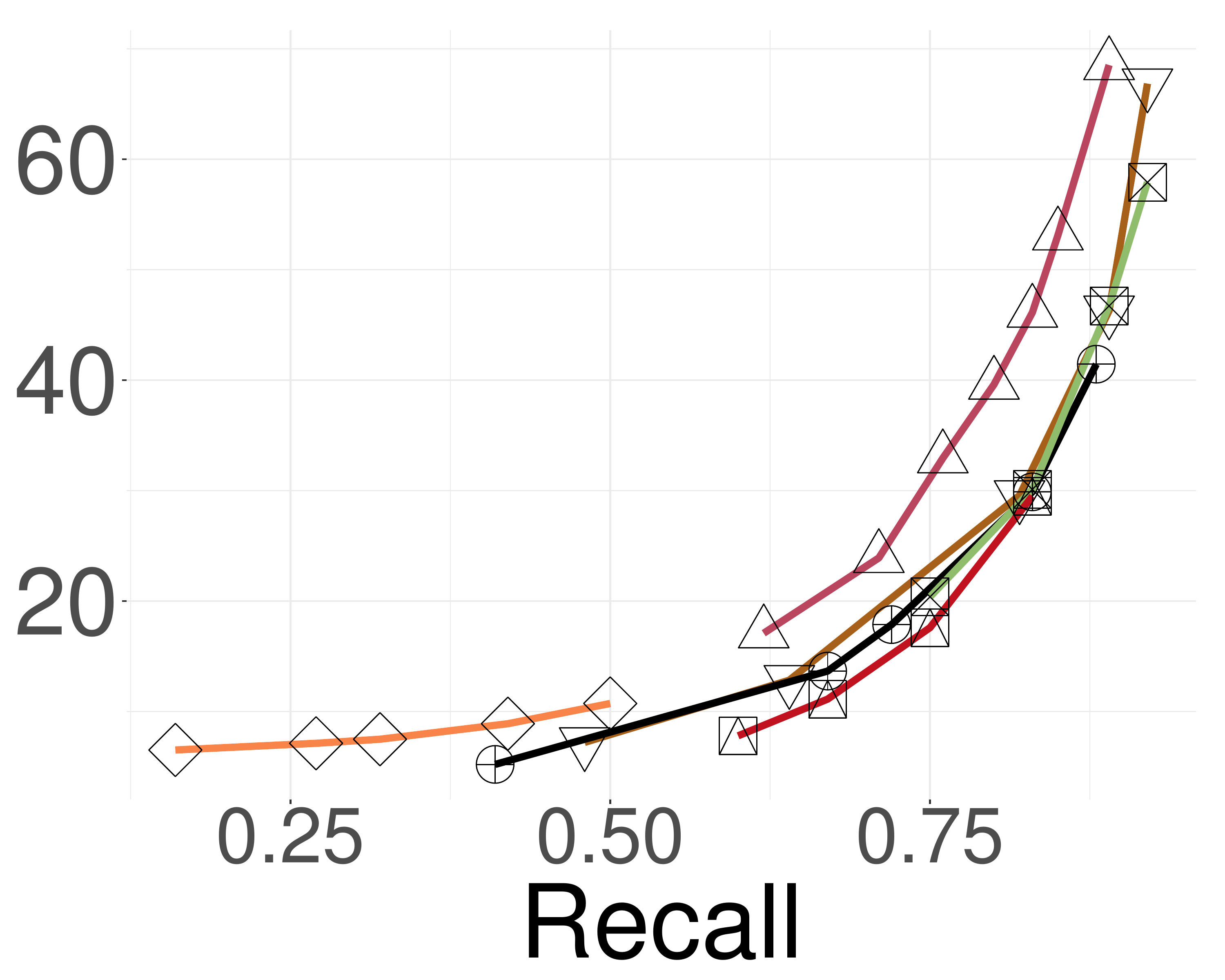}
		\caption{\karima{\textbf{RandPow50}}} 
		\label{fig:query:performance:25GB:rand:pow50:10NN}
	\end{subfigure}		
		\caption{{\karima{25GB datasets}}}
		\label{fig:elpis:query:performance:25GB}
	\end{minipage}
	\begin{minipage}{\soneMs\textwidth}
		\captionsetup{justification=centering}
		\captionsetup[subfigure]{justification=centering}
		\begin{subfigure}{\textwidth}
			\includegraphics[width=\textwidth]{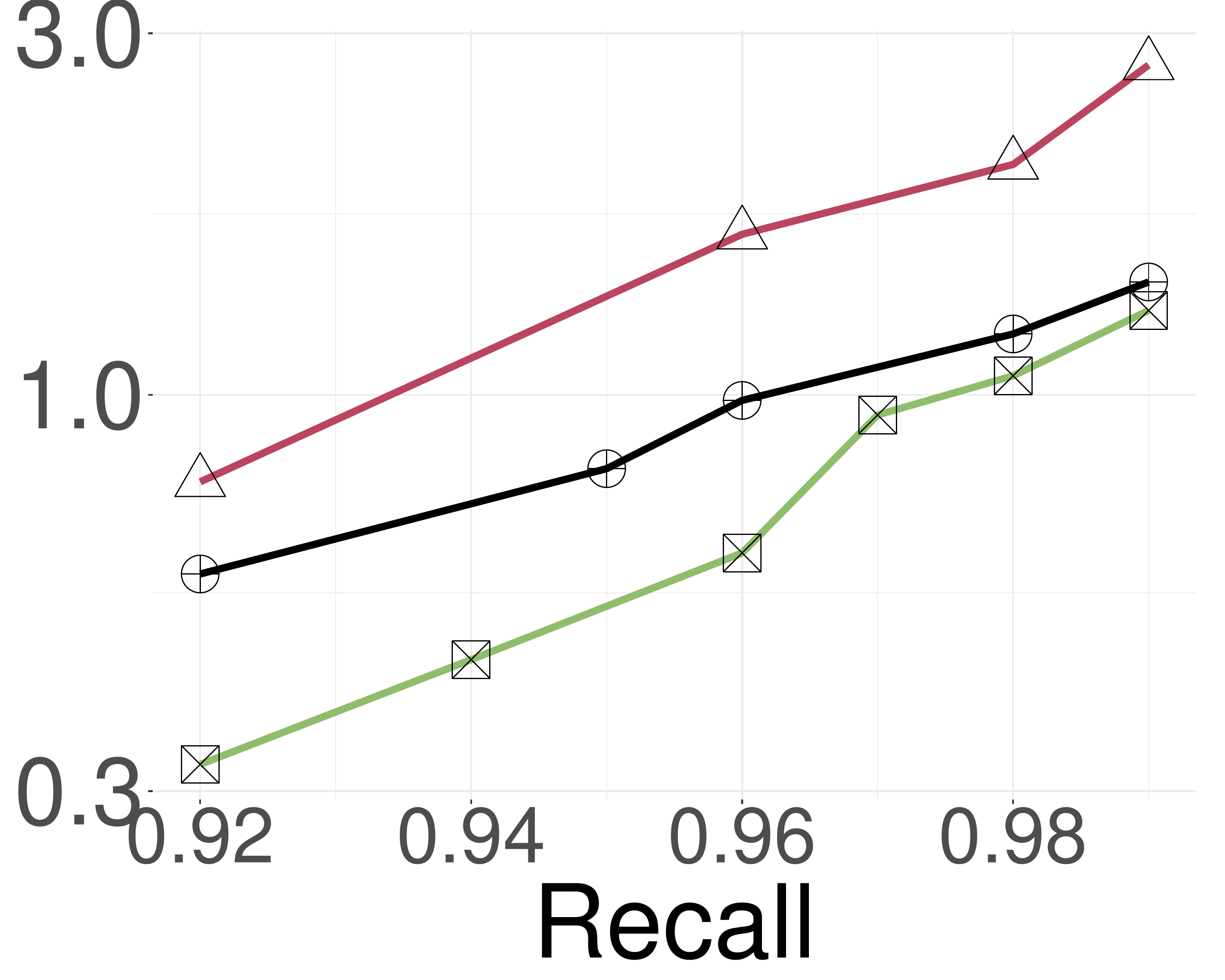}
			\caption{Deep}  
		\label{fig:elpis:query:performance:100GB:deep:10NN}
		\end{subfigure}
		\begin{subfigure}{\textwidth}
			\includegraphics[width=\textwidth]{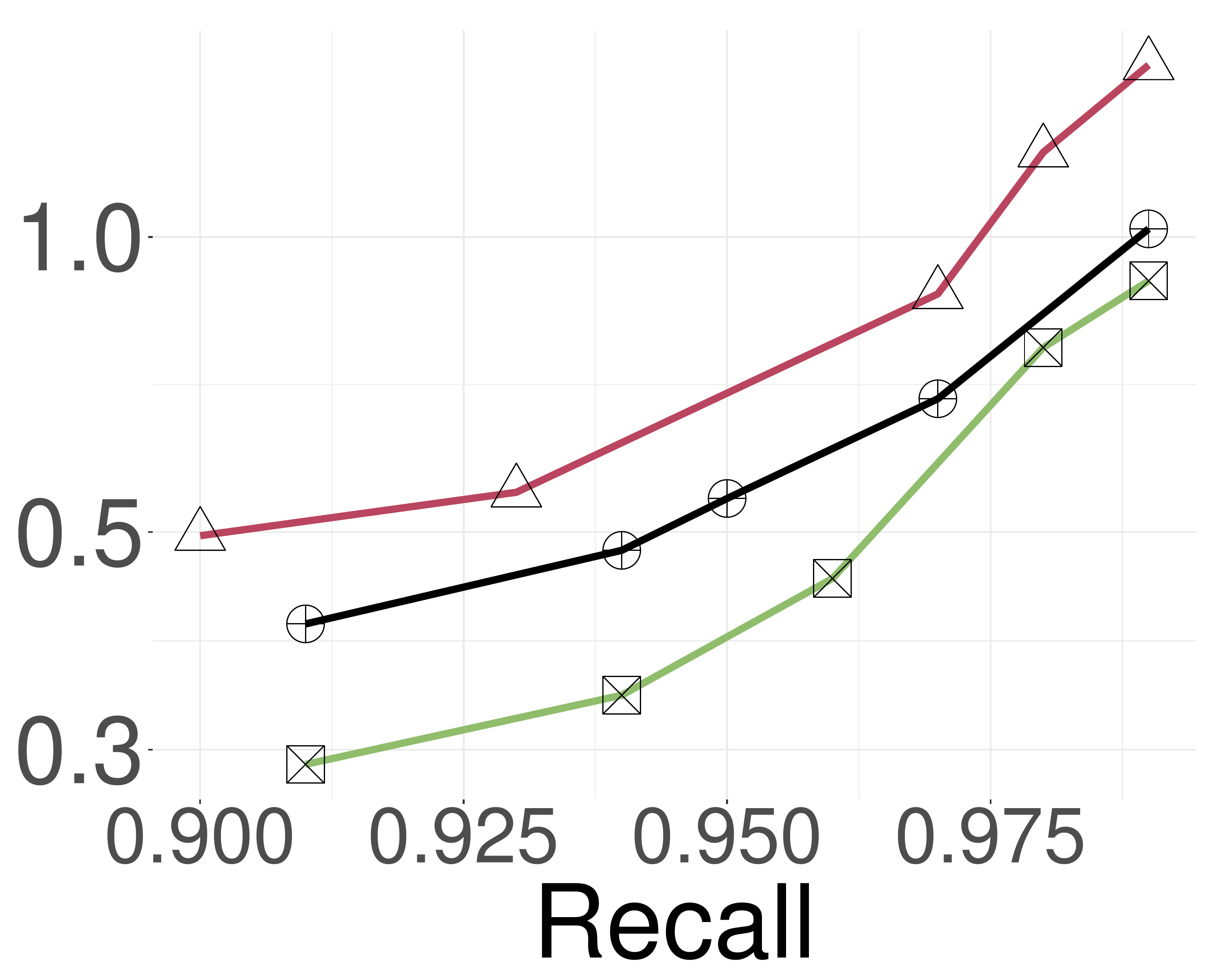}
			\caption{Sift}  
			\label{fig:elpis:query:performance:100GB:sift:10NN}
		\end{subfigure}
		\caption{{100GB datasets}}	
		\label{fig:elpis:query:performance:100GB}
	\end{minipage}
	\begin{minipage}{\soneMs\textwidth}
		\captionsetup{justification=centering}
		\captionsetup[subfigure]{justification=centering}
		\begin{subfigure}{\textwidth}
		\includegraphics[width=\textwidth]{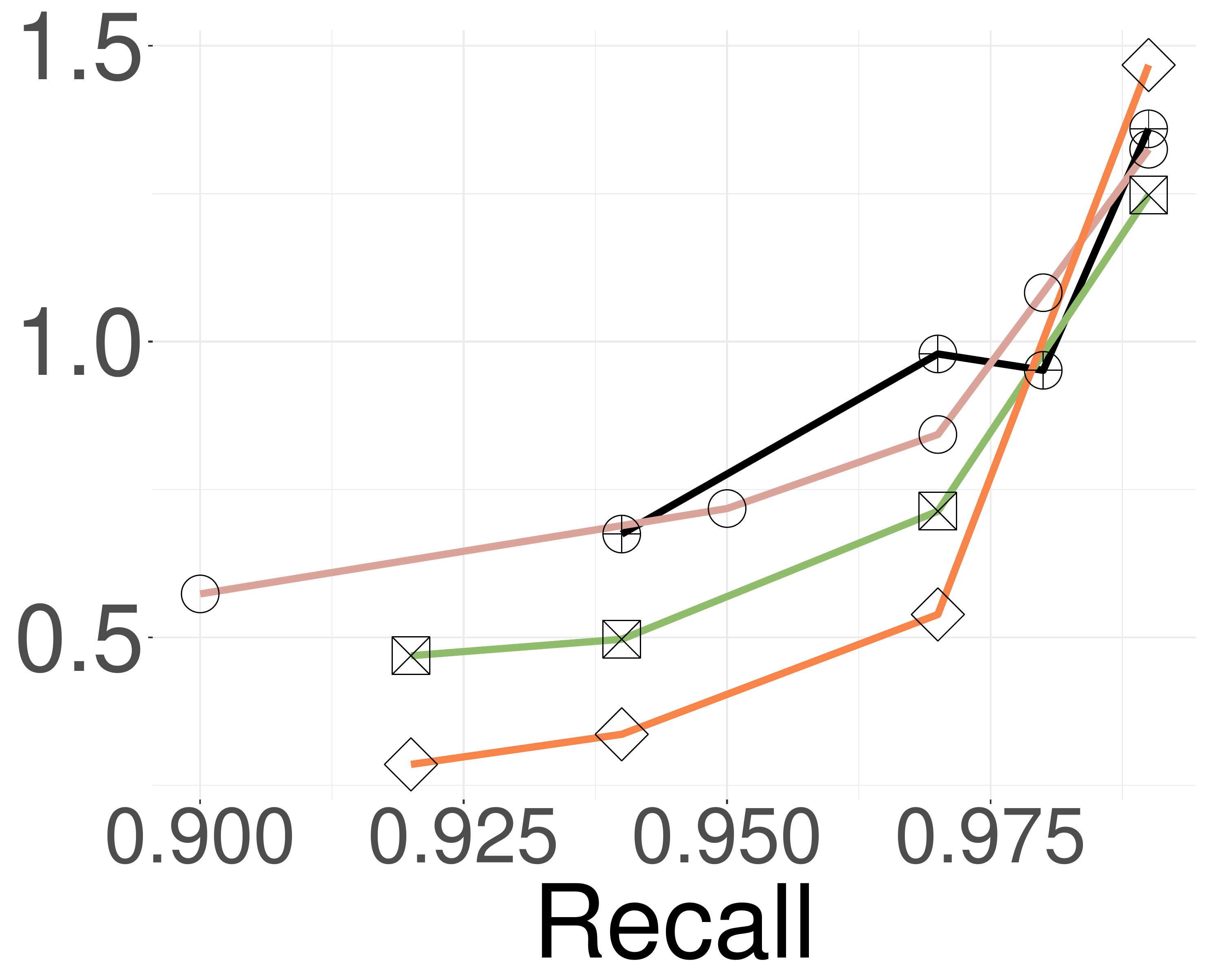}
		\caption{\textbf{\karima{1\% noise}}} 
		\label{fig:search:query:performance:25GB:hard:1p}
		\end{subfigure}
		\begin{subfigure}{\textwidth}
		\includegraphics[width=\textwidth]{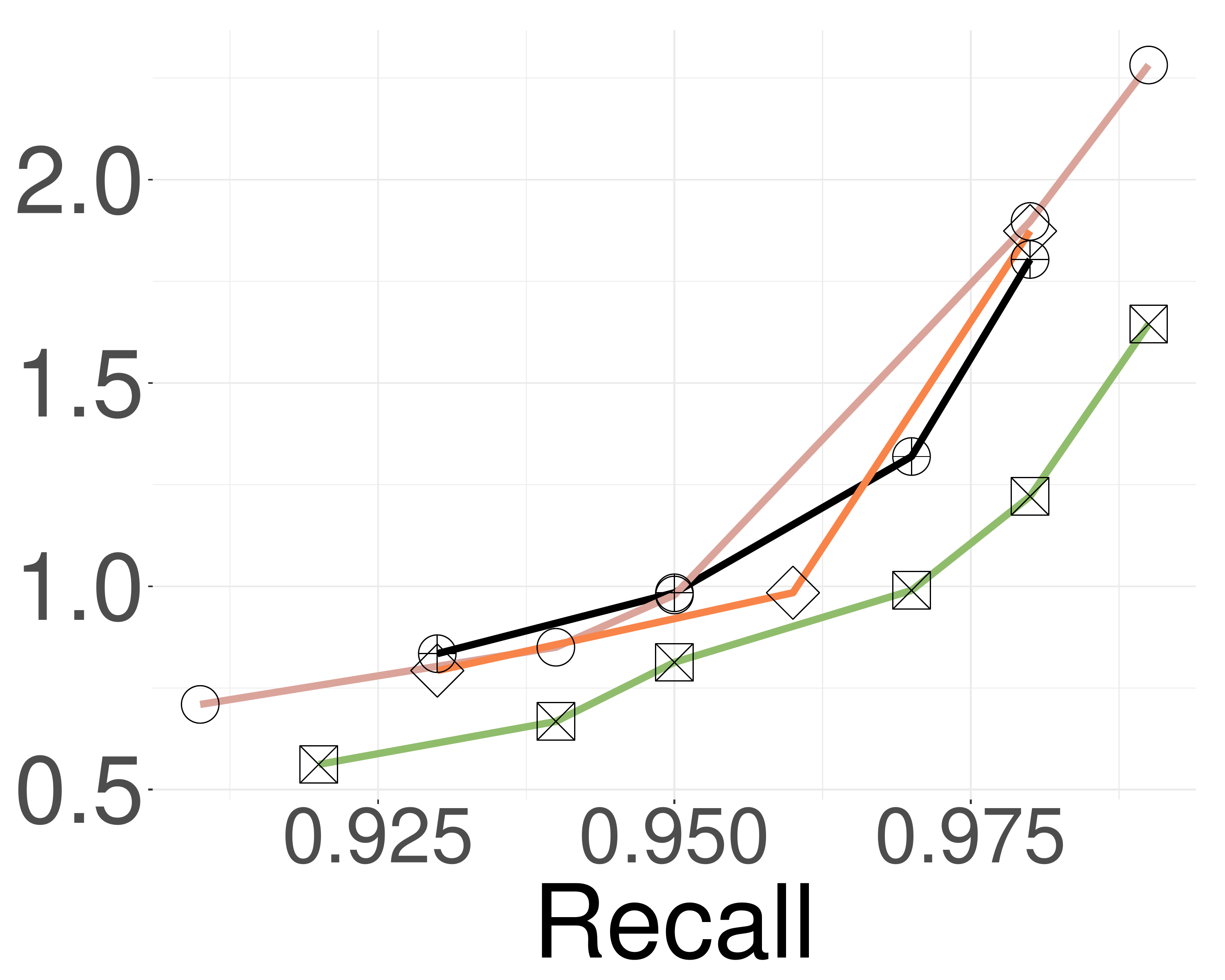}
		\caption{\textbf{\karima{10\% noise}}} 
		\label{fig:search:query:performance:25GB:hard:10p}
		\end{subfigure}
	\caption{\karima{Varying workloads}}
		\label{fig:search:query:performance:25GB:hard}
	\end{minipage}
\end{figure}

\karima{\noindent{\bf Implementation Impact.} 
We evaluate the performance of original implementations of the best performing methods on 1B experiments, i.e., Vamana, HNSW, and ELPIS against optimized methods from the ParlayANN library~\cite{parlayann} (Vamana\_Opt, HNSW\_Opt, and HCNNG\_Opt) on Deep1B.  
Figure~\ref{fig:optimized_impl} indicates that Vamana\_Opt and HNSW\_Opt are faster for recall below 0.97 compared to their original counterparts, due to more efficient data structures~\cite{parlayann, parlayanncode}. However, at higher recall, this advantage diminishes as distance computations dominate; 
HCNNG\_Opt is competitive with Vamana and HNSW, while ELPIS maintains a performance lead. 
}

%% file: src/discussion.tex
\section{Discussion}
\label{sec:discussion}
\karima{In the previous section, we presented the results of an extensive evaluation of twelve state-of-the-art graph-based vector search methods. Table~\ref{tab:comp} summarizes the evaluation across key criteria: 
for search, we assess efficiency, accuracy, and the number of tunable parameters; 
for indexing, we evaluate efficiency at high recall, memory footprint, and parameter tuning complexity. }

\karima{The best-performing methods, HNSW, VAMANA, and ELPIS, have the best search performance and index efficiency. However, ELPIS requires an extra parameter during both indexing (leaf size) and search (nprobes), whereas VAMANA requires an extra parameter to tune during indexing (alpha). NSG and SSG exhibit efficient query performance, but their indexing capability is hindered because of their base graph EFANNA, which similarly to KGraph, is tedious to tune and suffers from high indexing time and footprint. Both SPTAG (BKT) and NGT show satisfactory performance during search; however, they do not scale well to large datasets and require more tuning compared to the best methods. The assessment of HCNNG is based on the optimized parlayNN implementation which has shown competitive performance on large-scale datasets.} 

We now summarize the key insights and pinpoint promising research directions.

\begin{table}[tb]
\begin{minipage}{0.9\textwidth}
\hspace{4cm}
$\checkmark$ Good \quad
$\sim$ Medium \quad
$\times$ Bad
\end{minipage} \\[2pt]
\captionsetup{justification=centering}
\resizebox{0.8\columnwidth}{!}{
\begin{tabular}{lccc|ccc}
\toprule
\textbf{Method} & \multicolumn{3}{c|}{\textbf{Query Answering}} & \multicolumn{3}{c}{\textbf{Index Building}} \\
                & \textbf{Efficiency} & \textbf{Accuracy} & \textbf{Tuning} & \textbf{Efficiency} & \textbf{Footprint} & \textbf{Tuning} \\
\midrule
HNSW     & $\checkmark$ & $\checkmark$ & $\checkmark$ & $\checkmark$ & $\checkmark$ & $\checkmark$ \\
ELPIS    & $\checkmark$ & $\checkmark$ & $\sim$       & $\checkmark$ & $\checkmark$ & $\sim$ \\
VAMANA   & $\checkmark$ & $\checkmark$ & $\checkmark$ & $\checkmark$ & $\checkmark$ & $\sim$ \\
NSG      & $\checkmark$ & $\checkmark$ & $\checkmark$ & $\sim$       & $\sim$       & $\sim$ \\
SSG      & $\checkmark$ & $\checkmark$ & $\checkmark$ & $\sim$       & $\sim$       & $\sim$ \\
EFANNA   & $\times$     & $\sim$       & $\times$     & $\times$     & $\times$     & $\times$ \\
KGRAPH   & $\times$     & $\times$     & $\times$     & $\times$     & $\times$     & $\times$ \\
DPG      & $\times$     & $\sim$       & $\sim$       & $\sim$       & $\sim$       & $\sim$ \\
SPTAG    & $\sim$       & $\checkmark$ & $\times$     & $\times$     & $\checkmark$ & $\times$ \\
HCNNG    & $\checkmark$ & $\checkmark$ & $\checkmark$ & $\checkmark$ & $\checkmark$ & $\sim$ \\
LSHAPG   & $\times$     & $\sim$       & $\times$     & $\sim$       & $\checkmark$ & $\checkmark$ \\
NGT      & $\sim$       & $\sim$       & $\times$     & $\times$     & $\checkmark$ & $\times$ \\
SPTAG    & $\sim$       & $\sim$       & $\sim$       & $\times$     & $\checkmark$ & $\times$ \\
\bottomrule
\end{tabular}
}
\centering
\caption{\karima{Comparative Analysis}}
   \vspace*{-0.2cm}

\label{tab:comp}
\end{table}

\noindent{\bf Unexpected Results.} Our results lead to interesting observations that warrant further study. 

\noindent
\textit{(1) Stacked NSW:} while hierarchical layers of NSW graphs have shown promise in improving search performance on billion-scale datasets (Figure~\ref{fig:ss:search}), our experiments demonstrate that a simpler approach like K-random sampling can achieve better results on smaller and medium-sized datasets. 

\noindent
\textit{(2) Scalability of Graph Approaches:} while all graph-based vector search methods can efficiently build indexes on small datasets, most approaches face significant scalability challenges. 
Some methods (SPTAG, NGT, NSG, and SSG) demonstrate
impressive search performance on 1M and 25GB datasets (Figs. ~\ref{fig:elpis:query:performance:1M:sift:10NN}, ~\ref{fig:elpis:query:performance:1M:deep:10NN}, ~\ref{fig:elpis:query:performance:25GB:seismic:10NN},~\ref{fig:elpis:query:performance:25GB:deep:10NN}, and ~\ref{fig:elpis:query:performance:25GB:sald:10NN}) but their index construction could not scale to 100GB and billion-scale datasets. An important research direction is to improve the indexing scalability for these methods either by adopting summarization techniques during index construction or by using a scalable data structure to construct the base graph (i.e IVFPQ~\cite{faiss} to find the neighbors of nodes during insertion). 

\noindent
\textit{(3) DC-based approach for hard datasets and workloads:} an interesting finding was the superior performance of DC-based approaches compared to other methods like HNSW, NSG, and Vamana on challenging datasets/workloads such as Seismic, RandPow0, RandPow50 and Deep hard query workload for 1M and 25GB dataset sizes. We believe the DC strategy helps in this context because the graphs are built on clustered subsets of data, which facilitates beam search in retrieving more accurate nearest neighbors (NN), as opposed to running the search on the entire dataset, which results in lower accuracy (Figures~\ref{fig:elpis:query:performance:1M:seismic:10NN}, \ref{fig:elpis:query:performance:25GB:seismic:10NN}, \ref{fig:query:performance:25GB:rand:pow1:10NN}, and~\ref{fig:query:performance:25GB:rand:pow50:10NN}).
\karima{The optimized implementations from~\cite{parlayann} exhibit faster query times during search, but the gap narrows down with high recall. 
}

\noindent{\bf Neighborhood Diversification.} Adopting an ND strategy to sparsify the graph \textit{always} leads to better search performance, particularly as the dataset size grows

(Figures \ref{fig:elpis:query:performance:1M:seismic:10NN}, \ref{fig:elpis:query:performance:25GB:seismic:10NN}). 
Our experiments also show that RND and MOND lead to the best search performance overall (Figure \ref{fig:ND:search:real}). Nevertheless, we believe there is significant room for improvement in this direction. Besides, we argue that further theoretical studies are necessary to better understand the trade-offs between proximity and sparsity, and thus build graph structures 
 that are efficient during search and maintain good connectivity.

\noindent{\bf Seed Selection:} Our experiments demonstrate that the SS strategy plays a crucial role in enhancing not only search performance (Figure~\ref{fig:ss:search}) but also indexing efficiency (Table \ref{tab:ss:idx}). An important research direction is to develop novel, lightweight SS strategies. Such strategies could significantly improve the overall performance of graph-based vector search, both in terms of indexing and query-answering. Additionally, they could enhance the ability to handle out-of-distribution queries, particularly for large datasets where efficient seed selection becomes even more critical (Figures \ref{fig:ss:deep1b}, \ref{fig:ss:sift1b}).

\noindent{\bf Data-Adaptive Techniques.} Our experiments evaluate the performance of various graph-building paradigms within our taxonomy (SS, NP, II, ND, and DC). While NP-based methods perform the worst overall and are the least scalable, there is no clear winner across all dataset sizes and query workloads. \textit{(1) Scalability:} II-based approaches have superior efficiency during indexing and higher scalability in both querying and indexing.
\textit{(2) Query Answering:} ND-based methods have the best query performance overall. 
Meanwhile, DC-based approaches are superior on challenging datasets (High LID\&Low RC, Fig. \ref{fig:datacomp}) and hard query workloads \karima{(Fig. \ref{fig:elpis:query:performance:1B:t2i:10NN},
\ref{fig:elpis:query:performance:1M:seismic:10NN}, 
\ref{fig:elpis:query:performance:25GB:seismic:10NN}, 
\ref{fig:search:query:performance:25GB:hard},
\ref{fig:query:performance:25GB:rand:pow50:10NN},
\ref{fig:query:performance:25GB:rand:pow1:10NN})}. 
A promising research direction would be to develop techniques that adapt to dataset characteristics such as dataset size, dimensionality, RC and LID to excel both in indexing and query answering across a variety of query workloads and dataset sizes.

\noindent{\bf Hybrid Design.}  Most recent methods use a mix of paradigms. HNSW leverages II to scale index construction to large datasets and ND to support efficient query answering. ELPIS incorporates a DC-based strategy during both index building and search to further enhance the scalability of HNSW across varying dataset difficulty levels. Interestingly, Vamana, relying only on the ND paradigm, achieves good search performance and scalability, however its indexing time is prohibitive. 
A promising research direction is building hybrid approaches that combine the key strengths of different techniques, particularly II, ND and DC. Besides, devising novel base graphs, clustering and summarization techniques tailored for DC-based methods can further improve their performance.

\karima{\noindent{\bf Optimized Libraries.}  
Our experiments with the parlayNN library~\cite{parlayann} (Fig. \ref{fig:optimized_impl}) indicate the importance of such work. 
For instance, the parlayNN HCNNG\_Opt implementation was scalable to 1B datasets, whereas the non-optimized version could not scale beyond 25GB. 
Other methods could benefit from such optimizations to scale to large datasets, and further efforts are needed by the community to adopt and support libraries such as parlayNN.}

\begin{figure}[tb]
  \captionsetup{justification=centering}
    \includegraphics[width=0.55\columnwidth]{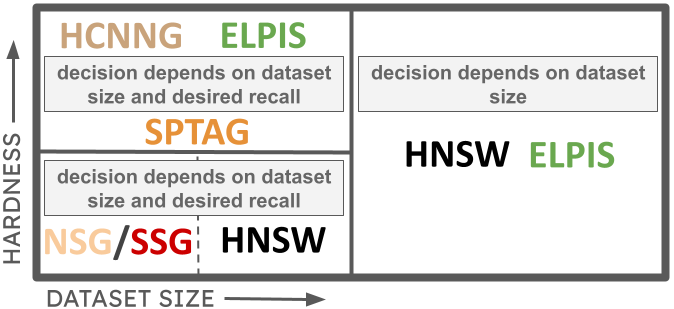}
    \vspace*{-0.2cm}
    \caption{\karima{Recommendations (Indexing + 10K queries)} }
   \vspace*{-0.2cm}
    \label{fig:recgann}
\end{figure}

\karima{\noindent{\bf Recommendations.}
Our study demonstrates varying performance trends across datasets of different sizes, query workloads of different hardness and desired recall values. Figure~\ref{fig:recgann} provides recommendations for methods based on these criteria.  For small to medium-sized datasets (25GB and below), HNSW, NSG, and its improved version, SSG, consistently demonstrate excellent performance on easier datasets (Fig.\ref{fig:elpis:query:performance:1M:deep:10NN}, \ref{fig:elpis:query:performance:1M:sift:10NN}, \ref{fig:elpis:query:performance:1M:imagenet:10NN}, \ref{fig:elpis:query:performance:1M:gist:10NN}). On harder datasets, DC-based methods like SPTAG, ELPIS, and HCNNG prove more efficient (Figs. \ref{fig:elpis:query:performance:1M:seismic:10NN} \ref{fig:elpis:query:performance:25GB:seismic:10NN}, \ref{fig:elpis:query:performance:1M:sald:10NN}, \ref{fig:elpis:query:performance:25GB:sald:10NN}, \ref{fig:search:query:performance:25GB:hard:1p}, \ref{fig:search:query:performance:25GB:hard:10p}). 
On large datasets (100GB and above), HNSW and ELPIS consistently rank as top choices (Figs.\ref{fig:elpis:query:performance:100GB}, \ref{fig:elpis:query:performance:1B}). 
}

%% file: src/conclusions.tex
\section{Conclusions}
\label{sec:conclusions}
In this paper, we conduct a survey of the SotA graph-based methods for in-memory $ng$-approximate vector search, proposing a new taxonomy based on five key design paradigms.
The chronological development and inter-method influence are outlined, along with an evaluation of key design choices like seed selection and neighborhood diversification. 

Through extensive experimentation on datasets with up to 1B vectors, we highlight the scalability challenges faced by most methods, with incremental insertion methods showing the best scalability on datasets exceeding 100GB. 
We observe that light-weight hierarchical structures help select better seeds to start the search on billion-scale datasets, and that neighborhood diversification is a key contributor in improving the query answering performance, with RND and MOND being the best techniques. 
We also propose promising research directions.